\documentclass[11pt]{article}
\usepackage{a4}
\usepackage{ifthen,array}
\newcommand{\ams}{\usepackage{amsfonts,amssymb,amsmath}}

\ams
\allowdisplaybreaks[3]
\newlength{\textwidthorig}
\newlength{\oddsidemarginorig}
\newlength{\textheightorig}
\newlength{\topmarginorig}
\def\seitenlaengenabsolut#1 #2 #3 #4 {\setlength{\textwidth}{#1}
                                      \setlength{\oddsidemargin}{#2}
                                      \setlength{\textheight}{#3}
                                      \setlength{\topmargin}{#4}}
\def\seitenlaengenrelzustandard#1 #2 #3 #4 {\setlength{\textwidth}{\textwidthorig+#1}
                                            \setlength{\oddsidemargin}{\oddsidemarginorig+#2}
                                            \setlength{\textheight}{\textheightorig+#3}
                                            \setlength{\topmargin}{\topmarginorig+#4}}
\def\seitenlaengenrelzuvorher#1 #2 #3 #4 {\addtolength{\textwidth}{#1}
                                          \addtolength{\oddsidemargin}{#2}
                                          \addtolength{\textheight}{#3}
                                          \addtolength{\topmargin}{#4}}
\newcommand{\standardseite}{\seitenlaengenrelzuvorher2.2cm -0.8cm 1.8cm -1.5cm }   
\standardseite
\newcommand{\Wegdamit}[1]{}

\newlength{\laengespatium}

\newcommand{\nach}{\longrightarrow}      

\newcommand{\auf}{\longmapsto}           
\newcommand{\txtauf}[1]{\auf}            
\newcommand{\impliz}{\Longrightarrow}    
\newcommand{\aequ}{\Longleftrightarrow}  
 
\newcommand{\invimpliz}{\Longleftarrow}  
\newcommand{\gegen}{\rightarrow}         
\newcommand{\konvrunter}{\downarrow}     
\newcommand{\iso}{\cong}                 
\newcommand{\ident}{\equiv}              
\newcommand{\teilmenge}{\subseteq}       
\newcommand{\obermenge}{\supseteq}       
\newcommand{\aeqrel}{\sim}               
\newcommand{\rund}{\approx}              

\newcommand{\fueralle}{\hspace{1.7em}\forall}
\newcommand{\leeremenge}{\varnothing}    
\newcommand{\tensor}{\otimes}            
\newcommand{\kreuz}{\times}              

\newcommand{\keil}{\wedge}               
\newcommand{\dirsum}{\oplus}           

\newcommand{\bigtensor}{\bigotimes}      
\newcommand{\kp}{\odot}                  
\newcommand{\betraganpass}[1]%
           {\left| #1 \right|}           
\newcommand{\bigbetrag}[1]%
           {\bigl|{#1}\bigr|}            
\newcommand{\betrag}[1]%
           {|{#1}|}                      
\newcommand{\betragnichtanpass}[1]%
           {\mid #1 \mid}                
\newcommand{\norm}[1]%
           {{}{\parallel}#1{\parallel}{}}      
\newcommand{\erww}[1]%
           {\langle #1 \rangle}          
\newcommand{\skalprod}[2]%
           {\langle #1,#2 \rangle}       
\newcommand{\supnorm}[1]{{\norm{#1}_\infty}}        
\newcommand{\quer}{\overline}            
\newcommand{\dach}{\widehat}             
\newcommand{\inv}[1]{\frac{1}{#1}}       
\newcommand{\einhalb}{\inv{2}}           
\newcommand{\einsdurchn}{\inv{N}}        
\newcommand{\re}{\text{Re }}                           
\newcommand{\im}{\text{im\;}}                          
\newcommand{\tr}{\text{tr}}                           
\newcommand{\spann}{\text{span}}                       
\newcommand{\supp}{\text{supp }}                       
\newcommand{\vol}{\text{vol}\,}                        
\newcommand{\Ad}{{\text{Ad}}}                          
\newcommand{\elanz}{\#}                                
\newcommand{\del}{\partial}                            
\newcommand{\Hom}{\text{Hom}}                          
\newcommand{\Maps}{\text{Maps}}                        
\newcommand{\dd}{\text{d}}                             
\newcommand{\e}{\text{e}}                              
\newcommand{\I}{\text{i}}                              
\newcommand{\NULL}{\mathbf{0}}                         
\newcommand{\EINS}{{\boldsymbol{1}}}                   
\newcommand{\field}[1]{\mathbb{#1}}                    
\newcommand{\K}{{\field{K}}}                           
\newcommand{\C}{{\field{C}}}                           
\newcommand{\N}{{\field{N}}}                           
\newcommand{\R}{{\field{R}}}                           
\newcommand{\Z}{{\field{Z}}}                           
\newcommand{\Gl}{\text{Gl}}                            
\newcommand{\gl}{\mathfrak{gl}}                        
\newcommand{\rnkl}[2]{\raisebox{-0.4ex}{$#1$}%
\raisebox{-0.12ex}{{\large$\setminus$}}\,#2}   
\newcommand{\agb}{{\overline{{\cal A}/{\cal G}}}}      
\newcommand{\agbfact}[1][]{\text{$\agb/\!\aeqrel$}}    
\newcommand{\ag}{{\cal A}/{\cal G}}                    
\newcommand{\Ab}{{\overline{{\cal A}}}}                
\newcommand{\A}{{\cal A}}                              
\newcommand{\Gb}{{\overline{{\cal G}}}}                

\newcommand{\AbGb}{{\Ab/\Gb}}                          
\newcommand{\gen}{{\text{gen}}}
\newcommand{\Abgen}{{\Ab_\gen}}                        
\newcommand{\AbGbgen}{{\Abgen/\Gb}}                    
\newcommand{\abbAbGbagb}{{\boldsymbol \phi}}           
\newcommand{\qa}{{\quer{A}}}                           
\newcommand{\qg}{{\quer{g}}}                           
\newcommand{\hg}{{\cal HG}}                            
\newcommand{\ha}{{\cal HA}}                            

\newcommand{\holgr}{{\mathbf H}}                       
\newcommand{\bz}{{\mathbf B}}                          

\newcommand{\Cyl}{\text{Cyl}}                          

\newcommand{\GR}{\Gamma}                               
\newcommand{\Ver}{\mathbf{V}}                          
\newcommand{\Edg}{\mathbf{E}}                          
\newcommand{\gross}[1]{{\boldsymbol #1}}               
\newcommand{\ga}{\gross{\alpha}}                       
\newcommand{\gb}{\gross{\beta}}                        
\newcommand{\gc}{\gross{\gamma}}                       
\newcommand{\FW}{{\cal F}}                             
\newcommand{\innen}{{\text{innen}}}                    
\newcommand{\Pf}{{\cal P}}                             
\newcommand{\Haar}{{\text{Haar}}}                      
\newcommand{\LG}{{\mathbf{G}}}                         
\newcommand{\LN}{{\mathbf{N}}}                         
\newcommand{\Lieg}{{\mathfrak{g}}}                            
\newcommand{\aeqrelzush}[1][]{\sim}                    
\newcommand{\kopp}{{\text{g}}}                         

\newcommand{\nklza}[1][]{\ifthenelse{\equal{#1}{}}     
                                    {\rnkl{Z(\holgr_\qa)}{\LG}}        
                                   {\rnkl{Z(\holgr_{#1})}{\LG}}}       
\newcommand{\nkla}[1][]{\ifthenelse{\equal{#1}{}}      
                                    {\rnkl{\bz(\qa)}{\Gb}}        
                                    {\rnkl{\bz(#1)}{\Gb}}}       
\newcommand{\he}{{\text{he}}}                          
\newcommand{\ab}{{\text{ab}}}                          

\newcommand{\lzweihaar}[2]{(#1,#2)_{\Haar}}
\newcommand{\lzweihaarnorm}[1]{{{}{\parallel}#1{\parallel}_{\Haar}{}}}
\newcommand{\lzweispez}[3]{(#1,#2)_{#3}}
\newcommand{\lzweihaarn}[3][n]{\lzweispez{#2}{#3}{\Haar,#1}}
\newcommand{\killing}{\kappa}                          
\newcommand{\kmt}{\kappa}
\newcommand{\pskilling}{{\lambda}}                     
\newcommand{\natform}{{\varkappa}}                     
\newcommand{\natmt}{{\varkappa}}                       
\newcommand{\lapl}{\Delta}                             
\newcommand{\ueinsbasis}{{e}}                          
\newcommand{\darst}{{\phi}}                            
\newcommand{\allirrdarst}[2][]{{\cal D}_{#1}(#2)}      
\newcommand{\charakt}{\chi}                            
\newcommand{\wurzs}{{\Sigma}}                          
\newcommand{\vz}{{\vec z}}                             
\newcommand{\vnu}{{\vec{\mathbf{n}}}}                  
\newcommand{\vn}{{\vec n}}                             

\newcommand{\kmo}[2]{(#1,#2)^\kmt}                     
\newcommand{\kmonorm}[1]{{{}{\parallel}#1{\parallel}^{\kmt}{}}}
\newcommand{\hgew}{\Lambda}                            
\newcommand{\vr}{{\vec r}}                             
\newcommand{\matfkt}{{\cal M}}                         
\newcommand{\allchar}{\matfkt_\Ad}                     
\newcommand{\nmatf}{T}                                 
\newcommand{\nmatfkt}{\matfkt^n}                       
\newcommand{\nchar}{T}                                 
\newcommand{\ncharfkt}{\matfkt_\Ad^n}                  
\newcommand{\kontr}{{\cal C}}                          
\newcommand{\VR}{{V}}                                  
\newcommand{\VRbasis}{{B}}                             
\newcommand{\algnorm}[1]{\norm{#1}_\bullet}            
\newcommand{\lnws}{T}                                  
\newcommand{\lnwss}{t}                                 
\newcommand{\alllnws}{{\cal L}}                        
\newcommand{\alllnwss}{{\cal L}}                       
\newcommand{\FI}[1]{\betrag{#1}}                       
\newcommand{\FIL}[1]{\betrag{G_{#1}}}                  
\newcommand{\FIG}[2]{{G_{#1,#2}}}                      
\newcommand{\rh}{n}                                    
\newcommand{\vg}{\vec g}
\newcommand{\YM}{{\text{YM}}}                          
\newcommand{\YMueberschrift}{{\mathrm{YM}}}
\newcommand{\ymwirk}[1][]{\ifthenelse{\equal{#1}{}}{S_{\YM}}{S_{\YM,#1}}}
\newcommand{\ymd}{{\chi}}                              
\newcommand{\erw}{{\mathbf{E}}}                        
\newcommand{\vx}{{\vec x}}                             
\newcommand{\wuerfel}[3]{\text{W}^{#1}_{#3,#2}}
\newcommand{\charfkt}[1]{1_{#1}}                       
\newcommand{\karte}{\chi}
\renewcommand{\karte}{{\kappa}}             

\newcommand{\bmat}{\begin{pmatrix}}
\newcommand{\emat}{\end{pmatrix}}

\newcommand{\const}{\text{const}}                      
\newcommand{\laufi}{\nu}                               
\newcommand{\veci}{\vec\imath}                         
\newcommand{\vecj}{\vec\jmath}                         
\newcommand{\klammerunten}[2]{\underbrace{#1}_{\textstyle#2}}

\newcommand{\ListNullAbstaende}{\setlength{\topsep}{0pt}%
                                \setlength{\parskip}{0pt}%
                                \setlength{\partopsep}{0pt}%
                                \setlength{\itemsep}{0pt}%
                                \setlength{\parsep}{0pt}}
\newcommand{\ListNurAnstrichAbstand}{\setlength{\topsep}{0pt}%
                                     \setlength{\parskip}{0pt}%
                                     \setlength{\partopsep}{0pt}%
                                     \setlength{\parsep}{0pt}}
\newenvironment{StandardListe}[2]%
               {\begin{list}%
                      {#1}%
                      {\settowidth{\leftmargin}{M#1}%
                       \settowidth{\labelwidth}{#1}%
                       \settowidth{\labelsep}{M}%
                       #2%
                      }%
                }%
               {\end{list}}%
\newenvironment{EinfachListe}[1]%
               {\begin{StandardListe}{#1}{\ListNullAbstaende}}%
               {\end{StandardListe}}%
               {\begin{StandardListe}{#1}{\ListNurAnstrichAbstand}}%
               {\end{StandardListe}}%
\newcommand{\labelsatz}[1]{#1}
\newcounter{listennr}                      
\newlength{\hilfslaenge}
\newlength{\stdlabellaenge}
\newlength{\maximum}
\newcommand{\stdlabel}{}
\newcommand{\Maximum}{}
\newcommand{\iitem}[1][]{\ifthenelse{\equal{#1}{}}%
                           {\item \setlength{\hilfslaenge}{\stdlabellaenge}}%
                           {\item[\labelsatz{#1}\hfill]%
                            \settowidth{\hilfslaenge}{\labelsatz{#1}}}%
                         \ifthenelse{\lengthtest{\maximum < \hilfslaenge}}%
                           {\setlength{\maximum}{\hilfslaenge}%
                            \ifthenelse{\equal{#1}{}}%
                               {\renewcommand{\Maximum}{\stdlabel}}%
                               {\renewcommand{\Maximum}{#1}}}%
                           {}%
                      }      
\makeatletter
\newenvironment{AutoLabelLaengenListe}[2][]%
               {\begin{list}%
                      {\labelsatz{#1}\hfill}%
                      {\stepcounter{listennr}%
                       \settowidth{\leftmargin}{M\labelsatz{\ref{listnr\arabic{listennr}}}}%
                       \settowidth{\labelwidth}{\labelsatz{\ref{listnr\arabic{listennr}}}}%
                       \settowidth{\labelsep}{M}%
                       \settowidth{\stdlabellaenge}{\labelsatz{#1}}%
                       \renewcommand{\stdlabel}{#1}%
                       #2%
                       \renewcommand{\Maximum}{}%
                      }%
                }%
               {\renewcommand{\@currentlabel}{\Maximum}%
                \label{listnr\arabic{listennr}}%
                \end{list}%
                }%
\makeatother
\newenvironment{StandardEinrueckung}[2]%
               {\begin{list}%
                      {#1}%
                      {\settowidth{\leftmargin}{M#1}%
                       \settowidth{\labelwidth}{#1}%
                       \settowidth{\labelsep}{M}%
                       #2%
                      }%
                \item}%
               {\end{list}}%
\newenvironment{Einrueckungpur}[1]%
               {\begin{StandardEinrueckung}{#1}{\ListNullAbstaende}}%
               {\end{StandardEinrueckung}}%
\newenvironment{Einrueckung}[1]%
               {\begin{StandardEinrueckung}{#1}{\setlength{\parsep}{0pt}}}%
               {\end{StandardEinrueckung}}%
\newcommand{\EineZeileGleichung}[2][0.0ex]
           {
            
            \vspace{#1} 
            \noindent
            \hspace*{\fill}
            $\displaystyle{#2}$
            \hspace*{\fill}

            \vspace{#1} 
            
           }

\makeatletter
\newcommand{\EineNumZeileGleichung}[2][0.5ex]
           {
            
            \vspace{#1} 
            \noindent
            \stepcounter{equation}
            \renewcommand{\@currentlabel}{\arabic{equation}}%
            \phantom{(\arabic{equation})}\hspace*{\fill}
            $\displaystyle{#2}$
            \hspace*{\fill}
            (\arabic{equation})

            \vspace{#1} 
            
           }
\makeatother
\makeatletter
\newcommand{\EineErwNumZeileGleichung}[2][0.5ex]
           {
            
            \vspace{#1} 
            \noindent
            \stepcounter{equation}
            \renewcommand{\@currentlabel}{\arabic{equation}}%
            \phantom{(\arabic{equation})}\hspace*{\fill}
            #2 
            \hspace*{\fill}
            (\arabic{equation})

            \vspace{#1} 
            
           }
\makeatother
\newcommand{\breitrel}[1]{\hspace*{\tabcolsep} #1 \hspace*{\tabcolsep}}
\newlength{\abstaug}              
\newenvironment{AllgUnnumGleichung}[2][1.0ex]
               {
  
                \setlength{\abstaug}{#1}
                \vspace{\abstaug}
                \hspace*{\fill}
                $\begin{array}[t]{#2}
                }%
               {\end{array}$
                \hspace*{\fill}
  
                \vspace{\abstaug}

                }%
\newenvironment{AllgNumGleichung}[2][0.0ex]
               {
  
                \setlength{\abstaug}{#1}
                \vspace{\abstaug}
                $\begin{tabular*}{\textwidth}[t]{#2}
                }%
               {\end{tabular*}$

                \vspace{\abstaug}

               }%
\newenvironment{StandardUnnumGleichungKlein}[1][0ex]
               {\renewcommand{\s}{\\[#1] }%
                \begin{AllgUnnumGleichung}{rcl}}%
               {\end{AllgUnnumGleichung}}%
\newenvironment{StandardUnnumGleichung}[1][0ex]
               {\renewcommand{\s}{\\[#1] }%
                \begin{AllgUnnumGleichung}{>{\displaystyle}rc>{\displaystyle}l}}%
               {\end{AllgUnnumGleichung}}%
\newenvironment{XrelYZNumGleichung}[1][0ex]
               {\renewcommand{\s}{\\[#1] }%
                \begin{AllgNumGleichung}{rcll}}%
               {\end{AllgNumGleichung}}%
\newcommand{\pktplatz}{\phantom{.}}            
\newcommand{\erl}[1]{\hfill\mbox{\hspace*{1.5em}\small (#1)}}

\newcommand{\erllang}[2][0.5\textwidth]%
              {\hfill\hspace*{1.5em}%
               \begin{minipage}[t]{#1}{\small%
                          \begin{list}{(}{\ListNullAbstaende%
                                          \settowidth{\leftmargin}{(}%
                                          \settowidth{\labelwidth}{(}%
                                          \settowidth{\labelsep}{}%
                                         }%
                          \item#2)%
                          \end{list}}%
               \end{minipage}\\[-0.9ex]
              }%
\newcommand{\DefBemUmgeb}[1]
           {\newenvironment{#1}[1][]%
                           {\begin{Einrueckung}{{\bf #1}}%
                            \ifx##1\empty\else{{\bf ##1}
                            
                                                        }\fi%
                            }%
                           {\end{Einrueckung}}}
\newcommand{\DefSBemUmgeb}[2]
           {\newenvironment{#1}[1][]%
                           {\begin{Einrueckung}{{\bf #2}}%
                            \ifx##1\empty\else{{\bf ##1}
                            
                                                        }\fi%
                            }%
                           {\end{Einrueckung}}}
\makeatletter
\newcommand{\DefBspUmgeb}[3]
           {\newcounter{#2}[#3]%
            \newenvironment{#1}[1][]%
                           {\stepcounter{#2}%
                            \renewcommand{\ZaehlerMarke}{\arabic{#2}}%
                            \renewcommand{\Einzugsname}{{\bf #1 \ZaehlerMarke}}%
                            \begin{Einrueckung}{\Einzugsname}
                            \ifx##1\empty\else{{\bf ##1}\\}\fi%
                            \renewcommand{\@currentlabel}{\ZaehlerMarke}%
                            }%
                           {\end{Einrueckung}}}
\makeatother
\newcommand{\ZaehlerbisEbene}{section}
\newcommand{\Ebenea}{section}
\newcommand{\Ebeneb}{subsection}

\newcommand{\Abschnittnummer}{%
            \ifx\ZaehlerbisEbene\Ebenea{\arabic{section}}%
             \else{%
              \ifx\ZaehlerbisEbene\Ebeneb{\arabic{section}.\arabic{subsection}}%
               \else{\arabic{section}.\arabic{subsection}.\arabic{subsubsection}}%
              \fi}%
            \fi}     
\newcommand{\Abschnittnummerpunkt}{\Abschnittnummer.}     
\newcommand{\Einzugsname}{}
\newcommand{\ZaehlerMarke}{}
\makeatletter
\newcommand{\DefThmUmgeb}[3]
           {\newcounter{#1}[#3]%
            \newenvironment{#1}[1][]%
                           {\stepcounter{#2}%
                            \setcounter{#1}{\value{#2}}%
                            \renewcommand{\ZaehlerMarke}{\Abschnittnummerpunkt\arabic{#1}}%
                            \renewcommand{\Einzugsname}{{\bf #1 \ZaehlerMarke}}%
                            \begin{Einrueckung}{\Einzugsname}
                            \ifx##1\empty\else{{\bf ##1}
                            
                                                        }\fi%
                            \renewcommand{\@currentlabel}{\ZaehlerMarke}%
                            }%
                           {\end{Einrueckung}}}
\makeatother
\makeatletter
\newcommand{\DefSThmUmgeb}[4]
           {\newcounter{#1}[#3]%
            \newenvironment{#1}[1][]%
                           {\stepcounter{#2}%
                            \setcounter{#1}{\value{#2}}%
                            \renewcommand{\ZaehlerMarke}{\Abschnittnummerpunkt\arabic{#1}}%
                            \renewcommand{\Einzugsname}{{\bf #4 \ZaehlerMarke}}
                            \begin{Einrueckung}{\Einzugsname}
                            \ifx##1\empty\else{{\bf ##1}

                                                        }\fi%
                            \renewcommand{\@currentlabel}{\ZaehlerMarke}%
                            }%
                           {\end{Einrueckung}}}
\makeatother
\makeatletter
\newcommand{\DefUnterNumThmUmgeb}[5]
           {\newcounter{#1}[#3]%
            \newcounter{#4}%
            \newenvironment{#1}[1][]%
                           {\ifx##1\empty\else{\stepcounter{#2}\setcounter{#4}{0}}\fi%
                            \stepcounter{#4}%
                            \setcounter{#1}{\value{#2}}%
                            \renewcommand{\ZaehlerMarke}{\Abschnittnummerpunkt\arabic{#1}\alph{#4}}%
                            \renewcommand{\Einzugsname}{{\bf #5 \ZaehlerMarke}}
                            \begin{Einrueckung}{\Einzugsname}
                            \renewcommand{\@currentlabel}{\ZaehlerMarke}%
                            }%
                           {\end{Einrueckung}}}
\makeatother
\newenvironment{Beweis}[1][]%
               {\begin{Einrueckung}{{\bf Beweis}}%
                \ifx#1\empty\else{{\bf #1}

                                            }\fi%
                }%
               {\end{Einrueckung}%
                }%
\newenvironment{Proof}[1][]%
               {\begin{Einrueckung}{{\bf Proof}}%
                \ifx#1\empty\else{{\bf #1}

                                            }\fi%
                }%
               {\end{Einrueckung}%
                }%
               {\begin{Einrueckung}{{\bf \glqq Beweis\grqq}}%
                \ifx#1\empty\else{{\bf #1}
                
                                            }\fi%
                }%
               {\end{Einrueckung}%
                }%
               {\begin{Einrueckung}{{\bf Begr"undung}}%
                \ifx#1\empty\else{{\bf #1}
                
                                            }\fi%
                }%
               {\end{Einrueckung}%
                }%
\newenvironment{Hinrichtung}%
               {\begin{Einrueckungpur}{$\impliz$}}%
               {\end{Einrueckungpur}}%
\newenvironment{Rueckrichtung}%
               {\begin{Einrueckungpur}{$\invimpliz$}}%
               {\end{Einrueckungpur}}%
               {\begin{Einrueckungpur}{\glqq$\teilmenge$\grqq}}%
               {\end{Einrueckungpur}}%
               {\begin{Einrueckungpur}{\glqq$\obermenge$\grqq}}%
               {\end{Einrueckungpur}}%
               {\begin{Einrueckungpur}{"$\teilmenge$"}}%
               {\end{Einrueckungpur}}%
               {\begin{Einrueckungpur}{"$\obermenge$"}}%
               {\end{Einrueckungpur}}%
\newcommand{\qed}{\nopagebreak\hspace*{2em}\hspace*{\fill}{\bf qed}}
\newcommand{\ARabic}{\arabic}
\newcommand{\Nummerntypa}{\arabic}   
\newcommand{\Nummerntypb}{\alph}
\newcommand{\Nummerntypc}{\roman}
\newcommand{\Nummerntypd}{\Alph}

\newcommand{\Nra}{\Nummerntypa{Nummera}}            
\newcommand{\Nrb}{\Nummerntypb{Nummerb}}            
\newcommand{\Nrc}{\Nummerntypc{Nummerc}}                
\newcommand{\Nrd}{\Nummerntypd{Nummerd}}                
\newcommand{\ZeichenzuNrTyp}[1]%
           {\ifx#1\ARabic {.}\else{)}%
                  \fi}                              
\newcommand{\NrZeicha}{\ZeichenzuNrTyp{\Nummerntypa}}
\newcommand{\NrZeichb}{\ZeichenzuNrTyp{\Nummerntypb}}
\newcommand{\NrZeichc}{\ZeichenzuNrTyp{\Nummerntypc}}
\newcommand{\NrZeichd}{\ZeichenzuNrTyp{\Nummerntypd}}
\newcommand{\ListMarkea}%
           {\Nra\NrZeicha}
\newcommand{\ListMarkeb}%
           {\Nra\NrZeicha\Nrb\NrZeichb}
\newcommand{\ListMarkec}%
           {\Nra\NrZeicha\Nrb\NrZeichb\Nrc\NrZeichc}
\newcommand{\ListMarked}%
           {\Nra\NrZeicha\Nrb\NrZeichb\Nrc\NrZeichc\Nrd\NrZeichd}
\newcommand{\Anfangszeichen}{}
\newcommand{\Anfangspunkt}{}
\newcounter{Schachtelebene}
\newcounter{Hilfszaehler}
\newcommand{\Hilfsbefehl}{}
\newcommand{\Schachtelebene}{\alph{Schachtelebene}}
\makeatletter
\newenvironment{AllgNumerierteListe}[2][]
               {\addtocounter{Schachtelebene}{1}%
		\setcounter{Hilfszaehler}{#2}%
                \renewcommand{\Anfangszeichen}%
                             {\renewcommand{\Hilfsbefehl}{\csname Nummerntyp\Schachtelebene \endcsname}%
                              \Hilfsbefehl{Hilfszaehler}}%
                \renewcommand{\Anfangspunkt}%
                             {\csname NrZeich\Schachtelebene \endcsname}%
                \begin{list}%
                      {\stepcounter{Nummer\Schachtelebene}%
                       \csname Nr\Schachtelebene \endcsname
                       \csname NrZeich\Schachtelebene \endcsname
                       }%
                      {\settowidth{\leftmargin}{M\Anfangszeichen\Anfangspunkt}%
                       \settowidth{\labelwidth}{\Anfangszeichen\Anfangspunkt}%
                       \settowidth{\labelsep}{M}%
                       \setlength{\topsep}{0pt}%
                       \setlength{\parskip}{0pt}%
                       \setlength{\partopsep}{0pt}%
                       \setlength{\itemsep}{0pt}%
                       \setlength{\parsep}{0pt}%
                      }%
                \renewcommand{\@currentlabel}{\csname ListMarke\Schachtelebene \endcsname}%
                }%
               {\ifthenelse{\equal{}{}}{\setcounter{Nummer\Schachtelebene}{0}}{}
                \addtocounter{Schachtelebene}{-1}%
                \end{list}}
\makeatother
\newenvironment{NumerierteListe}[1]
               {\begin{AllgNumerierteListe}{#1}}
               {\end{AllgNumerierteListe}}
\newenvironment{WeiterNumerierteListe}[1]
               {\begin{AllgNumerierteListe}[Weiter]{#1}}
               {\end{AllgNumerierteListe}}

\newcommand{\UnnumAnfangszeichen}{}
\newcounter{UnnumSchachtelebene}
\newcommand{\UnnumSchachtelebene}{\alph{UnnumSchachtelebene}}
\makeatletter
\newenvironment{UnnumerierteListe}%
               {\addtocounter{UnnumSchachtelebene}{1}%
                \renewcommand{\UnnumAnfangszeichen}%
                             {\csname UnnumZeich\UnnumSchachtelebene \endcsname}%
                \begin{list}%
                      {\UnnumAnfangszeichen}%
                      {\settowidth{\leftmargin}{M\UnnumAnfangszeichen}%
                       \settowidth{\labelwidth}{\UnnumAnfangszeichen}%
                       \settowidth{\labelsep}{M}%
                       \setlength{\topsep}{0pt}%
                       \setlength{\parskip}{0pt}%
                       \setlength{\partopsep}{0pt}%
                       \setlength{\itemsep}{0pt}%
                       \setlength{\parsep}{0pt}%
                      }%
                }%
               {\addtocounter{UnnumSchachtelebene}{-1}%
                \end{list}}
\makeatother
\newlength{\fktdefhilfslaenge}
\newcommand{\ohnefktdef}[4]
           {\hspace*{\fill}
            $\begin{array}[t]{ccc}%
            #1 & \nach & #2 \\
            #3 & \auf  & #4
            \end{array}$
            \hspace*{\fill}}
\newcommand{\fktdef}[5]
           {\hspace*{\fill}
            $\begin{array}[t]{cccc}%
            #1: & #2 & \nach & #3 \\    
                & #4 & \auf  & #5
            \end{array}$
            \settowidth{\fktdefhilfslaenge}{$#1$:}
            \hspace*{0.6 \fktdefhilfslaenge}  
            \hspace*{\fill}}
\newcommand{\fktdefpur}[5]
           {$\begin{array}[t]{cccc}%
            #1: & #2 & \nach & #3 \\    
                & #4 & \auf  & #5
            \end{array}$}
\newcommand{\fktdefabgesetztpur}[5]
           {
            
            $\begin{array}[t]{cccc}%
            #1: & #2 & \nach & #3 \\    
                & #4 & \auf  & #5
            \end{array}$
            \settowidth{\fktdefhilfslaenge}{$#1$:}
            \hspace*{0.6 \fktdefhilfslaenge}
            
           }
\newcommand{\fktdefabgesetzt}[5]
           {
           
            \hspace*{\fill}
            $\begin{array}[t]{cccc}%
            #1: & #2 & \nach & #3 \\    
                & #4 & \auf  & #5
            \end{array}$
            \settowidth{\fktdefhilfslaenge}{$#1$:}
            \hspace*{0.6 \fktdefhilfslaenge}  
            \hspace*{\fill}
            
            }
\newcommand{\ohnefktdefabgesetzt}[4]
           {      

            \hspace*{\fill}
            $\begin{array}[t]{ccc}%
            #1 & \nach & #2 \\
            #3 & \auf  & #4
            \end{array}$
            \hspace*{\fill}

            }
\newcommand{\doppelohnefktdefabgesetzt}[6]
           {

            \hspace*{\fill}
            $\begin{array}[t]{ccccc}%
            #1 & \nach & #2 & \nach & #3\\
            #4 & \auf  & #5 & \auf  & #6
            \end{array}$
            \hspace*{\fill}

            }
\newcommand{\anhang}%
           {\appendix
            \sectioninh{Anhang}
            \renewcommand{\Abschnittnummer}{%
                  \ifx\ZaehlerbisEbene\Ebenea{\Alph{section}}%
                  \else{%
                        \ifx\ZaehlerbisEbene\Ebeneb{\Alph{section}.\arabic{subsection}}%
                        \else{\Alph{section}.\arabic{subsection}.\arabic{subsubsection}}%
                        \fi}%
                  \fi}%
            \renewcommand{\Abschnittnummerpunkt}{\Abschnittnummer.}     
            }            
\newcommand{\anhangengl}%
           {\appendix
            \sectioninh{Appendix}
            \renewcommand{\Abschnittnummer}{%
                  \ifx\ZaehlerbisEbene\Ebenea{\Alph{section}}%
                  \else{%
                        \ifx\ZaehlerbisEbene\Ebeneb{\Alph{section}.\arabic{subsection}}%
                        \else{\Alph{section}.\arabic{subsection}.\arabic{subsubsection}}%
                        \fi}%
                  \fi}%
            \renewcommand{\Abschnittnummerpunkt}{\Abschnittnummer.}     
            }

\newcounter{wdhlstufe}
\newcommand{\sectioninh}[1]%
           {\section*{#1}%
            \addcontentsline{toc}{section}{#1}}
\newcommand{\bezeichnung}[3]
           {\begin{Einrueckungpur}{\hbox to 6em{#1}\hbox to 2.4em{\hfill#2}}
            #3
            \end{Einrueckungpur}}

\newcommand{\doppelteinfach}{e}

\newcommand{\ifdoppelt}[1]{\ifthenelse{\equal{\doppelteinfach}{d}}{#1}{}}
\newcommand{\ifeinfach}[1]{\ifthenelse{\equal{\doppelteinfach}{e}}{#1}{}}

\newlength{\querfhilfsl}              

\newlength{\hll}

\DefThmUmgeb{Theorem}{Theorem}{\ZaehlerbisEbene}
\DefThmUmgeb{Definition}{Definition}{\ZaehlerbisEbene}
\DefThmUmgeb{Satz}{Theorem}{\ZaehlerbisEbene}
\DefThmUmgeb{Proposition}{Theorem}{\ZaehlerbisEbene}
\DefThmUmgeb{Lemma}{Theorem}{\ZaehlerbisEbene}
\DefThmUmgeb{Folgerung}{Theorem}{\ZaehlerbisEbene}
\DefThmUmgeb{Corollary}{Theorem}{\ZaehlerbisEbene}
\DefThmUmgeb{Vorschrift}{Definition}{\ZaehlerbisEbene}
\DefThmUmgeb{Construction}{Definition}{\ZaehlerbisEbene}
\DefSThmUmgeb{FormSatz}{Theorem}{\ZaehlerbisEbene}{\glqq Satz\grqq} 
\DefThmUmgeb{Vermutung}{Theorem}{\ZaehlerbisEbene}
\DefThmUmgeb{Conjecture}{Theorem}{\ZaehlerbisEbene}
\DefThmUmgeb{Konvention}{Definition}{\ZaehlerbisEbene}
\DefThmUmgeb{Feststellung}{Theorem}{\ZaehlerbisEbene}
\DefUnterNumThmUmgeb{DefinitionZusatzNum}{Definition}{\ZaehlerbisEbene}{DefZN}{Definition}
\DefBspUmgeb{Beispiel}{Beispiel}{subsubsection}
\DefBspUmgeb{Example}{Example}{subsubsection}
\DefBspUmgeb{Frage}{Frage}{section}
\DefBspUmgeb{Question}{Question}{section}
\DefBspUmgeb{Aufgabe}{Aufgabe}{section}
\DefBspUmgeb{Ziel}{Ziel}{section}
\DefBemUmgeb{Bemerkung}
\DefBemUmgeb{Remark}
\DefSBemUmgeb{OffeneFrage}{Offene Frage}
\newcommand{\bdf}{\begin{Definition}}
\newcommand{\edf}{\end{Definition}}
\newcommand{\bvorsch}{\begin{Vorschrift}}
\newcommand{\evorsch}{\end{Vorschrift}}
\newcommand{\bconst}{\begin{Construction}}
\newcommand{\econst}{\end{Construction}}
\newcommand{\bthm}{\begin{Theorem}}
\newcommand{\ethm}{\end{Theorem}}
\newcommand{\bsatz}{\begin{Satz}}
\newcommand{\esatz}{\end{Satz}}
\newcommand{\bprop}{\begin{Proposition}}
\newcommand{\eprop}{\end{Proposition}}
\newcommand{\blem}{\begin{Lemma}}
\newcommand{\elem}{\end{Lemma}}
\newcommand{\bfolg}{\begin{Folgerung}}
\newcommand{\efolg}{\end{Folgerung}}
\newcommand{\bcorr}{\begin{Corollary}}
\newcommand{\ecorr}{\end{Corollary}}
\newcommand{\bfest}{\begin{Feststellung}}
\newcommand{\efest}{\end{Feststellung}}
\newcommand{\bbew}{\begin{Beweis}}
\newcommand{\ebew}{\end{Beweis}}
\newcommand{\bpf}{\begin{Proof}}
\newcommand{\epf}{\end{Proof}}
\newcommand{\bwnum}{\begin{WeiterNumerierteListe}}
\newcommand{\ewnum}{\end{WeiterNumerierteListe}}
\newcommand{\bdfzn}{\begin{DefinitionZusatzNum}}
\newcommand{\edfzn}{\end{DefinitionZusatzNum}}
\newcommand{\bbem}{\begin{Bemerkung}}
\newcommand{\ebem}{\end{Bemerkung}}
\newcommand{\brem}{\begin{Remark}}
\newcommand{\erem}{\end{Remark}}
\newcommand{\bnum}{\begin{NumerierteListe}}
\newcommand{\enum}{\end{NumerierteListe}}
\newcommand{\bunum}{\begin{UnnumerierteListe}}
\newcommand{\eunum}{\end{UnnumerierteListe}}
\newcommand{\bbsp}{\begin{Beispiel}}
\newcommand{\ebsp}{\end{Beispiel}}
\newcommand{\bex}{\begin{Example}}
\newcommand{\eex}{\end{Example}}
\newcommand{\bfrag}{\begin{Frage}}
\newcommand{\efrag}{\end{Frage}}
\newcommand{\bquest}{\begin{Question}}
\newcommand{\equest}{\end{Question}}
\newcommand{\baufg}{\begin{Aufgabe}}
\newcommand{\eaufg}{\end{Aufgabe}}
\newcommand{\bof}{\begin{OffeneFrage}}
\newcommand{\eof}{\end{OffeneFrage}}
\newcommand{\bverm}{\begin{Vermutung}}
\newcommand{\everm}{\end{Vermutung}}
\newcommand{\bconj}{\begin{Conjecture}}
\newcommand{\econj}{\end{Conjecture}}
\newcommand{\bkonv}{\begin{Konvention}}
\newcommand{\ekonv}{\end{Konvention}}
\newcommand{\bglklein}{\begin{StandardUnnumGleichungKlein}}
\newcommand{\eglklein}{\end{StandardUnnumGleichungKlein}}
\newcommand{\bgl}{\begin{StandardUnnumGleichung}}
\newcommand{\egl}{\end{StandardUnnumGleichung}}
\newcommand{\bglrtext}{\begin{XrelYZNumGleichung}}
\newcommand{\eglrtext}{\end{XrelYZNumGleichung}}
\newcommand{\zgl}{\EineZeileGleichung}

\newcommand{\zglklein}[1]{\zgl{\textstyle#1}}
\newcommand{\znumgl}{\EineNumZeileGleichung}

\newcommand{\berlgl}{\begin{StandardUnnumGleichung}}
\newcommand{\eerlgl}{\end{StandardUnnumGleichung}}
\newcommand{\beinrueck}{\begin{Einrueckungpur}} 
\newcommand{\eeinrueck}{\end{Einrueckungpur}}
\newcommand{\beinflist}{\begin{EinfachListe}} 
\newcommand{\eeinflist}{\end{EinfachListe}}
\newcommand{\beq}{\begin{equation}}
\newcommand{\eeq}{\end{equation}}
\newcommand{\bhin}{\begin{Hinrichtung}}
\newcommand{\ehin}{\end{Hinrichtung}}
\newcommand{\brueck}{\begin{Rueckrichtung}}
\newcommand{\erueck}{\end{Rueckrichtung}}
\newcommand{\bvl}{\begin{AutoLabelLaengenListe}{\ListNullAbstaende}}
\newcommand{\evl}{\end{AutoLabelLaengenListe}}
\newcommand{\df}[1]{{\bf #1}}
\newcommand{\zglnum}[2]{\znumgl{#1\label{#2}}}
\renewcommand{\he}{{\text{ss}}}                          
\chardef\tempcat=\the\catcode`\@
\catcode`\@=11

\def\@gobble#1{}
\def\@testgrave{\`}
\def\@stressit{\futurelet\chartest\@stresschar }

\def\@stresschar#1{%
  \ifx #1y\def\result{\futurelet\chartest\@yligature}%
  \else \ifx #1Y\def\result{\futurelet\chartest\@Yligature}%
  \else \ifx\chartest\@testgrave \def\result{\accent"26 }%
  \else \def\result{\accent"26 #1}%
  \fi \fi \fi
  \result }

\def\@yligature{%
  \ifx a\chartest \def\result{\accent"26 \char"1F \@gobble}%
  \else \ifx u\chartest \def\result{\accent"26 \char"18 \@gobble}%
  \else \def\result{\accent"26 y}%
  \fi \fi
  \result }

\def\@Yligature{%
  \ifx a\chartest \def\result{\accent"26 \char"17 \@gobble}%
  \else \ifx A\chartest \def\result{\accent"26 \char"17 \@gobble}%
  \else \ifx u\chartest \def\result{\accent"26 \char"10 \@gobble}%
  \else \ifx U\chartest \def\result{\accent"26 \char"10 \@gobble}%
  \else \def\result{\accent"26 Y}%
  \fi \fi \fi \fi
  \result }

\def\!{\ifmmode \mskip-\thinmuskip \fi}

\def\cyracc{\chardef\i="10%
  \def\cydot{{\kern0pt}}%
  \def\cprime{\char"7E }\def\Cprime{\char"5E }%
  \def\cdprime{\char"7F }\def\Cdprime{\char"5F }%
  \def\dbar{dj}\def\Dbar{Dj}%
  \def\dz{\char"1E }\def\Dz{\char"16 }%
  \def\dzh{\char"0A }\def\Dzh{\char"02 }%
  \def\'##1{\if c##1\char"0F %
    \else \if C##1\char"07 %
    \else \accent"26 ##1\fi \fi }%
  \def\`##1{\if e##1\char"0B %
    \else \if E##1\char"03 %
    \else \errmessage{accent \string\` not defined in cyrillic}%
        ##1\fi \fi }%
  \def\=##1{\if e##1\char"0D %
    \else \if E##1\char"05 %
    \else \if \i##1\char"0C %
    \else \if I##1\char"04 %
    \else \errmessage{accent \string\= not defined in cyrillic}%
        ##1\fi \fi \fi \fi }%
  \def\u##1{\if \i##1\accent"24 i%
    \else \accent"24 ##1\fi }%
  \def\"##1{\if \i##1\accent"20 \char"3D %
    \else \if I##1\accent"20 \char"04 %
    \else \accent"20 ##1\fi \fi }%
  \def\!{\ifmmode \def\result{\mskip-\thinmuskip}%
    \else \def\result{\@stressit}\fi \result}}

\def\keep@cyracc{\let\cyr=\relax \let\i=\relax
        \let\ubar=\relax \let\cydot=\relax
        \let\cprime=\relax \let\Cprime=\relax
        \let\cdprime=\relax \let\Cdprime=\relax
        \let\dbar=\relax \let\Dbar=\relax
        \let\dz=\relax \let\Dz=\relax
        \let\dzh=\relax \let\Dzh=\relax
        \let\'=\relax \let\`=\relax \let\==\relax
        \let\u=\relax \let\"=\relax \let\!=\relax }

\DeclareFontEncoding{OT2}{\cyracc}{}

\catcode`\@=\tempcat

\newlength{\adressabstand}
\begin{document}
\title{{On the Support of Physical Measures in Gauge Theories}}
\author{Christian Fleischhack\thanks{e-mail: 
            Christian.Fleischhack@itp.uni-leipzig.de {\it or}    
            Christian.Fleischhack@mis.mpg.de} \\   
        \\
        {\normalsize\em Max-Planck-Institut f\"ur Mathematik in den
                        Naturwissenschaften}\\[\adressabstand]
        {\normalsize\em Inselstra\ss e 22-26}\\[\adressabstand]
        {\normalsize\em 04103 Leipzig, Germany}
        \\[-25\adressabstand]      
        {\normalsize\em Institut f\"ur Theoretische Physik}\\[\adressabstand]
        {\normalsize\em Universit\"at Leipzig}\\[\adressabstand]
        {\normalsize\em Augustusplatz 10/11}\\[\adressabstand]
        {\normalsize\em 04109 Leipzig, Germany}
        \\[-25\adressabstand]}      
\date{September 21, 2001}
\maketitle
\begin{abstract}
It is proven that the physical measure for the two-dimensional
Yang-Mills theory is purely singular with respect to the 
kinematical Ashtekar-Lewandowski measure. For this, an explicit 
decomposition of the gauge orbit space into supports of these
two measures is given. Finally, the results are extended to more
general (e.g.\ confining) theories. Such a singularity implies, 
in particular, that the standard method of determining the 
physical measure via
``exponential of minus the action times kinematical measure''
is not applicable.
\end{abstract}
\newpage

\section{Introduction}
The functional integral approach to quantum field theories
consists of two basic steps: first the determination
of a ``physical'' Euclidian measure on the configuration space and second
the reconstruction of the quantum theory via an Osterwalder-Schrader
procedure. 
The latter issue has been treated rigorously in several approaches~--
first by Osterwalder and Schrader \cite{OS1, OS2} for scalar fields, 
recently by Ashtekar et al.\ \cite{b12} for diffeomorphism invariant theories.
However, in contrast to this, the former step
kept a problem that has been solved completely
only for some examples.

One of the most promising attempts to overcome this problem in a rather
general context is the Ashtekar approach to gauge field theories. 
It is motivated by the observation that the first step above consists
not only of the determination of the physical measure, but also of the 
preceding determination of the configuration space of the theory.
Originally, in standard (pure) gauge field theories this space contains
all smooth gauge fields modulo smooth gauge transformations.
However, such a space has a very difficult mathematical structure~-- 
it is typically non-compact, non-affine, not finite-dimensional 
and not a manifold. This makes measure theory very complicated. 
How to get rid of this? First, Faddeev and Popov \cite{FP}
tried to use gauge fixings to transfer the problem 
from the gauge orbit space to the much simpler affine space
of all gauge fields. However, this failed because of the Gribov problem,
i.e.\ the non-existence of global gauge fixings \cite{f15,f6}.
Next, it is well-known that the quantization of a theory is typically 
accompanied with a loss of smoothness. This motivated the enlargement
of the configuration space by Sobolev (i.e.\ non-smooth) gauge fields and
gauge transforms \cite{f1,f9,f5}. This way, wide success has been made in the
investigation of the geometry of the (enlarged) configuration space. 
It has been shown that the gauge transform action obeys a slice theorem
which yields a stratification \cite{f11,f12}. Recently, all occurring gauge orbit
types have been classified for certain models \cite{f17}. But, there
is no nontrivial measure known on the total gauge orbit space.
Third, the lattice theory has been developped. For this, one first 
reduces the degrees of freedom to a finite (floating) lattice and hopes
for a reconstruction of the continuum theory by some continuum limit.
Although several physical properties like confinement \cite{Wilson}
have been explained within this approach, the full continuum limit
remains in general an open problem.

The Ashtekar approach, in a sense, brings together the two last issues~--
the enlargement of the configuration space and the lattice theories.
Its basic idea goes as follows: The continuum gauge theory is known
as soon as its restrictions to all finite floating lattices are known.
This means, in particular, that the expectation values of all observables
that are sensitive only to the degrees of freedom of a certain lattice can be 
calculated by the corresponding 
integration over these finitely many degrees of freedom.
Examples for those observables are the Wilson loop variables 
$\tr\:h_\beta$,
where $\beta$ is some loop in the space or space-time and $h_\beta$
is the holonomy along that loop.

The above idea has been implemented rigorously for compact structure
groups $\LG$ as follows:
First the original configuration 
space of all smooth gauge fields (modulo gauge transforms) has been enlarged
by distributional ones \cite{a72}. 
This way the configuration space became compact and
could now be regarded as a so-called projective limit of the lattice 
configuration spaces \cite{a30}. 
These, on the other hand, consist as in ordinary 
lattice gauge theories of all possible assignments of parallel transports
to the edges of the considered
floating lattices (again modulo gauge transforms). 
Since every parallel transport is an element of $\LG$,
the Haar measure on $\LG$ yields a natural measure for the lattice theories.
Now the so-called Ashtekar-Lewandowski measure $\mu_0$ \cite{a48}
is just that continuum 
measure whose restrictions to the lattice theories coincide
with these natural lattice Haar measures. It serves as a canonical
kinematical measure.

Due to the compactness both of the space $\Ab$ of these generalized 
gauge fields (or, mathematically, connections) and of the group $\Gb$ 
of generalized gauge transforms, the geometry of the factor space $\AbGb$
is well-understood. As in the Sobolev case a slice theorem has been proven,
a stratification has been found and the occurring gauge orbit types
w.r.t.\ the action of $\Gb$ have been determined (here completely 
for all space-times and all compact structure groups) \cite{paper2+4}.
Moreover, it has been shown that the so-called non-generic connections
\cite{paper2+4,paper5}
form a $\mu_0$-zero subset of $\Ab$. Additionally, as for smooth connections,
typically (i.e.\ for $\LG = SU(N)$ and some other groups) a Gribov problem
arises in the sense that there is no continuous gauge fixing in $\Ab$.
However, here one can find a $\mu_0$-zero subset in $\Ab$
such that after its removal there is a continuous gauge fixing \cite{paper5}.
This implies that the Faddeev-Popov determinant equals $1$ almost everywhere.
Therefore no problems arise when integrating over $\AbGb$ using 
such (almost complete) gauge fixings~-- at least on the kinematical level.

\subsection*{Problems Considered in this Article}
In this article we are going to study the physical relevance of these
rather mathematical structures. Our considerations are motivated by the 
following two, obviously connected problems.
\bquest
What is the impact of non-generic connections?
\equest
\bquest
How severe is the Gribov problem?
\equest
In this generality both questions, of course, can hardly be answered.
Therefore we will first analyze them by means of a concrete example.
Unfortunately, within the Ashtekar approach we have only two theories
at our disposal that are investigated in detail:
the quantum gravity
(in particular, the canonical quantization \cite{a70,a25} 
and the quantum geometry \cite{a69,e30,d10,a12,a13,e20,a14,e33,b5,b6}) 
and the two-dimensional Yang-Mills theory \cite{b11,a6,dipl,paper1,b7}.
Beyond these two there are only attempts for the 
treatment of matter fields \cite{e35}, heat-kernel measures or measures
coming from knot theory \cite{a30} or from Chern-Simons theory \cite{d9,d2}.
Recently, the Fock space formulation has been connected to the
Ashtekar framework \cite{b19,b20,d46,b18}.

However, in the field of quantum gravity the problem is still a bit unclear.
This is due to the canonical quantization used there \cite{a25,a70,a79,a77,b8}.
Its starting point is a classical phase space (hence for quantum gravity
a symplectic space whose position variables are just the 
Ashtekar connections) with certain constraints
(here, e.g., Gau\ss\ constraint and diffeomorphism constraint). 
Afterwards, some algebra of functions on this phase space 
is associated an algebra of operators by naive quantization,
such that Poisson brackets correspond to operator commutators, and
then some Hilbert space is chosen where these operators are represented.
For quantum gravity this Hilbert space is just the space 
$L^2(\AbGb)$ with the Ashtekar-Lewandowski measure $\mu_0$ 
where $\LG = SU(2)$.
For $\mu_0$, however, the Gribow problem and the impact of non-generic 
connections has already been investigated \cite{paper2+4,paper5}. 
Consequently, we can consider the questions above answered. 
But, of course, one can take the view that in any case $L^2(\AbGb)$ 
is only an auxiliary tool. Then these questions are not at issue
because up to now it is not clear how the physical Hilbert space 
of quantum gravity looks like.

Therefore we will focus on the example of the two-dimensional 
quantum Yang-Mills theory ($\YM_2$).
As mentioned in the beginning, the central point here is the determination
of a physical interaction measure $\mu_\YM$ on $\AbGb$.
Typically --~neglecting mathematical problems~-- such a measure 
is defined by multiplying some kinematical measure with $\e^{-S}$,
where $S$ is the action of the physical theory. 
The natural kinematical measure in the Ashtekar approach 
is the Ashtekar-Lewandowski measure; but the action
$S(A) \ident \ymwirk(A) = \inv4 \int_M \tr\: F_{\mu\nu} F^{\mu\nu} \: \dd x$ 
is only defined in the case $\ag$ and not for $\AbGb$.
This is obvious because products of space-time derivatives of distributional
connections cannot be defined in general.
This problem has been solved first by Thiemann \cite{b11} and
Ashtekar et al.\ \cite{a6} for $SU(N)$ and $U(1)$:
They used the fact that by the Riesz-Markov theorem 
the knowledge of all Wilson-loop expectation values
is sufficient for the determination of $\mu_\YM$ and calculated
these expectation values by means of a lattice regularization of $\ymwirk$.
More precisely, they chose on every quadratic lattice $\GR$ the Wilson action
$\ymwirk[\text{reg}\GR] (A):= 
  \frac{N}{\kopp^2 a^2} \sum_{\square}(1-\einsdurchn\re\tr\: h_\square(A))$, 
where $\square$ runs over all plaquettes of the lattice with lattice spacing
$a$ and side lengths $L_x$ and $L_y$ \cite{Wilson}.
This function can be extended in a natural way to $\AbGb$. 
Then the Wilson-loop expectation values 
are defined by exchanging limit and integral:
\zgl{\erww{\tr\: h_{\alpha_1} \cdots \tr\: h_{\alpha_n}} :=
     \lim_{a\gegen 0, L_x,L_y\gegen\infty}
      \inv{Z_{a,L_x,L_y}}
      \int_{\AbGb} 
      \e^{-\ymwirk[\text{reg}\GR]} \: \tr\: h_{\alpha_1} \cdots \tr\: h_{\alpha_n}
         \: \dd\mu_0,}\noindent
where $Z_{a,L_x,L_y}$ only normalized $\erww{1}$ to $1$.
The usage of a fixed quadratic lattice remained a disadvantage
because it only permitted the consideration of 
loops fitting in such a lattice; but this is per se not sufficient for a
rigorous determination of $\mu_\YM$.
This drawback has been removed in \cite{paper1,dipl}
where not the loops are adapted to the regularization,
but the regularization is adapted to the given loops. 
So for an arbitrary graph 
first the sum over all plaquettes has been replaced by the sum over all
interior domains and second $a^2$ simply by the area of the corresponding 
domain.
Moreover, the limiting process now instead of  
$a\gegen 0, L_x,L_y\gegen\infty$
consists of all possible refinements of the graph built by the $\alpha_i$.
This way, $\mu_\YM$ has been defined rigorously.

However, properties of $\mu_\YM$ are almost unknown. Only the invariance
w.r.t.\ area-preserving diffeomorphisms has been shown \cite{a6}.
Regarding to the two questions above there is a very interesting
\bquest
Is $\mu_\YM$ absolutely continuous w.r.t.\ $\mu_0$?
\equest
If we were able to answer this question with ``yes'', we would
have proven that the set of all non-generic connections has not only 
Ashtekar-Lewandowski measure, but also Yang-Mills measure $0$, and
the Gribov problem remains harmless as well.
Moreover, such an absolute continuity would guarantee the existence
of a non-negative $L^1(\mu_0)$-function $\ymd$ in $\AbGb$
with $\dd\mu_\YM = \ymd \: \dd\mu_0$. This function could be considered  
as $\e^{-S_\YM}$ for some generalized Yang-Mills action
$S_\YM$. However~-- $\mu_\YM$ is {\em not}\/ absolutely continuous 
w.r.t.\ $\mu_0$. We will even be able to prove that 
$\mu_\YM$ is {\em purely singular}\/ w.r.t.\ $\mu_0$,
this means that the support of $\mu_\YM$ is contained in a $\mu_0$-zero 
subset.
This, on the other hand, does {\em not}\/ mean, that for instance the 
non-generic connections need have a Yang-Mills measure different from $0$.
This comes from the fact that despite of the singularity of $\mu_\YM$ 
w.r.t.\ $\mu_0$ on $\AbGb$ the corresponding lattice measures 
are always absolutely continuous w.r.t.\ the lattice Haar measures. 
Since both the non-generity and the almost global triviality of the 
generic stratum being responsible for the relevance of the Gribow-Problem
can be described already on the level of graphs,
we will get for the Yang-Mills measure similar answers to the first
two questions as we did for the Ashtekar-Lewandowski measure.
However, since we will observe a certain concentration of the Yang-Mills
measure ``near'' non-generic connections, such strata should not simply 
be neglected.

\subsection*{Outline of the Article}
The outline of the present artice is as follows:
\bunum
\item
First after fixing the notations 
we will provide some theorems from the Fourier analysis on arbitrary
compact Lie groups that will be needed for the investigation of the 
Radon-Nikodym derivatives $\dd\mu_{\YM,\GR}/\dd\mu_{0,\GR}$ on the 
lattice levels and for the singularity theorem afterwards.
\item
Next we will review the construction of the Yang-Mills measure $\mu_\YM$
\cite{paper1,a6} in terms of loop-network states 
introduced by Thiemann \cite{b7} and give a proof for the well-definedness
of $\mu_\YM$ for arbitrary compact structure groups $\LG$.
\item
Third we will investigate the lattice Radon-Nikodym derivatives
and prove the inequivalence between the continuum Yang-Mills 
and the Ashtekar-Lewandowski measure by studying the support of the
Yang-Mills measure. As by-products we get that the non-generic connections
are contained in a $\mu_\YM$-zero subset, that the Gribov problem
is again harmless, and that the regular (smooth) gauge orbits are 
again contained in a zero subset.
\item
Finally, we will indicate how these results can be generalized 
to other models \cite{paper7}.
We will see, e.g., that analogous support properties are shared typically by
theories describing confinement.
\eunum

\section{Preliminaries}
We recall the basic notations and results about generalized connections
\cite{a72,a48,a42,a30,paper2+4,paper3,diss,paper5}.

Let $M$ be some at least two-dimensional manifold, $m$ be fixed in $M$ and
$\LG$ be a connected compact (real) Lie group. 
$\Pf$ denotes the groupoid of all paths in $M$, $\hg$ the group of all
paths starting and ending in $m$.
The set $\Ab$ of generalized connections $\qa$ is defined by 
$\Ab:=\varprojlim_\GR \Ab_\GR \ident \varprojlim_\GR \LG^{\elanz\Edg(\GR)} 
 = \Hom(\Pf,\LG)$. Here $\GR$ runs
over all (finite) graphs in $M$. $\Edg(\GR)$ is the set of edges in $\GR$,
$\Ver(\GR)$ will be that of all vertices. The canonical projections
from $\Ab$ to the spaces $\Ab_\GR$ of lattice connections
are denoted by $\pi_\GR$.
Given $\Ab$ the projective limit topology, it becomes compact Hausdorff.
The group $\Gb$ of generalized gauge transforms $\qg$ is defined by 
$\Gb:=\varprojlim_\GR \Gb_\GR \ident \varprojlim_\GR \LG^{\elanz\Ver(\GR)} 
= \Maps(M,\LG)$.
It is compact as well and acts continuously on $\Ab$ via 
$h_{\qa\circ\qg}(\gamma) = g_{\gamma(0)}^{-1} h_\qa(\gamma) g_{\gamma(1)}$
where the path $\gamma$ is in $\Pf$ 
and $h_\qa$ is the homomorphism corresponding to $\qa$. The
projections are again denoted by $\pi_\GR$.
Analogously to the definition of $\pi_\GR$ we set
$\pi_\gc: \Ab \nach \Ab_\gc \iso \LG^{\elanz\gc}$, $h \auf h(\gc)$ etc.\
for all finite subsets $\gc$ of $\Pf$.
The projections $\pi_{\GR_1}^{\GR_2} : \Ab_{\GR_2} \nach \Ab_{\GR_1}$
are given similarly for all $\GR_1 \leq \GR_2$, whereas the last notation
means that every edge of $\GR_1$ is a product of edges in $\GR_2$.  
Moreover, we set $\agb := \varprojlim_\GR \Ab_\GR/\Gb_\GR$.
If the paths in $\Pf$ are restricted to the piecewise analytic
category, there is a natural homeomorphism $\abbAbGbagb:\AbGb\nach\agb$.

Every self-consistent family $(\mu_\GR)_\GR$ 
of normalized regular Borel measures on the $\Ab_\GR$, i.e.\  
$\mu_{\GR_1} = (\pi_{\GR_1}^{\GR_2})_\ast \mu_{\GR_2}$ for all 
$\GR_1 \leq \GR_2$,
defines a unique normalized regular Borel measure $\mu$ on $\Ab$, such that
$\mu_\GR = (\pi_\GR)_\ast \mu$.
Conversely, every such $\mu$ defines via $\mu_\GR := (\pi_\GR)_\ast \mu$
a self-consistent family. If one chooses for $\mu_\GR$ always
the Haar measure on $\LG^{\elanz\Edg(\GR)}$, one gets the 
Ashtekar-Lewandowski measure $\mu_0$.

Finally, we call a generating system $\ga\teilmenge\hg$ of 
the fundamental group $\pi_1(\GR)$ of a connected graph $\GR$ 
weak fundamental system iff there is a maximal tree $T$ in $\GR$ 
such that for every path $\alpha_i \in \ga$ there is an edge
$e_i$ in $\GR \setminus (T \cup \{e_1,\ldots,e_{i-1}\})$ 
such that $\alpha_i$ is a product of $e_i$ and certain edges
in $T \cup \{e_1,\ldots,e_{i-1}\}$.
A weak fundamental system $\ga$ is always a free generating system 
and fulfills $(\pi_\ga)_\ast \mu_0 = \mu_\Haar^{\elanz\ga}$.

\section{Fourier Analysis}
\label{sect:fourieranalysis}
On compact Lie groups, integration is strongly related to Fourier analysis.  
The crucial connecting links are the integration formulae and
the Peter-Weyl theorem. However, in contrast to the extensively investigated
case of functions on $U(1)$ (or simply $2\pi$-periodical functions on $\R$),
general
results about the convergence of Fourier series beyond the Peter-Weyl theorem
are very rare and widespread. There are only few original articles such as
\cite{Taylor} or results for special cases like
smooth functions (see \cite{BourbakiLie9Russ}) or
expansions of heat-kernels (see \cite{Stein}). Some results
presented in the sequel (in particular, in the 
subsections \ref{uabschn:fourier:dim(darst)} till \ref{uabschn:convcrit(four)})
seem to be folklore in part; however, we were not able to
find the proofs in the literature.
Therefore we briefly collect in this section the facts needed in the following,
and provide the proofs if they are -- to the best of our knowledge -- 
unknown or non-standard. A more detailled treatment is given in \cite{diss}.

\subsection{Representations of Compact Lie Groups}
For every connected compact Lie group there is \cite{Encycl5} 
a simply connected
semisimple compact Lie group $\LG_\he$, some natural number $k$ and
some finite Lie subgroup $\LN\teilmenge Z(\LG_\he) \kreuz U(1)^k$,
such that 
\zgl{\LG \iso (\LG_\he \kreuz U(1)^k)/\LN.}\noindent
Here, $Z(\LG_\he)$ is the center of $\LG_\he$.
We set $l$ to be the rank of $\LG_\he$, i.e.\ 
the dimension of a maximal torus in its Lie algebra $\Lieg_\he$. 
The set of all (equivalence classes of) irreducible unitary 
representations of $\LG$ is denoted by $\allirrdarst\LG$. 
It is well-known that every representation $\darst\in\allirrdarst\LG$ of 
$\LG = (\LG_\he \kreuz U(1)^k)/\LN$ can be identified with 
a uniquely determined irreducible representation of $\LG_\he \kreuz U(1)^k$
and consequently \cite{BrtD} with a tensor product
$\darst_\he \tensor \darst_\ab$ of irreducible representations 
of $\LG_\he$ and of $U(1)^k$, respectively. Hence it can be viewed as an element
$(\vn,\vz) \in \N^l \kreuz \Z^k$.
Here, $\vn\in\N^l$ characterizes the heighest
weight $\hgew_\vn := \sum n_i\hgew_i$ of the representation $\darst_\he$ 
and $\vz\in\Z^k$ identifies the representation 
$\vg\auf (g_i^{z_i})_i$ of the torus part.
Typically, we will simply write $\vnu$ instead of $(\vn,\vz)$ and 
use $\darst_\vnu$ or even simpler $\vnu$ to denote the corresponding 
representation. Finally, we denote by $d_\vnu$ (or $d_\darst$) the dimension
of the representation $\vnu$ (or $\darst$).

\subsection{Peter-Weyl Theorem}
For every irreducible representation $\darst$ of $\LG$ we fix a basis
on the corresponding representation space $\VR$.
By $\darst(g) \in\Gl_\C(\VR)$ we can view every $\darst(g)$ as some matrix. 
In the following $\darst^{ij}(g) \in \C$ denotes the matrix element of
$\darst(g)$ belonging to the $i$-th column and the $j$-th row.%
\footnote{In order to assign the same matrix element to equivalent 
representations, we choose the bases on the vector spaces $\VR$ 
``consistently''. More precisely, we fix in every equivalence class
$[\darst]\in\allirrdarst\LG$ some representation $\darst$ and choose
on the corresponding representation space $\VR$ a basis $\VRbasis_\darst$.
Now, for the other $\darst'\in[\darst]$ there is an isomorphism
$A:\VR\nach\VR'$ with $\darst'(g) = A \darst(g) A^{-1}$. We choose 
$\VRbasis_{\darst'} := A \VRbasis_\darst$ as a basis on $\VR'$.}
We call the elements of
$\matfkt:=\{\sqrt{\dim\darst} \: \darst^{ij} \mid 
            [\darst]\in\allirrdarst\LG, \: i,j=1,\ldots,\dim\darst\}
          \teilmenge C^\infty(\LG)$ 
elementary matrix functions.       
The set 
$\{\charakt_\darst \mid [\darst]\in\allirrdarst\LG\} 
  \teilmenge C_\Ad^\infty(\LG)$
of all characters $\charakt_\darst$ 
of irreducible representations is denoted by 
$\allchar$. 
\bprop
\label{prop:ON-matrixelem}
Let $\darst_1$ and $\darst_2$ be irreducible unitary representations 
of $\LG$. Then \cite{Barut}
\zgl{\int_\LG \quer{\darst^{i_1 j_1}_1(g)} \: \darst^{i_2 j_2}_2(g) \: \dd\mu_\Haar
     \ident \lzweihaar{\darst^{i_1 j_1}_1}{\darst^{i_2 j_2}_2}
     = \inv{\dim\darst_1} \: \delta_{i_1 i_2} \delta_{j_1 j_2} 
                          \delta_{\darst_1\darst_2}.}\noindent
Here, $\delta_{\darst_1 \darst_2} = 1$, if
$\darst_1 \iso \darst_2$, and $\delta_{\darst_1 \darst_2} = 0$ else.
\eprop
\bcorr
\label{corr:ON-charakt}
Under the assumptions of the preceding proposition we have
\zgl{\lzweihaar{\charakt_{\darst_1}}{\charakt_{\darst_2}} = \delta_{\darst_1 \darst_2}.}
\ecorr
\bthm[Peter-Weyl Theorem {\rm\cite{Barut}}]
\label{thm:peter+weyl}
\bnum{2}
\item
\bnum{2}
\item
$\matfkt$ is a complete orthonormal system in $L^2(\LG)$.
\item
$\spann_\C \matfkt$ is dense in $C(\LG)$.
\enum
\item
\bnum{2}
\item
$\allchar$ is a complete orthonormal system in $L^2_\Ad(\LG)$. 
\item
$\spann_\C \allchar$ is dense in $C_\Ad(\LG)$.
\enum
\enum
\ethm
Here, $L^2_\Ad(\LG)$ contains precisely the conjugation invariant
$L^2$-functions on $\LG$.
Analogously, $C_\Ad(\LG)$ collects the conjugation invariant continuous
functions on $\LG$.
\bcorr
\label{corr:fourier_L2(LG)}
For all $f\in L^2(\LG)$ we have
\zgl{f = \sum_{[\darst]\in\allirrdarst\LG} \sum_{i,j=1}^{\dim\darst}
         \dim \darst \: \lzweihaar{\darst^{ij}}{f} \darst^{ij}.}
Analogously, for all $f\in L^2_\Ad(\LG)$ we have
\zgl{f = \sum_{[\darst]\in\allirrdarst\LG}
         \lzweihaar{\charakt_\darst}{f} \charakt_\darst.}
\ecorr

\subsection{Laplace-Beltrami and Casimir Operator}
Let $\{X_i\}$ be a basis of the Lie algebra $\Lieg$ of $\LG$.
The left-invariant vector field on $\LG$ corresponding to $X_i$ 
is denoted by $\widetilde X_i$. 
\bdf
\label{def:lapl}
Let $A = A^{ij} \in\R^{\dim\LG,\dim\LG}$ be some matrix.

Then 
$\lapl_A := A^{ij} \widetilde X_i \widetilde X_j : C^n(\LG) \nach C^{n-2}(\LG)$,
$n\in\N$, $n \geq 2$,
is called \df{Laplace-Beltrami operator} for $A$ and $\{X_i\}$.
\edf
Let $\Lieg = \Lieg_\he \dirsum \Lieg_\ab$ be the splitting of 
the Lie algebra $\Lieg$ into its semisimple and abelian part.
Using the Killing form $\killing$ on $\Lieg_\he$ and the pseudo-Killing form
$\pskilling$ on $\Lieg_\ab$ we define by 
$\natform((X_\he, X_\ab),(Y_\he, Y_\ab)) := 
  \killing(X_\he,Y_\he) + \pskilling(X_\ab,Y_\ab)$
a non-degenerate, symmetric, negative-definite bilinear form $\natform$
on $\Lieg$ -- the so-called \df{natural} bilinear form. Here, 
the pseudo-Killing
form is defined by 
$\pskilling(\ueinsbasis_i,\ueinsbasis_j) := -\delta_{ij}$, where
$\ueinsbasis_j = (0,\ldots,0,\I,0,\ldots,0)$ with $\I$ on the $j$-th slot
gives the canonical basis of $\Lieg_\ab \iso \I\R \dirsum \ldots \dirsum \I\R$.

\bdf
\label{def:nat_lapl}
$\lapl := \lapl_{\natmt^{-1}}$ is called \df{Casimir operator} on $\LG$, where
the matrix $\natmt$ is defined by $\natmt_{ij} := \natform(X_i,X_j)$.
\cite{Stein}
\edf
One immediately sees that $\lapl$ does not depend on the choice 
of the basis $\{X_i\}$. 
Moreover, $\lapl$ is symmetric on $C^2(\LG) \teilmenge L^2(\LG)$.
\cite{Stein} Now, we have 
\bprop
\label{prop:natlapl=casimir}
For every irreducible representation $\darst$ of $\LG$ there is 
a non-negative real number $c_\darst$, such that
$\lapl\darst^{ij} = c_\darst \: \darst^{ij}$ for every elementary 
matrix function $\darst^{ij}$ of $\darst$. \cite{Stein}
\eprop
$c_\darst$ is also called Casimir eigenvalue.
From $\charakt_\darst(g) = \tr\: \darst(g) = \sum_i \darst^{ii}(g)$
we get
\bcorr
\label{corr:natlapl=casimir}
The character $\charakt_\darst$ of $\darst$ fulfills the eigenvalue equation
$\lapl\charakt_\darst = c_\darst \: \charakt_\darst$.
\ecorr
We will frequently use the following properties of the Casimir eigenvalues:
\bprop
\label{prop:absch(casimir-eigenwerte)}
\bunum
\item
$c_\vnu = 0$ $\aequ$ $\vnu = \vec 0$ $\aequ$ $\darst_\vnu$ is trivial.
\item
There are positive real numbers $c_-$ and $c_+$,
such that 
\zgl{c_- \norm\vnu^2 \leq c_\vnu \leq c_+ \norm\vnu^2}\noindent
for all $\vnu \ident (\vn,\vz) \in \N^l \kreuz \Z^k$, whereas
$\norm\vnu^2 := \norm\vn^2 + \norm\vz^2$ gives the standard norm on $\R^{l+k}$.
\eunum
\eprop
Here, $c_\vnu$ is simply the Casimir eigenvalue for the irreducible
representation $\darst_\vnu$.

\subsection{Dimension of Representations}
\label{uabschn:fourier:dim(darst)}
We estimate the dimension of irreducible representations.
\bprop
\label{prop:absch(dim(darst))}
For every $\LG$ there are positive constants $\const_\LG$ and $\const'_\LG$,
such that we have for all $\vn\in\N^l$ and $\vz\in\Z^k$:
\bnum{3}
\item
$d_{\vn,\vz} \leq \const'_\LG \: (\norm\vn^{\einhalb(\dim\LG_\he-l)} + 1)$,
\item
$d_{\vn,\vz} \leq \const_\LG \: \norm\vn^{\einhalb(\dim\LG_\he-l)}$ 
if $\vn \neq \vec 0$ and
\item
$d_{\vn,\vz} \leq \const_\LG \: \norm{\vn,\vz}^{\einhalb(\dim\LG_\he-l)}$ 
if $(\vn,\vz) \neq (\vec 0,\vec 0)$.
\enum
Here, $\dim\LG_\he$ equals the dimension 
of the semisimple part $\Lieg_\he$ of $\Lieg$.
\eprop
\bpf
\bunum
\item
By the Weyl formula \cite{Barut,Cornwell2} the dimension $d_{\vn,\vz}$ of the 
irreducible representation $\darst_{\vn,\vz}$ equals
\zgl{
    d_{\vn,\vz} 
 := \dim \darst_{\vn,\vz}
  = \prod_{\alpha\in\wurzs^+} 
    \frac{\kmo{\alpha}{\hgew_\vn+\hgew_\vr}}{\kmo{\alpha}{\hgew_\vr}}.}\noindent
Here, $\wurzs^+$ denotes the system of positive roots of $\Lieg_\he$,
$\hgew_\vn$ is the highest weight of $\darst_\vn$, 
$\hgew_\vr := \einhalb \sum_{\alpha\in\wurzs^+} \alpha$ is the so-called
Weyl vector and $\kmo{\cdot}{\cdot}$ denotes the symmetric non-degenerate
bilinear form on the root space induced by the Killing form.
Using the standard properties of $\kmo\cdot\cdot$
we get
\bgl
{\kmo{\alpha}{\hgew_\vn+\hgew_\vr}}
 & \leq & \kmonorm\alpha c'_+(\norm{\vn} + \norm{\vr}),
\egl 
for some constant $c'_+$, hence
\zgl{
d_{\vn,\vz} 
      =   \prod_{\alpha\in\wurzs^+}
          \frac{\kmo{\alpha}{\hgew_\vn+\hgew_\vr}}{\kmo{\alpha}{\hgew_\vr}} \\
    \leq  \Bigl(\prod_{\alpha\in\wurzs^+} c'_+
               \frac{\kmonorm{\alpha}}{{\kmo{\alpha}{\hgew_\vr}}}\Bigr)
          \bigl(\norm\vn + \norm\vr\bigr)^{\einhalb(\dim\LG_\he-l)}
}\noindent
because of $\elanz\wurzs^+ = \einhalb(\dim\LG_\he-l)$.
\item
Since for arbitrary $d\in\R$ and $s\in\N$ 
the function $x\auf\frac{(x+d)^s}{x^s+1}$ is bounded on $[0,\infty)$,
we have
\bgl
d_{\vn,\vz}
   & \leq & \Bigl(\prod_{\alpha\in\wurzs^+} c'_+
                  \frac{\kmonorm{\alpha}}{{\kmo{\alpha}{\hgew_\vr}}}\Bigr)
	    \frac{\bigl(\norm\vn + \norm\vr\bigr)^{\einhalb(\dim\LG_\he-l)}}
	         {\norm\vn^{\einhalb(\dim\LG_\he-l)}+1}
	    \bigl(\norm\vn^{\einhalb(\dim\LG_\he-l)}+1\bigr)\\
   & \leq & \const'_\LG
	    \bigl(\norm\vn^{\einhalb(\dim\LG_\he-l)}+1\bigr)
\egl
for all $\vn\in\N^l$ and $\vz\in\Z^k$.
\item
The remaining cases are proven analogously.
\qed
\eunum
\epf
\subsection{Asymptotic Behaviour of Fourier Coefficients}
As we know from Corollary \ref{corr:fourier_L2(LG)},
the Fourier series
$\sum_{[\darst]\in\allirrdarst\LG} \sum_{i,j=1}^{\dim\darst}
     \dim\darst\:\lzweihaar{\darst^{ij}}{f} \darst^{ij}$
of an arbitrary function $f\in L^2(\LG)$
converges to $f$ in the $L^2$-sense. 
For studying when this series 
even converges in the space of continuous functions,
i.e.\ uniformly, we need estimates about the asymptotic
behaviour of the Fourier coefficients.
\bprop
\label{prop:asymp(fourierkoeff)}
Let $f \in C^{2s}(\LG)$ be a $2s$-times continuously differentiable
function on $\LG$. 
Then we have for all nontrivial irreducible representations $\darst$
and for all $i,j=1,\ldots,\dim\darst$

\zgl{\betrag{\lzweihaar{\darst^{ij}}{f}} \leq 
               \inv{\sqrt{\dim\darst}}\frac{\const_{s,f}}{c_\darst^s}}\noindent
and
\zgl{\betrag{\lzweihaar{\charakt_\darst}{f}} \leq 
               \frac{\const_{s,f}}{c_\darst^s}.}\noindent
Here, $\const_{s,f} := \lzweihaarnorm{\lapl^s f} < \infty$ does {\em not}\/
depend on $\darst$, but only on $s$ and $f$.
\eprop
\bpf
\bunum
\item
Let $\darst$ be some nontrivial representation.
By Proposition \ref{prop:absch(casimir-eigenwerte)}
the eigenvalue $c_\darst$ of the Casimir operator is positive.
Hence 
\bgl
       \lzweihaar{\darst^{ij}}{f}
 & = & c_\darst^{-s} \: \lzweihaar{\lapl^s\darst^{ij}}{f}
       \erl{Proposition \ref{prop:natlapl=casimir}}\pktplatz \\
 & = & c_\darst^{-s} \: \lzweihaar{\darst^{ij}}{\lapl^s f}
       \erl{Symmetry of $\lapl$}.
\egl
Using the Schwarz inequality and 
$\lzweihaarnorm{\darst^{ij}} = (\dim\darst)^{-\einhalb}$ 
(cf.\ Proposition \ref{prop:ON-matrixelem})
we get 
\bgl
          \betrag{\lzweihaar{\darst^{ij}}{f}}
 &  =   & c_\darst^{-s} \: \betrag{\lzweihaar{\darst^{ij}}{\lapl^s f}}\\
 & \leq & c_\darst^{-s} \: \lzweihaarnorm{\darst^{ij}}
                               \: \lzweihaarnorm{\lapl^s f}\\
 &  =   & \const_{s,f} \: (\dim\darst)^{-\einhalb} \: c_\darst^{-s}
\egl
with $\const_{s,f} = \lzweihaarnorm{\lapl^s f} < \infty$.
\item
The proof for the characters is completely analogous. Note only
$\lzweihaarnorm{\charakt_\darst} = 1$.
\qed
\eunum
\epf

\subsection{Convergence Criterion for Fourier Series}
\label{uabschn:convcrit(four)}
For the proof of the uniform convergence we need the following lemmata:
\blem
\label{lem:krit(glm_konv)}
Let $X$ be a metric space, $Y$ a Banach space over $\K$ and let
$f_\laufi\in C(X,Y)$ for all $\laufi\in \N$. Then we have:

If $\sum_{\laufi\in\N} \supnorm{f_\laufi}$ converges, then
$\sum_{\laufi\in\N} f_\laufi$ converges absolutely and uniformly on $X$ 
to some $f\in C(X,Y)$.
\elem
\blem
\label{lem:sum_dim+casimir}
For all $\mu,\nu\in\R$ with $\mu \geq 0$ and
$\einhalb(\dim\LG_\he - l)\mu + 2 \nu \leq -(k+l+1)$,
\zgl{\sum_{\vnu\in\allirrdarst\LG, \vnu\neq\vec 0} d_\vnu^\mu \: c_\vnu^\nu}\noindent
converges.
\elem
\bpf
By Proposition \ref{prop:absch(casimir-eigenwerte)} and 
Proposition \ref{prop:absch(dim(darst))}
we have for $\nu\geq 0$
\bgl
\sum_{\vnu\in\allirrdarst\LG, \vnu\neq\vec 0} d_\vnu^\mu \: c_\vnu^\nu
 & \leq & \sum_{\vnu\in\N^l\kreuz\Z^k, \vnu\neq\vec 0} d_\vnu^\mu \: c_\vnu^\nu \\
 & \leq & \sum_{\vnu\in\N^l\kreuz\Z^k, \vnu\neq\vec 0} 
            \const_\LG^\mu \norm\vnu^{\einhalb(\dim\LG_\he - l)\mu}
	    c_+^{\nu} \norm\vnu^{2\nu} \\
 & \leq & \const_\LG^\mu c_+^{\nu}
          \sum_{\vnu\in\N^l\kreuz\Z^k, \vnu\neq\vec 0} 
             \norm\vnu^{-(k+l+1)}.
\egl
In the last step the assumption 
$\einhalb(\dim\LG_\he - l)\mu + 2 \nu \leq -(k+l+1)$ and $\norm\vnu\geq 1$
have been used.
The convergence is implied by Corollary \ref{corr:konvkritNZ_poly}.

For $\nu<0$ the argumentation is completely analogous. Just replace
$c_+$ by the constant $c_-$.
\qed
\epf

\bprop
\label{prop:glm+abs_konv(fourier)}
Let $f\in C^{2s}(\LG)$ be a $2s$-times continuously differentiable function 
on $\LG$ with $2s \geq \dim\LG+1$.
Then the Fourier series
\zgl{\sum_{[\darst]\in\allirrdarst\LG} \sum_{i,j=1}^{\dim\darst}
         \dim \darst \: \lzweihaar{\darst^{ij}}{f} \darst^{ij}}\noindent
of $f$ converges absolutely and uniformly to $f$.%
\footnote{There is an even stronger result: Taylor \cite{Taylor} 
proved using Sobolev techniques that the Fourier series converges
for every $f\in C^{2s}(\LG)$ if $s\in\N$ is larger than 
$\inv4\dim\LG$ (cf.\ the review \cite{TaylorKomm}).}
\eprop
\brem
Even
\zgl{\sum_{[\darst]\in\allirrdarst\LG} \sum_{i,j=1}^{\dim\darst}
         \dim \darst \: \lzweihaar{\darst^{ij}}{f} D\darst^{ij}}\noindent
converges absolutely and uniformly to $Df$ for all smooth
differential operators $D$,
whose order is not larger than 
$2s - \dim\LG+1$.
\erem
We remember that instead of
$\sum_{[\darst]\in\allirrdarst\LG}$ we can simply write
$\sum_{\vnu\in\allirrdarst\LG\teilmenge\N^l\kreuz\Z^k}$ 
and replace $\darst$ correspondingly by $\darst_\vnu$ or just $\vnu$.
\bpf
First we show that 
$\sum_{\vnu\in\allirrdarst\LG\teilmenge\N^l\kreuz\Z^k} \sum_{i,j=1}^{d_\vnu}
       d_\vnu\lzweihaar{\darst_\vnu^{ij}}{f} \darst_\vnu^{ij}$ 
converges uniformly and absolutely.
\bunum
\item
By the Schwarz inequality and the unitarity of $\darst_\vnu$ we have
\zglklein{
\sum_{i,j=1}^{d_\vnu} \betrag{\darst_\vnu^{ij}(g)}
 \leq \sqrt{\sum_{i,j=1}^{d_\vnu} 1} 
      \sqrt{\sum_{i,j=1}^{d_\vnu} \betrag{\darst_\vnu^{ij}(g)}^2}
   =  d_\vnu \sqrt{\tr\: \darst_\vnu^+ \darst_\vnu}
   = d_\vnu^{\frac32}}\noindent for all $g\in\LG$.
Hence by the asymptotics of the Fourier coefficients
we get for $\vnu\neq\vec 0$
\bgl
\sum_{i,j=1}^{d_\vnu} 
 \supnorm{d_\vnu \lzweihaar{\darst_\vnu^{ij}}{f} \darst_\vnu^{ij}}
 & =    & \sum_{i,j=1}^{d_\vnu} d_\vnu\:\betrag{\lzweihaar{\darst_\vnu^{ij}}{f}} \:
           \supnorm{\darst_\vnu^{ij}}\\ 
 & \leq & d_\vnu \: d_\vnu^{-\einhalb} \: \const_{s,f} c_\vnu^{-s} \: 
          d_\vnu^{\frac32} \\
 & =    & \const_{s,f} \: d_\vnu^2 \:  c_\vnu^{-s}.
\egl
Moreover,
$\supnorm{d_{\vec 0} \lzweihaar{\darst_{\vec 0}^{11}}{f} \darst_{\vec 0}^{11}}
 \leq \lzweihaarnorm f < \infty$.

Now, 
$\sum_{\vnu\in\allirrdarst\LG} \sum_{i,j=1}^{d_\vnu} 
   \supnorm{d_\vnu\lzweihaar{\darst_\vnu^{ij}}{f} \darst_\vnu^{ij}}$
converges by Lemma \ref{lem:sum_dim+casimir} and by
$\einhalb(\dim\LG_\he - l)\cdot 2 + 2 \cdot (-s)
\leq \dim\LG_\he - l - \dim\LG - 1 = -(k+l+1)$.
\item
By Lemma \ref{lem:krit(glm_konv)},
$\sum_{\vnu\in\allirrdarst\LG\teilmenge\N^l\kreuz\Z^k} \sum_{i,j=1}^{d_\vnu} 
         d_\vnu\lzweihaar{\darst_\vnu^{ij}}{f} \darst_\vnu^{ij}$
converges absolutely and uniformly to some $\dach f\in C(\LG)$.	 
\eunum
By the Peter-Weyl theorem $\dach f$ and $f$ coincide as 
$L^2$-functions. The continuity yields $\dach f \ident f$.
\qed
\epf
\bcorr
\label{corr:glm+abs_konv(fourier)}
Let $f\in C^{2s}_\Ad(\LG)$ be a $2s$-times continuously differentiable
conjugation invariant function on $\LG$
and let $2s \geq \dim\LG+1$.
Then
\zgl{\sum_{[\darst]\in\allirrdarst\LG}
         \lzweihaar{\charakt_\darst}{f} \charakt_\darst}\noindent
converges absolutely and uniformly to $f$.
\ecorr
The proof is straightforward.

\subsection{Fourier Series on $\LG^n$ and $\LG^n/{\mathrm{Ad}}$}
\label{uabschn:fourierLGn}
Later on we are mostly concerned not with functions on $\LG$, but on
$\LG^n$. To treat them we need complete orthonormal systems
on $\LG^n$ and $\LG^n/\Ad$. The first case is simple; one  
gets such systems due to $L^2(\LG^n) = \bigtensor^n L^2(\LG)$ by tensoring
orthonormal bases on $L^2(\LG)$. $\Ad$-invariant functions are more
complicated, since not every 
$\Ad$-invariant function on $\LG^n$ can be written as a tensor product 
of $\Ad$-invariant functions on $\LG$.
The solution of this problem comes from the theory of the so-called
loop-network states introduced by Thiemann \cite{b7}. 

Let $n\in\N_+$ be fixed and denote by
$\vec\darst\in\allirrdarst\LG^n$ some $n$-tuple of irreducible representations.
Analogously, $\veci$ and $\vecj$ are $n$-tuples of natural numbers.
We will call the functions
\zgl{
\nmatf^{\veci\vecj}_{\vec\darst} 
 \ident \nmatf^{i_1\ldots i_n \: j_1 \ldots j_n}_{\darst_1\ldots\darst_n}
 := \bigtensor_{\nu=1}^n \sqrt{\dim\darst_\nu} \: \darst_\nu^{i_\nu j_\nu} 
 : \LG^n \nach \C}\noindent
with $i_\nu, j_\nu = 1,\ldots,\dim\darst_\nu$
\df{elementary $n$-matrix functions} and
the $\Ad$-invariant functions
\zgl{
\nchar_{\vec\darst,\darst} 
 := \inv{\sqrt{\dim\darst}}
        \sum_{\veci,\vecj} \nmatf^{\veci\vecj}_{\vec\darst} 
                           \kontr^{\vecj\veci}_{\vec\darst,\darst} 
    : \LG^n \nach \C}\noindent
\df{$n$-characters}.
Here, $\darst$ is an irreducible representation contained 
in $\bigtensor_\nu \darst_\nu$ and
$\kontr_{\vec\darst,\darst} : 
  \bigtensor \VR_{\darst_\nu} \nach \VR_{\darst} 
  \teilmenge \bigtensor \VR_{\darst_\nu}$ 
is the corresponding projection matrix.
The set of all elementary $n$-matrix functions is denoted by $\nmatfkt$,
that of $n$-characters by $\ncharfkt$.

Now we have the generalized Peter-Weyl theorem
\bthm
\label{thm:peter+weyl_verallg}
\bnum{2}
\item
\bnum{2}
\item
$\nmatfkt$ is a complete orthonormal system in $L^2(\LG^n)$.
\item
$\spann_\C \nmatfkt$ is dense in $C(\LG^n)$.
\enum
\item
\bnum{2}
\item
$\ncharfkt$ is a complete orthonormal system in $L^2_\Ad(\LG^n)$.
\item
$\spann_\C \ncharfkt$ is dense in $C_\Ad(\LG^n)$.
\enum
\enum
\ethm
The proof is not very difficult, but quite technical and is therefore 
skipped here. It can be found in \cite{diss}.

\section{Determination of the Yang-Mills Measure}
\label{abschn:ym2:masz}
In this section we review the definition 
of the measure $\mu_\YM$ for the
two-dimensional quantum Yang-Mills theory, first proposed
by Thiemann \cite{b11} and Ashtekar et al.\ \cite{a6} for loops
in a quadratic lattice and later extended to the general case \cite{paper1}. 
However, in the last reference only Wilson loops have been used for the 
calculation. Although this is sufficient for unitary $\LG$, for arbitrary
groups we have to resort to the loop networks of Thiemann \cite{b7}
that will be introduced in a slightly modified
version in the next subsection. Afterwards we describe the chosen
regularization and quote the basic results from \cite{paper1} about
flag worlds. Next we review the definition of the Yang-Mills measure
in a formulation that (after proving the existence of a certain limit)
can directly be reused for other models.
Finally, the necessary expectation values of the Yang-Mills measure
are given.

From now on $M$ equals\footnote{Our considerations 
can quite easily be transferred
to the case of an arbitrary (compact) Riemannian surface. 
In the classical approach this has been performed by 
Fine \cite{d28}, Witten \cite{d26} and Sengupta \cite{d29,d30}.
Within the Ashtekar approach Ashtekar et al.\ \cite{a6}
were able to compute at least certain expectation values 
for $M = S^2$ or -- with $\LG = U(1)$ -- for $M = S^1 \kreuz S^1$ as well. 
The general case has been discussed by 
Aroca and Kubyshin \cite{iosa,iosa2}.} 
$\R^2$ and we restrict ourselves to the 
case of piecewise analytic paths.
Moreover, until the end of Section \ref{abschn:sing(YM2)}
we mean by ``graphs'' always connected simple graphs, i.e.\ graphs
whose interior domains are bounded by Jordan curves only.
This is not a severe restriction, since every graph can be 
refined to such a graph \cite{paper1,dipl}.
Simple domains are just domains enclosed by Jordan curves.
Finally we denote the Ashtekar-Lewandowski measure pushed forward by 
the homeomorphism $\abbAbGbagb:\AbGb\nach\agb$ from $\AbGb$ to $\agb$ 
again by $\mu_0$.

\subsection{Loop-Network States}
\label{uabschn:ym2:loopnetw}
For the determination of a regular Borel measure on a compact Hausdorff
space $X$ it is by the Riesz-Markov theorem sufficient to define 
a positive, linear and continuous functional $F$ on $C(X)$.
Since even this is quite difficult in general, one only determines
the restriction of $F$ to some (if possible, easily controllable)
dense subset $D$ of $C(X)$ and gets $F$ then by continuous extension.
In the case of $X = \agb$ one typically \cite{a6,paper1} chooses for $D$
the so-called holonomy algebra $\ha$ generated by the Wilson loops 
$T_\alpha = \tr\: h_\alpha : \agb \nach \C$.
However, here we will use the loop-network states introduced 
by Thiemann \cite{b7,b11}. Those indeed span (in a certain interpretation)
a dense subalgebra in $C(\agb)$ for {\em every}\/ $\LG$. 
For the holonomy algebra such a result is only known 
for $SU(N)$, $U(N)$, $SO(2N+1)$ and $O(N)$. 
For connected structure groups this problem has not been solved;
for non-connected ones there are counterexamples mentioned in the 
paper of Sengupta \cite{d25} based on investigations of Burnside. 

Before we come to the loop-network states, we recall the definition
of cylindrical functions on $\agb$.
\bdf
A function $f\in C(\agb)$ is called \df{cylindrical function},
if there is a graph $\GR$ and a function $f_\GR \in C(\Ab_\GR/\Gb_\GR)$ 
such that $f = f_\GR \circ \pi_\GR$.

The set of all cylindrical functions is denoted by $\Cyl(\agb)$.
\edf
Since we deal here with piecewise analytic 
graphs only, we have \cite{a48} 
\blem
\label{lem:cyl(agb)=dicht}
$\Cyl(\agb)$ is a dense $\ast$-subalgebra in $C(\agb)$.
\elem
\bdf
\bunum
\item
The triple $(\ga,\vec\darst,\darst)$ is called 
\df{loop-network} iff
\bunum
\item
there is a connected graph $\GR$ with $m\in\Ver(\GR)$ which 
$\ga$ is a weak fundamental system for,
\item
$\vec\darst = (\darst_1,\ldots,\darst_{\elanz\ga})$ 
is a $\elanz\ga$-tuple of (equivalence classes of) irreducible representations
of $\LG$ and
\item
$\darst$ is some irreducible representation contained 
in $\bigtensor_{i=1}^{\elanz\ga} \darst_i$.
\eunum
\item
Every loop-network 
$(\ga,\vec\darst,\darst)$ is assigned a function 
\zgl{
\lnws_{(\ga,\vec\darst,\darst)} 
   := \nchar_{\vec\darst,\darst} \circ \pi_\ga : \Ab \nach \C,}\noindent
where 
$\nchar_{\vec\darst,\darst} : \LG^{\elanz\ga} \nach \C$ is the
$\elanz\ga$-character to $(\vec\darst,\darst)$.

$\lnws_{(\ga,\vec\darst,\darst)}$ is called \df{loop-network state}.
\eunum
\edf
We note that our definition makes that of Thiemann \cite{b7} a bit more
general because here the impact of the choice of the generating system
is taken into account.
\blem
Let $(\ga,\vec\darst,\darst)$ be a loop-network and
$\GR$ be the graph spanned by $\ga$.
Then there is a unique continuous function 
$\lnwss_{(\ga,\vec\darst,\darst)} \in C(\Ab_\GR/\Gb_\GR)$ with
\zgl{\lnwss_{(\ga,\vec\darst,\darst)} \circ \pi_\GR 
                                      \circ \abbAbGbagb \circ \pi =
     \lnws_{(\ga,\vec\darst,\darst)}.}
\elem
\bpf
We set 
\zgl{\lnwss_{(\ga,\vec\darst,\darst)} 
     := (\iota_\ga^\GR)^\ast
         \bigl( (\pi_\Ad^\ast)^{-1} \nchar_{\vec\darst,\darst} \bigr),}\noindent
whereas 
$\pi_\Ad$ denotes the canonical projection from $\LG^n$ to $\LG^n/\Ad$
and $\iota_\ga^\GR$ is the homeomorphism $[h] \auf [h(\ga)]_\Ad$
between $\Ab_\GR/\Gb_\GR$ and $\LG^{\dim\pi_1(\GR)}/\Ad$.
$\lnwss_{(\ga,\vec\darst,\darst)}$ is well-defined 
by the $\Ad$-invariance of $\nchar_{\vec\darst,\darst}$ and 
we have
\noindent\bglklein
 \lnwss_{(\ga,\vec\darst,\darst)} \circ \pi_\GR \circ \abbAbGbagb \circ \pi
 & = & \bigl( (\pi_\Ad^\ast)^{-1} \nchar_{\vec\darst,\darst} \bigr) 
              \circ \iota_\ga^\GR \circ \pi_\GR \circ \abbAbGbagb \circ \pi \\
 & = & \bigl( (\pi_\Ad^\ast)^{-1} \nchar_{\vec\darst,\darst} \bigr) 
                   \circ \iota_\ga^\GR \circ \pi_{\Gb_\GR} \circ \pi_\GR \\
 & = & \bigl( (\pi_\Ad^\ast)^{-1} \nchar_{\vec\darst,\darst} \bigr) 
              \circ \pi_\Ad \circ \pi_\ga
 \breitrel= \nchar_{\vec\darst,\darst} \circ \pi_\ga
 \breitrel= \lnws_{(\ga,\vec\darst,\darst)}.
\eglklein\noindent
Here, $\pi_{\Gb_\GR}$ is the canonical projection 
$\Ab_\GR \nach \Ab_\GR/\Gb_\GR$.
\qed
\epf
Sometimes we call $\lnwss_{(\ga,\vec\darst,\darst)}$
loop-network state as well.
\bdf
The set of all functions $\lnwss_{(\ga,\vec\darst,\darst)}$ 
belonging to some $\ga$ is denoted by $\alllnwss_\ga$.
\edf
\bprop
\label{prop:lnwss=dicht}
Let $\GR$ be a graph and $\ga$ be some of its weak fundamental systems.

Then $\spann_\C \alllnwss_\ga$ is dense in $C(\Ab_\GR/\Gb_\GR)$.
\eprop
\bpf
Let $n := \elanz\ga$.
By the generalized Peter-Weyl theorem (Theorem \ref{thm:peter+weyl_verallg}) 
the set $\ncharfkt$ of all $n$-characters
$\nchar_{\vec\darst,\darst}$ spans a dense subspace in the space of all
$\Ad$-invariant continuous functions in $C(\LG^n)$.
The assertion now comes from the fact that 
$(\pi_\Ad^\ast)^{-1} : C_\Ad(\LG^n) \nach C(\LG^n/\Ad)$ 
and $(\iota_\ga^\GR)^\ast : C(\LG^n/\Ad) \nach C(\Ab_\GR/\Gb_\GR)$ 
are norm-preserving isomorphisms.
\qed
\epf
\bprop
\label{neuprop:alllnwss=dicht}
Choose for every graph $\GR$ some weak generating system $\ga(\GR)$
and put 
\zgl{\alllnwss := \bigcup_{\GR} \pi_\GR^\ast(\alllnwss_{\ga(\GR)}).}\noindent

Then $\spann_\C \alllnwss$ is dense in $C(\agb)$.
\eprop
\bpf
Let $f\in C(\agb)$. 
By the denseness of $\Cyl(\agb)$ in $C(\agb)$ 
(see Lemma \ref{lem:cyl(agb)=dicht})
there is a graph $\GR$ and an $f_\GR'' \in C(\Ab_\GR/\Gb_\GR)$
with $\supnorm{f_\GR''\circ\pi_\GR - f} < \einhalb\varepsilon$. 
By Proposition \ref{prop:lnwss=dicht} there is some 
$f_\GR' \in \spann_\C \alllnwss_{\ga(\GR)}$
with $\supnorm{f_\GR' - f_\GR''} < \einhalb\varepsilon$, hence with
$     \supnorm{f - f_\GR' \circ \pi_\GR} 
 \leq \supnorm{f - f_\GR''\circ \pi_\GR} +
      \supnorm{f_\GR'' \circ \pi_\GR - f_\GR' \circ \pi_\GR}
 \leq  \supnorm{f - f_\GR''\circ \pi_\GR} + 
      \supnorm{f_\GR'' - f_\GR'}
  < \varepsilon$. 
\qed
\epf
\brem
In contrast to \cite{b7} we do not claim that the set $\alllnws$ 
(after removing all loop-network states being pull-backs of other ones)
is an orthonormal system for $L^2(\agb)$. 
\erem
By means of the proposition above every regular Borel measure
on $\agb$ can be reconstructed from the corresponding expectation values 
of loop-network states. However, this does not mean
that every assignment of real numbers to loop-networks 
indeed corresponds to expectation values of some measure.

\subsection{Continuum Limit and Regularization of the Yang-Mills Action}
First we define our version of the continuum limit
$\GR \gegen \R^2$. (Actually, this definition includes both the continuum limit
and the thermodynamical limit.)
\bdf
\bnum{2}
\item
We say that a sequence $(\GR_n)$ of graphs
converges to $\R^2$ (shortly $\GR_n \gegen \R^2$) iff
for $n\gegen\infty$
\bunum
\item
the supremum of the diameters of all interior domains of $\GR_n$ 
goes to zero and
\item
the supremum of the diameters of all circles in $\R^2$ having center $m$
and being disjoint to the exterior domain of $\GR_n$ goes to infinity.
\eunum
\item
Let $\GR$ be a graph and $z_{\GR'} \in \C$ for all $\GR'\geq\GR$.

We say $\lim_{\GR\leq\GR'\gegen\R^2} z_{\GR'} = z$
iff $\lim_{n\gegen\infty} z_{\GR_n} = z$
for all sequences $(\GR_n)$ of graphs with $\GR_n\geq\GR$ and 
$\GR_n\gegen\R^2$.
\enum
\edf
Now we come to the regularization of the Yang-Mills action.
Neglecting convergence problems we have for all 
sufficiently regular partitions $\{G_\alpha\}$ of $\R^2$:
\noindent\bgl
 S_\YM(A) 
      & = & \inv4 \int_{\R^2} \tr\: F_{\mu\nu} F^{\mu\nu} \: \dd x \\
      & = & \inv2 \sum_\alpha \int_{G_\alpha} \tr\: F_{12} F^{12} 
                               \: \dd x^1 \keil \dd x^2 \\
  & \rund & \inv2 \sum_\alpha \FIL\alpha 
                              \tr\: \frac{(h_A(\alpha) - \EINS)^2}{\FIL\alpha^2} 
                  \erl{Proposition \ref{prop:kl_holo}}\pktplatz \\
  & \rund & \sum_\alpha \frac{N}{\FIL\alpha}
                        \bigl(1 - \einsdurchn \re \tr\: h_A(\alpha)\bigr) \\
  &       &           
                  \erl{$\tr(g-\EINS)^2 \rund -2\re\tr(g-\EINS)$
                       for small $\algnorm{g-\EINS}$}. \\
\egl\noindent\noindent
Including the coupling constant $\kopp\in (0,\infty)$ 
of the theory we have (cf.\ \cite{Wilson,a6,paper1,iosa,iosa2})
\bdf
For every graph $\GR$ the \df{regularized Yang-Mills action} 
$\ymwirk[\GR] : \Ab \nach \R$ is defined by
\zgl{
  \ymwirk[\GR] (\qa)
    = \sum_{G_I\in L_\innen(\GR)} 
        \frac{N}{\kopp^2}\inv{\FI{G_I}}
          \Bigl(1-\einsdurchn\re\tr\: h_\qa(\alpha_I)\Bigr).}\noindent
Here, $\alpha_I$ is some boundary loop of the domain
$G_I$ in $\GR$, $N$ is the natural number given by the embedding 
$\LG \teilmenge U(N)$ and $L_\innen(\GR)$ collects the interior domains
of $\GR$.
\edf
Obviously, $\ymwirk[\GR]$ is well-defined and a gauge-invariant function.
This way, we get by
$(\pi^\ast)^{-1} \ymwirk[\GR] \circ \abbAbGbagb^{-1}:\agb\nach\R$
a regularized Yang-Mills action on
$\agb$
that will be denoted by $\ymwirk[\GR]$ as well. 

One could now try to define via
$\ymwirk[\text{reg}](\qa) :=  
 \lim_{m\leq\GR\gegen\R^2} \ymwirk[\GR](\qa)$ a
generalized Yang-Mills action on $\Ab$.
Although then --~as indicated above~--
we would have $\ymwirk[\text{reg}](A) = \ymwirk(A)$ for smooth connections 
(if necessary under additional regularity assumptions);
for generalized connections the 
limit $\ymwirk[\text{reg}]$ however does typically {\em not}\/ exist.
It is easy to see that $\ymwirk[\GR](\qa)$ still converges in $[0,\infty]$
for every limiting process 
$\GR_n \gegen \R^2$ where $\GR_{n+1}$ is a refinement of $\GR_n$,
but this limit depends crucially on the considered limiting process.
There are even connections that yield sometimes
$0$ and sometimes $\infty$ in the limit \cite{paper1}.

\subsection{Flag Worlds}
\label{uabschn:ym2:fw}
In this section we recall the most important facts about the 
flag worlds and adapt them to our formulation. 
Their introduction has been necessary 
because the Wilson regularization 
has been extended from quadratic to floating lattices. 
For a detailed discussion we refer to \cite{paper1,diss}.

\blem
For every simple domain $G$ in $\GR$ and every 
$v\in\Ver(\GR)\cap \del G$ there is a unique
loop $\alpha_{G,v}$ in $\GR$ with base point $v$ and
without self-intersections, such that
\bunum
\item
$\im \alpha_{G,v}$ equals the boundary $\del G$ of $G$ and
\item
$\alpha_{G,v}$ surrounds $G$ counterclockwise.
\eunum
\elem
We call $\alpha_{G,v}$ boundary loop of $G$ with base point $v$.
\bdf
\label{neuvarfahne}
\label{neufahne}
Let $G$ be a simple domain in a graph $\GR$.

A closed path $f_{G,v}$ in $\GR$ is called 
\df{flag} for the domain $G$ iff
\bunum
\item
$f=\gamma_{mv} \alpha_{G,v} \gamma_{mv}^{-1}$,
\item
$\alpha_{G,v}$ is the boundary loop of $G$ with base point $v$ and
\item
$\gamma_{mv}$ is a path in $\GR\setminus\GR_G$ from $m$ to $v$.
\eunum
\edf
Here, $\GR_G$ denotes the subgraph of $\GR$ built by the boundary of $G$.
If we later say a path to be a flag, we will simply assume the 
existence of some graph $\GR$ that contains this path as a flag.
For instance, every flag itself spans a graph with exactly one 
interior domain. One easily sees that for every simple domain $G$ in $\GR$ 
and every $v\in\Ver(\GR_G)$ there is a flag $f_{G,v}$.
Additionally, two flags in a simple graph $\GR$ are called 
non-overlapping iff the domains enclosed by them are disjoint.

The most important examples of sets of non-overlapping flags are the
flag worlds.
\bdf
\label{neuflag world}
\label{neuvarFW}
Let $\GR$ be a simple graph.
\bunum
\item
A set $\FW$ of flags in $\GR$ is called \df{flag world} for $\GR$
iff $\FW = \{f_G\mid G \text{ interior domain of $\GR$}\}$, where 
$f_G$ in each case is some flag enclosing the (simple) 
interior domain $G$.
\item
A flag world of $\GR$ is called \df{moderately independent} iff
it is a weak fundamental system for $\GR$.
\eunum
\edf
\bprop
\label{prop:ex(mu_fw)}
In every simple graph there is a moderately independent flag world.
\eprop

For the investigation of the limiting process in the calculation of
the Wilson-loop expectation values that runs over all refinements of a
given graph, we have to study the behaviour of flag worlds under refining
the graph under consideration.
\bdf
\label{def:verf(fw)}
Let $\GR$ and $\GR'$ be simple graphs with $\GR' \geq \GR$ and
$\FW$ and $\FW'$ be flag worlds for $\GR$ and $\GR'$, respectively.

$\FW'$ is called \df{refinement} of $\FW$ iff for every 
interior domain $G_I$ of $\GR$ the flag $f_I \in \FW$
corresponding to $G_I$ can be written as a product
$f_{I,1} \cdots f_{I,\lambda_I}$ of just those flags $f_{I,i_I} \in \FW'$ 
that belong to the interior domains $G_{I,i_I}\teilmenge G_I$ of $\GR'$.
\edf
Obviously the refinement relation is transitive.
\bprop
\label{prop:ex(verf_mu_fw)}
Let $\GR$ and $\GR'$ be simple graphs with $\GR' \geq \GR$.

Then for every moderately independent flag world $\FW$ of $\GR$ there 
is a moderately independent flag world $\FW'$ of $\GR'$ being a 
refinement of $\FW$.
\eprop
The rather technical proof is analogous to that given in \cite{paper1}
for a similar proposition.

Finally we express the Yang-Mills action in terms of flag worlds.
\blem
\label{neulem:ymwirk(fw)}
For every graph $\GR$ and every moderately independent flag world
$\FW$ in $\GR$ there is a unique continuous function
$\ymwirk[\GR]^\FW : \LG^{\rh(\GR)} \nach \R$
with
$\ymwirk[\GR]^\FW \circ \pi_\FW = \ymwirk[\GR]$.

Then we have \zgl{
  \ymwirk[\GR]^\FW (\vec g)
    = \sum_{G_I\in L_\innen(\GR)} 
        \frac{N}{\kopp^2}\inv{\FI{G_I}}
          \Bigl(1-\einsdurchn\re\tr\: g_I\Bigr).}\noindent
\elem
$\rh(\GR)$ denotes the rank of the fundamental group of $\GR$.

\subsection{Expectation Values}
Since the limit $\lim_{m\leq\GR\gegen\R^2} \ymwirk[\GR]$ 
on $\agb$ is known to be not well-defined, a 
definition of the expectation values via
\zgl{
  \inv Z \int_\agb 
         \e^{-\lim_{m\leq\GR\gegen\R^2} \ymwirk[\GR]}
                     f \: \dd\mu_0}\noindent
is not meaningful. Now, the trick of Thiemann 
was simply to exchange limit and integration.
A priori it is unclear whether this is possible, in particular, because of the
customary non-existence of the limit $\ymwirk[\text{reg}]$.
However, just this (mathematically not justifiable)
operation enables the calculation of the expectation values.
Moreover, the results become completely independent of the choice of 
the limiting process.

Now we 
\bnum{4}
\item
set for all graphs $\GR$, all continuous functions $f_\GR \in C(\Ab_\GR/\Gb_\GR)$
and all refinements $\GR'$ of $\GR$
\zglnum{
\hspace*{-5em}\phantom{\frac{\int_\agb \e^{-\ymwirk[\GR']} 
                     f_\GR \circ \pi_\GR \: \dd\mu_0}
          {\int_\agb \e^{-\ymwirk[\GR']} 
                     \phantom{f_\GR \circ \pi_\GR} \: \dd\mu_0},}
\erw_{\GR'}^\GR (f_\GR) 
  \breitrel{:=} \frac{\int_\agb \e^{-\ymwirk[\GR']} 
                     f_\GR \circ \pi_\GR \: \dd\mu_0}
          {\int_\agb \e^{-\ymwirk[\GR']} 
                     \phantom{f_\GR \circ \pi_\GR} \: \dd\mu_0},
\phantom{\erw_{\GR'}^\GR (f_\GR)}
}{gl:erww1}\noindent
\item
set for all graphs $\GR$ and all continuous functions $f_\GR \in C(\Ab_\GR/\Gb_\GR)$
\zglnum{
\hspace*{-5em}\phantom{\lim_{\GR\leq\GR'\gegen\R^2} \erw_{\GR'}^\GR (f_\GR),}
\erw^\GR (f_\GR) 
  \breitrel{:=} \lim_{\GR\leq\GR'\gegen\R^2} \erw_{\GR'}^\GR (f_\GR),
\phantom{\erw^\GR (f_\GR)}}{gl:erww2}\noindent
\item
set for all cylindrical functions $f\in\Cyl(\agb)$
\zglnum{\hspace*{-5em}\phantom{\erw^\GR (f_\GR),}
\erw(f) \breitrel{:=} \erw^\GR (f_\GR), \phantom{\erw(f)}
}{gl:erww3}\noindent
where $\GR$ is some graph and $f_\GR\in C(\Ab_\GR/\Gb_\GR)$ some 
function fulfilling $f = f_\GR \circ \pi_\GR$, and
\item
extend $\erw : \Cyl(\agb) \nach \C$ to a functional
\zglnum{\erw \breitrel: C(\agb) \nach \C.}{gl:erww4}\noindent
\enum
Finally, this functional $\erw$ determines the desired regular Borel measure
$\mu_\YM$ on $\agb$ via the Riesz-Markov theorem.
The most (and only) important step in this procedure is 
the proof of the following
\bprop
\label{prop:ex(erww_lim)}
For every graph $\GR$, for every moderately independent 
flag world $\FW$ in $\GR$ and for every
loop-network state 
$\lnwss_{(\FW,\vec\darst,\darst)}\in\alllnwss_\FW$ the limit
\eqref{gl:erww2} exists.
\eprop

Namely, this proposition implies
\bthm
The functional $\erw : C(\agb) \nach \C$ 
is well-defined, linear, continuous and positive.

Hence there is a unique normalized 
regular Borel measure $\mu_\YM$ on $\agb$
with $\erw(f) = \int_\agb f \: \dd\mu_\YM$ for all $f\in C(\agb)$.
\ethm
\bpf
\bnum{5}
\item
Well-definedness of $\erw^\GR$

The linearity of $\erw_{\GR'}^\GR$ in \eqref{gl:erww1}
implies the existence of the limit $\erw^\GR$ in \eqref{gl:erww2}
for all $f_\GR \in \spann_\C \alllnwss_\FW$. 

By
$\betrag{\erw_{\GR'}^\GR (f_\GR)} \leq \supnorm{f_\GR\circ\pi_\GR} 
                                  \leq \supnorm{f_\GR}$
for all $f_\GR \in C(\Ab_\GR/\Gb_\GR)$ we have 
$\betrag{\erw^\GR (f_\GR)} \leq \supnorm{f_\GR}$
on $\spann_\C \alllnwss_\FW$.
Hence $\erw^\GR$ is a linear, continuous and 
(as $\erw_{\GR'}^\GR$) positive functional on $\spann_\C \alllnwss_\FW$.

Since $\spann_\C \alllnwss_\FW$
is dense in $C(\Ab_\GR/\Gb_\GR)$ 
(Proposition \ref{prop:lnwss=dicht}), we can extend $\erw^\GR$ continuously
to a linear, continuous and positive functional $\quer\erw{}^\GR$
on the whole $C(\Ab_\GR/\Gb_\GR)$. 
Moreover, obviously
$\quer\erw{}^\GR (f_\GR) 
  = \lim_{\GR\leq\GR'\gegen\R^2} \erw_{\GR'}^\GR (f_\GR)
  =: \erw^\GR (f_\GR)$
for all $f_\GR\in C(\Ab_\GR/\Gb_\GR)$.
Hence $\quer\erw{}^\GR = \erw^\GR$.
\item
Well-definedness of $\erw$ on $\Cyl(\agb)$

Let $\GR\geq\GR_0$ and $f_{\GR_0} \in C(\agb_{\GR_0})$.
We have 
$\erw_{\GR'}^{\GR} (f_{\GR_0} \circ \pi_{\GR_0}^\GR) =
 \erw_{\GR'}^{\GR_0} (f_{\GR_0})$ for all $\GR'\geq\GR$
by \eqref{gl:erww1} and $\pi_{\GR_0} = \pi_{\GR_0}^\GR \circ \pi_\GR$.
Thus,
\noindent\bglklein
\erw^{\GR} (f_{\GR_0} \circ \pi_{\GR_0}^\GR) 
 & = & \lim_{\GR\leq\GR'\gegen\R^2}  
           \erw_{\GR'}^\GR (f_{\GR_0} \circ \pi_{\GR_0}^\GR) \\
 & = & \lim_{\GR\leq\GR'\gegen\R^2}  
           \erw_{\GR'}^{\GR_0} (f_{\GR_0}) \\
 & = & \lim_{\GR_0\leq\GR'\gegen\R^2}  
           \erw_{\GR'}^{\GR_0} (f_{\GR_0}) \\
 & = & \erw^{\GR_0} (f_{\GR_0}),
\eglklein\noindent
because the limit of a subsequence of a convergent sequence equals the 
limit of the total sequence.

Let now $\GR_1$ and $\GR_2$ be two graphs and $f_{\GR_1}$ and $f_{\GR_2}$
two continuous functions on $\agb_{\GR_1}$ and $\agb_{\GR_2}$, resp.,
with $f_{\GR_1} \circ \pi_{\GR_1} = f = f_{\GR_2} \circ \pi_{\GR_2}$.
For every refinement $\GR$ of $\GR_1$ and $\GR_2$
we have $f = f_\GR \circ \pi_\GR$ with 
$f_{\GR_1} \circ \pi_{\GR_1}^{\GR} = f_\GR = f_{\GR_2} \circ \pi_{\GR_2}^{\GR}$.
($f_\GR$ is well-defined by the surjectivity of $\pi_\GR$.)
As just seen, we have
\zgl{
\erw^{\GR_1} (f_{\GR_1}) = 
 \erw^{\GR} (f_{\GR_1} \circ \pi_{\GR_1}^\GR) =
 \erw^{\GR} (f_{\GR_2} \circ \pi_{\GR_2}^\GR) =
 \erw^{\GR_2} (f_{\GR_2}),}\noindent
i.e., $\erw$ is well-defined.
\item
Linearity, positivity and continuity of $\erw$ on $\Cyl(\agb)$

Linearity and positivity follow immediately from the corresponding
properties of $\erw^\GR$.
For the continuity let $f = f_\GR \circ \pi_\GR \in\Cyl(\agb)$.
Then $\erw(f) = \erw^\GR(f_\GR) \leq \supnorm{f_\GR}
   = \supnorm{f_\GR\circ\pi_\GR} = \supnorm{f}$ 
by the surjectivity of $\pi_\GR$ and by $\norm{\erw^\GR} \leq 1$.
\enum
Since the cylindrical functions form a dense subspace of $C(\agb)$,
the extension of $\erw$ to $C(\agb)$ is well-defined, linear,
continuous and positive.
Hence, by the Riesz-Markow theorem there is a unique
regular Borel measure $\mu_\YM$ on $\agb$
with $\erw(f) = \int_\agb f \: \dd\mu_\YM$ for all $f\in C(\agb)$.
The normalization comes from $\mu_\YM(\agb) = \erw(1) = 1$.
\qed
\epf
\brem
In the proof above we implicitely used the fact, that any 
graph can be refined to a simple connected graph.
This way we can find, in particular, for every cylindrical function $f$
a simple connected graph $\GR$ such that 
$f = f_\GR \circ \pi_\GR$ for some continuous $f_\GR$.
\erem
Note finally, that in this subsection it actually does not matter whether 
we speak about two-dimensional Yang-Mills theory or any other gauge theory.
We could simply substitute the regularized Yang-Mills actions
$\ymwirk[\GR]$ by some other actions. 
The above theorem remains valid for all theories; we have 
to guarantee ``only'' that Proposition \ref{prop:ex(erww_lim)} is valid.

\subsection{Determination of the Expectation Values}
Let us now take care of Proposition \ref{prop:ex(erww_lim)} (of course,
only in the $\YM_2$-case).

Let $\GR$ be some graph and $\FW$ be a moderately independent
flag world in $\GR$.
Furthermore, let $\GR'$ be a refinement of $\GR$ and $\FW'$ be
a corresponding refinement of $\FW$ according to 
Proposition \ref{prop:ex(verf_mu_fw)}.
Finally, $(\FW,\vec\darst,\darst)$ be some loop-network.
Then
\zgl{
     \lnwss_{(\FW,\vec\darst,\darst)} \circ \pi_\GR \circ \abbAbGbagb \circ \pi
  =  \lnws_{(\FW,\vec\darst,\darst)}
  =  \nchar_{\vec\darst,\darst} \circ \pi_\FW
  =  \nchar_{\vec\darst,\darst} \circ \pi_\FW^{\FW'} \circ \pi_{\FW'}.
}\noindent
Thus --~remember that $\mu_0$ and $\ymwirk[\GR']$
denote both objects on $\Ab$ and on $\agb$~--
\noindent\bgl
&   & \int_\agb \e^{-\ymwirk[\GR']} \:\:
         \lnwss_{(\FW,\vec\darst,\darst)} \circ \pi_\GR \: \dd\mu_0 \\
& = & \int_\Ab  \e^{-\ymwirk[\GR']} \:\:
         \lnwss_{(\FW,\vec\darst,\darst)} \circ \pi_\GR \circ \abbAbGbagb \circ \pi 
	  \: \dd\mu_0 \\
& = & \int_\Ab  \e^{-\ymwirk[\GR']^{\FW'} \circ \pi_{\FW'}} \:\:
         \nchar_{\vec\darst,\darst} \circ \pi_\FW^{\FW'} \circ \pi_{\FW'}
         \: \dd\mu_0  
      \erl{$\ymwirk[\GR']^{\FW'} \circ \pi_{\FW'} = \ymwirk[\GR']$,
           see Lemma \ref{neulem:ymwirk(fw)}} \\
& = & \int_{\LG^{\rh(\GR')}}  \e^{-\ymwirk[\GR']^{\FW'}} \:\:
         \nchar_{\vec\darst,\darst} \circ \pi_\FW^{\FW'}
         \: \dd\mu_\Haar^{\rh(\GR')}. 
\egl\noindent
In the last step we used that per definitionem
every moderately independent flag world is a weak fundamental system
and that therefore \cite{paper5} 
the projection of $\mu_0$ w.r.t.\ $\pi_{\FW'}$ 
equals the Haar measure.

Let us denote by $G_I$, $I = 1, \ldots, \rh(\GR)$, the interior domains of
$\GR$ and by $G_{I,i_I}$ those of $\GR'$. Here we assume that 
every $G_I$ is just refined into the set $\{G_{I,1},\ldots,G_{I,\lambda_I}\}$. 
Then we have
\zgl{
  \ymwirk[\GR']^{\FW'} (\vec g)
    = \sum_{G_{I,i_I} \in L_\innen(\GR')} 
        \frac{N}{\kopp^2}\inv{\FI{G_{I,i_I}}}
          \Bigl(1-\einsdurchn\re\tr\: g_{I,i_I}\Bigr)}\noindent
and due to the special relation between 
flags in $\FW$ and flags in $\FW'$ (cf. Definition \ref{def:verf(fw)})
\noindent\bgl
(\nchar_{\vec\darst,\darst}\circ \pi_\FW^{\FW'})(\vec g) 
 & = & \inv{\sqrt{\dim\darst}}
       \sum_{\vec p,\vec q} 
         \kontr^{\vec q\vec p}_{\vec\darst,\darst}
         \prod_{I=1}^{\rh(\GR)} 
             \sqrt{\dim\darst_I} \: 
             \darst_I^{p_I q_I}(g_{I,1} \cdots g_{I,\lambda_I}) \\
 & = & \inv{\sqrt{\dim\darst}}
       \sum_{\vec p,\vec q} 
         \kontr^{\vec q\vec p}_{\vec\darst,\darst}
         \prod_{I=1}^{\rh(\GR)} 
             \sqrt{\dim\darst_I} \: 
             \delta^{p_I}_{r_{I,0}} \delta^{q_I}_{r_{I,\lambda_I}} \:
	     \prod_{{i_I}=1}^{\lambda_I}
	       \darst_I^{r_{I,{i_I}-1}\:r_{I,{i_I}}}(g_{I,{i_I}}).
\egl\noindent
Since
$\e^{-\ymwirk[\GR']^{\FW'}}
  \:\: \nchar_{\vec\darst,\darst} \circ \pi_\FW^{\FW'}$
is a product of functions in the $g_{I,i_I}$,
the integral over $\LG^{\rh(\GR')}$ factorizes into the single 
$\LG$-integrations. Additionally,
\zgl{
\mu_{\Haar,\FI G} := \e^{-\frac{N}{\kopp^2}\inv{\FI G}
                      (1-\einsdurchn\re\tr\: \cdot)} \kp \mu_\Haar}\noindent
is an $\Ad$-invariant regular Borel measure on $\LG$; hence
\noindent\bgl
\hspace*{-3.0em}
 &   & \int_{\LG^{\rh(\GR')}}  \e^{-\ymwirk[\GR']^{\FW'}} \:\:
         \nchar_{\vec\darst,\darst} \circ \pi_\FW^{\FW'}
         \: \dd\mu_\Haar^{\rh(\GR')} \\
\hspace*{-3.0em}
 & = & \inv{\sqrt{\dim\darst}}
       \sum_{\vec p,\vec q} 
         \kontr^{\vec q\vec p}_{\vec\darst,\darst}
         \prod_{I > \rh(\GR)} \Biggl(
           \int_\LG 
                    \: \dd\mu_{\Haar, \FI{G_{I,1}}}\Biggr) \\
\hspace*{-3.0em}
 &   & \hspace*{0.19cm}
       \phantom{
       \inv{\sqrt{\dim\darst}}
       \sum_{\vec p,\vec q} 
         \kontr^{\vec q\vec p}_{\vec\darst,\darst}}
         \prod_{I=1}^{\rh(\GR)} 
             \Biggl(\sqrt{\dim\darst_I} \: 
             \delta^{p_I}_{r_{I,0}} \delta^{q_I}_{r_{I,\lambda_I}} \:
             \prod_{{i_I}=1}^{\lambda_I}
              \int_\LG
               \darst_I^{r_{I,{i_I}-1}\:r_{I,{i_I}}}(g_{I,{i_I}})
               \: \dd\mu_{\Haar,\FI{G_{I,i_I}}}(g_{I,{i_I}})\Biggr) \\
\hspace*{-3.0em}
 & = & \sqrt{\dim\darst} \:
         \prod_{I > \rh(\GR)} \Biggl(
           \int_\LG \: \dd\mu_{\Haar, \FI{G_{I,1}}}\Biggr) \\
\hspace*{-3.0em}
 &   & \hspace*{0.19cm}
       \phantom{
       \sqrt{\dim\darst} \:}       
       \prod_{I=1}^{\rh(\GR)} 
             \Biggl( \sqrt{\dim\darst_I}
             \prod_{{i_I}=1}^{\lambda_I}
              \int_\LG
               \inv{\dim\darst_I}
	       \charakt_{\darst_I}(g_{I,{i_I}})
               \: \dd\mu_{\Haar,\FI{G_{I,i_I}}}(g_{I,{i_I}})\Biggr). 
	       \\
\egl\noindent
In the last step we used the fact that first the  
integrals over matrix functions can be reduced to the integrations 
of the corresponding characters \cite{a6}
and second 
$\kontr_{\vec\darst,\darst}$ is the projection of
$\bigtensor_I \VR_{\darst_I}$ to $\VR_\darst$ and hence has
trace $\dim\darst$.

Hence, we have
\zglnum{
\erw_{\GR'}^\GR (\lnwss_{(\FW,\vec\darst,\darst)}) 
  =  \sqrt{\dim\darst} \:
         \prod_{I=1}^{\rh(\GR)} 
             \Biggl( 
             \sqrt{\dim\darst_I}
             \prod_{{i_I}=1}^{\lambda_I}
              \frac{\int_\LG
                    \inv{\dim\darst_I}
	            \charakt_{\darst_I}
                    \: \dd\mu_{\Haar,\FI{G_{I,i_I}}}}
                   {\int_\LG
                    \phantom{\inv{\dim\darst_I}}
	            \phantom{\charakt_{\darst_I}}
                    \: \dd\mu_{\Haar,\FI{G_{I,i_I}}}}
             \Biggr).
}{gl:erwwlnws}\noindent
We see, in particular, that integrals for domains outside of $\GR$
($I > \rh(\GR)$) do not contribute by the normalization. 
Thus, the only terms in \eqref{gl:erwwlnws} depending on $\GR'$
are the products $\prod_{i_I}\frac\cdots\cdots$ over
all refinements of the interior domains of $\GR$ into interior domains 
of $\GR'$.
Consequently, to prove convergence of 
$\lim_{\GR\leq\GR'\gegen\R^2} 
    \erw_{\GR'}^\GR (\lnwss_{(\FW,\vec\darst,\darst)})$
we only have to show that
\zglnum{
 \prod_{i=1}^{\lambda}
  \frac{\int_\LG \inv{\dim\darst_I} \charakt_{\darst_I}
                 \: \dd\mu_{\Haar,\FI{G_i}}}
       {\int_\LG \phantom{\inv{\dim\darst_I}} \phantom{\charakt_{\darst_I}}
                 \: \dd\mu_{\Haar,\FI{G_i}}}
}{gl:erwwJquot}\noindent
for $\{G_i\}$ with $\sum_{i=1}^\lambda \FI{G_i} = \FI G = \const$ always
goes to one and the same value if $\sup \FI{G_i}$ goes to zero.
This proof is not very difficult, but technically strenuous.
Therefore we simply refer to \cite{dipl}. The proof given there for 
$\LG = SU(N)$ and $\LG = U(1)$ can be quite easily extended to general 
compact $\LG$.\footnote{Similar proofs are already contained 
in articles about ``ordinary'' lattice Yang-Mills theory 
(see, e.g., \cite{g6}) written before the Ashtekar approach was born.
However, they were restricted to the case of quadratic lattices.}
It gives immediately
\bprop
\label{prop:erwwlnws}
Let $\GR$ be a graph having interior domains $G_I$ and
$\FW$ be a moderately independent flag world in $\GR$. 
Moreover, let $(\FW,\vec\darst,\darst)$ be a loop-network and 
$c_{\darst_I}$ be the corresponding eigenvalue
of the Casimir operator of the representation $\darst_I$. Then we have
\zgl{
\erw(\lnwss_{(\FW,\vec\darst,\darst)} \circ \pi_\GR) 
  = \erw^\GR (\lnwss_{(\FW,\vec\darst,\darst)}) 
  = \sqrt{\dim\darst} \:
         \prod_{I=1}^{\rh(\GR)} 
             \Bigl( 
             \sqrt{\dim\darst_I} \:
             \e^{-\einhalb \kopp^2 c_{\darst_I} \FI{G_I}}
             \Bigr).
}\noindent
If $\GR$ has precisely one interior domain $G$ 
(i.e.\ $\GR$ is a flag $\beta$), 
we have for all irreducible representations $\darst$
\zgl{
\erw(\lnwss_{(f,\darst,\darst)}\circ\pi_\GR) 
  = \dim\darst \:\: \e^{-\einhalb \kopp^2 c_\darst \FI G}.
\hspace*{0.74cm}
}\noindent
\eprop
This concludes the proof of the existence of the 
physical measure $\mu_\YM$ for the two-dimensional Euclidian
quantum Yang-Mills theory within the Ashtekar approach.
Moreover, all Wilson-loop expectation values given here 
coincide with those 
of other approaches \cite{e31,g11a,g11b,g15,g19,e23,e24,e22}.

\section{Radon-Nikodym Derivatives}
In this section we show that the push-forwards of the 
Yang-Mills measure to the lattice theories are absolutely continuous
w.r.t.\ the lattice Haar measures and study the 
properties of the corresponding Radon-Nikodym derivatives,
in particular, of their Fourier expansions. 

\label{abschn:ym2:ymd}
\nocite{Bellman} 
By means of the homeomorphism $\abbAbGbagb$ between $\AbGb$ and $\agb$ 
we can regard $\mu_\YM$ as a measure on $\AbGb$.
Moreover, it is well-known
that the canonical projection $\pi:\Ab\nach\AbGb$ yields a natural 
bijection between the $\Gb$-invariant measures on $\Ab$ and the measures
on $\AbGb$. 
Hence, there is a unique normalized $\Gb$-invariant Borel measure
$\mu_{\Ab,\YM}$ on $\Ab$, whose image measure on $\agb$ equals $\mu_\YM$.
Sometimes we will write instead of $\mu_{\Ab,\YM}$ simply $\mu_\YM$.
We get 
\bcorr
\label{corr:erwwlnws}
Under the assumptions of Proposition \ref{prop:erwwlnws} we have
\zglnum{
\int_\Ab \nchar_{\vec\darst,\darst} \circ \pi_\FW \: \dd\mu_{\Ab,\YM}
     = \sqrt{\dim\darst} \:
         \prod_{I=1}^{\rh(\GR)} 
             \Bigl( 
             \sqrt{\dim\darst_I} \:
             \e^{-\einhalb \kopp^2 c_{\darst_I} \FI{G_I}}
             \Bigr)
}{gl:erwwlnwsexplizit}\noindent
and 
\zglnum{\hspace*{-2.74cm}\hspace*{-0.14cm}
\int_\Ab \charakt_\darst \circ \pi_\beta \: \dd\mu_{\Ab,\YM}
       = \dim\darst \:\: \e^{-\einhalb \kopp^2 c_\darst \FI G}.
}%
{gl:erwwlnwsexplizitfahne}\noindent
\ecorr

The theory of measures on projective limits \cite{Ya}
allows to map $\mu_\YM$ consistently to the lattice gauge theories.
\bdf
Let $\GR$ be some (again connected) graph 
and $\ga$ be a weak fundamental system for $\GR$.
Then we denote by
\bunum
\item
$\mu_{0,\ga} := \pi_\ga{}_\ast \mu_0$ 
the image measure of $\mu_0$ on 
$\Ab_\ga \iso \LG^{\elanz\ga} \ident \LG^{\rh(\GR)}$ and 
\item
$\mu_{\YM,\ga} := \pi_\ga{}_\ast \mu_{\Ab,\YM}$ 
the image measure of $\mu_{\Ab,\YM}$ 
on $\LG^{\rh(\GR)}$.
\eunum
\edf
Since $\ga$ is a weak fundamental system, we have 
$\mu_{0,\ga} = \mu_\Haar^{\elanz\ga}$.
The continuity of $\pi_\ga$ implies
\blem
For every graph $\GR$ and every weak fundamental system (hence, in particular
every moderately independent flag world) $\ga$ of $\GR$,
the measure $\mu_{\YM,\ga}$ on $\LG^{\elanz\ga}$  
is $\Ad$-invariant, regular and Borel.
\elem

We know that $\mu_\YM$ and $\mu_0$ can be reconstructed from the 
corresponding self-consistent families
$(\mu_{\YM,\ga})_\ga$ and $(\mu_{0,\ga})_\ga$, respectively.%
\footnote{This is true indeed, because every arbitrary 
(not necessarily connected and simple) graph
can be refined to a connected and simple graph.}
To study the relation between $\mu_\YM$ and $\mu_0$ below,
we first investigate the relations between
$\mu_{\YM,\ga}$ and $\mu_{0,\ga}$.
The easiest case is that of a single flag $\beta$.%
\footnote{In order to avoid confusion of a flag $f$ with functions $f$,
we denote flags (as general closed paths) by $\alpha$ or $\beta$
and write $\ga$ or $\gb$ instead of $\FW$ for flag worlds analogously.}
If $\mu_{\YM,\beta}$ were absolutely continuous w.r.t.\ $\mu_{0,\beta}$,
i.e. $\mu_{\YM,\beta} = \ymd_\beta \kp \mu_{0,\beta}$ for some
appropriate $\ymd_\beta : \LG \nach \C$, we would have
\zgl{
\lzweihaar{\charakt_\vnu}{\ymd_\beta} 
  = \int_\LG \quer\charakt_\vnu \: \ymd_\beta \: \dd\mu_{0,\beta}
  = \int_\LG \quer\charakt_\vnu \: \dd\mu_{\YM,\beta}
  = d_\vnu \: \e^{-\einhalb \kopp^2 c_\vnu \FI G},}\noindent
i.e., integrability assumed, 
$\ymd_\beta = \sum_\vnu d_\vnu \: \e^{-\einhalb \kopp^2 c_\vnu \FI G}
                        \charakt_\vnu$.
Indeed the existence of such a $\ymd_\beta$ has been proven in \cite{b7}.
Here we even show that $\ymd_\beta$ is continuous and grows at most
polynomially for vanishing $\FIL\beta$.
\bprop
\label{prop:ymd_fahne}
For every flag $\beta$ there is an $\Ad$-invariant 
continuous\footnote{One can even prove $\ymd_\beta\in C^\infty(\LG)$.}
function $\ymd_\beta:\LG\nach\C$ with 
$\mu_{\YM,\beta} = \ymd_\beta \kp \mu_{0,\beta}$.
Moreover, we have:
\bnum{2}
\item
The Fourier series
\zglnum{\sum_{\vnu\in\allirrdarst\LG} d_\vnu \: 
         \e^{-\einhalb \kopp^2 c_\vnu \FIL\beta} \: \charakt_\vnu}{gl:ymdfour}\noindent
converges absolutely and uniformly to $\ymd_\beta$.
\item
There are constants $\const_\nu$ that depend only on $\nu$ (and $\LG$), but not
on $\FIL\beta$, such that
\zgl{\supnorm{\ymd_\beta} \ident \sup_{g\in\LG} \betrag{\ymd_\beta(g)}
        = \ymd_\beta(e_\LG) 
     \leq 1 + \sum_{\nu=1}^{\dim\LG}\const_\nu \FIL\beta^{-\frac\nu2}.}\noindent
\enum
\eprop
The following proof requires some estimates that are contained 
for reasons of readability in Appendix \ref{appabschn:fourierreihen}.
\bpf
We set 
\zgl{\ymd_\beta := \sum_{\vnu\in\allirrdarst\LG} d_\vnu \: 
         \e^{-\einhalb \kopp^2 c_\vnu \FIL\beta} \: \charakt_\vnu.}\noindent
\bunum
\item
For all $\vnu\in\allirrdarst\LG$, $\vnu\neq\vec 0$, we have by
Proposition \ref{prop:absch(casimir-eigenwerte)} and \ref{prop:absch(dim(darst))}
\bgl
\supnorm{d_\vnu \: \e^{-\einhalb \kopp^2 c_\vnu \FIL\beta} \: \charakt_\vnu}
 &  =   & d_\vnu^2 \: \e^{-\einhalb \kopp^2 c_\vnu \FIL\beta}
          \erl{by 
               $\sup_{g\in\LG} \betrag{\charakt_\vnu(g)} = \charakt_\vnu(e_\LG)
                = d_\vnu$}\pktplatz \\
 & \leq & \const_\LG^2 \: \norm\vnu^{\dim\LG_\he - l} \:
              \: \e^{-\einhalb \kopp^2 c_- \FIL\beta \norm\vnu^2}.
\egl
Define $f(r) := \const_\LG^2 \: (r + \sqrt{k+l})^{\dim\LG_\he - l} \:
                   \: \e^{-\einhalb \kopp^2 c_- \FIL\beta r^2}$.
Remember that $k$ and $l$ are determined by
$\LG = (\LG_\he \kreuz U(1)^k)/\LN$ 
(see the beginning of Section \ref{sect:fourieranalysis}).
Here $l$ is the dimension of a maximal torus in $\LG_\he$.

Let $\vx\in\R^{k+l}_{\geq 0}$, $\vx\neq\vec 0$, be arbitrary. 
Moreover, let 
\zgl{\wuerfel-{k+l}{\vnu} := 
  \{\vx\in\R^l\kreuz\R^k\mid n_i-1 < x_i \leq n_i \: \forall i\}}\noindent
denote the semi-open cube with edge length $1$ in $\R^l\kreuz\R^k$, 
that is determined by the corners $\vn-\vec 1$ and $\vn$.
Now, choose some $\vnu\in\N^l\kreuz\N^k$ with $\vx\in\wuerfel-{k+l}\vnu$.
Then $\norm\vnu\geq\norm\vx$ and 
$\norm\vx+\sqrt{k+l} \geq \norm\vnu - \norm{\vnu-\vx} + \sqrt{k+l}
                     \geq \norm\vnu$,
hence $f(\norm\vx)\geq\const_\LG^2 \: \norm\vnu^{\dim\LG_\he - l} \:
              \: \e^{-\einhalb \kopp^2 c_- \FIL\beta \norm\vnu^2}$.
\item
Corollary \ref{corr:konvkritNZ_rho=0} yields
\bgl
\hspace*{-1.5em}
 &      & \sum_{\vnu\in\allirrdarst\LG} 
          \sup_{g\in\LG} \betrag{d_\vnu \: 
                         \e^{-\einhalb \kopp^2 c_\vnu \FIL\beta} \: \charakt_\vnu} \\
\hspace*{-1.5em}
 & \leq & 1 + 
          \sum_{\vnu\in\N^l\kreuz\Z^k, \vnu\neq\vec 0} 
          \sup_{g\in\LG} 
	     \betrag{d_\vnu \: 
                     \e^{-\einhalb \kopp^2 c_\vnu \FIL\beta} \: \charakt_\vnu} \\
\hspace*{-1.5em}
 & \leq & 1 + {}
\\ \hspace*{-3.5em} & &
          \hspace*{-0.3em} \int_0^\infty \hspace*{-0.9em}
            \klammerunten%
	    {\const_\LG^2 2^k \sum_{\laufi=1}^{k+l} 
             \binom{k+l}\laufi 
             \frac{\pi^{\frac \laufi 2}}{2^{\laufi-1} \Gamma(\frac \laufi 2)}
             \: r^{\laufi-1} \: (r + \sqrt{k+l})^{\dim\LG_\he - l}}
            {=:\text{ polynom } \sum_{\nu=0}^{\dim\LG - 1} p_\nu r^\nu}  \:
             \e^{-\einhalb \kopp^2 c_- \FIL\beta \: r^2}
	     \dd r 
             \hspace*{-4pt}
	     \\
\hspace*{-1.5em}
 &  =   & 1 + \sum_{\nu=0}^{\dim\LG-1} \const_\nu \: \FIL\beta^{-\frac{\nu+1}2}.
\egl
Here we have used
$\int_0^\infty r^{\nu-1} \: \e^{-a^2 r^2} \: \dd r 
 = \einhalb \Gamma(\frac\nu2) a^{-\nu}$. \cite{Z1}
\item
Hence, by Lemma \ref{lem:krit(glm_konv)} the Fourier series \eqref{gl:ymdfour}
is absolutely and uniformly convergent.
\item
Moreover, obviously, 
$\sup_{g\in\LG} \betrag{\ymd_\beta(g)} = \ymd_\beta(e_\LG)$.
\eunum
We are now left with the proof of 
$\mu_{\YM,\beta} = \ymd_\beta \kp \mu_{0,\beta}$.
\bunum
\item
Since $\mu_{\YM,\beta}$ and 
$\ymd_\beta \kp \mu_{0,\beta}$ are in each case $\Ad$-invariant regular
Borel measures,
we are to prove only
$\int_\LG f \: \dd\mu_{\YM,\beta} = 
           \int_\LG f\ymd_\beta \: \dd\mu_{0,\beta}$
for all $\Ad$-invariant $f\in C(\LG)$.

Since $\spann_\C \{\charakt_\vnu \mid \vnu\in\allirrdarst\LG\}$
is dense in $C_\Ad(\LG)$ by the Peter-Weyl theorem, this follows from 
\bgl
\int_\LG \charakt_{\vnu} \ymd_\beta \: \dd\mu_{0,\beta}
 & = & \lzweihaar{\ymd_\beta}{\charakt_{\vnu}} 
       \erl{$\ymd_\beta$ real (see below)}\pktplatz \\
 & = & d_\vnu \: \e^{-\einhalb \kopp^2 c_\vnu \FIL\beta}
       \erl{Definition of $\ymd_\beta$}\pktplatz \\
 & = & \int_\Ab \charakt_{\vnu} \circ \pi_\beta \: \dd\mu_{\Ab,\YM}
       \erl{Corollary \ref{corr:erwwlnws}}\pktplatz  \\
 & = & \int_\LG \charakt_{\vnu} \: \dd\mu_{\YM,\beta}
       \erl{Definition of $\mu_{\YM,\beta}$}\pktplatz 
\egl
for all $\vnu\in\allirrdarst\LG$.
\item
We still show that $\ymd_\beta$ is real:
Since the character of the dual
representaton equals the complex conjugate character
of the original representation, we have 
\zgl{
   d_{\darst^\ast} \: \e^{-\einhalb \kopp^2 c_{\darst^\ast} \FIL\beta}
 = \int_\LG \charakt_{\darst^\ast} \: \dd\mu_{\YM,\beta}
 = \quer{\int_\LG \charakt_{\darst} \: \dd\mu_{\YM,\beta}}
 = \quer{d_{\darst} \: \e^{-\einhalb \kopp^2 c_{\darst} \FIL\beta}},}\noindent
i.e. the imaginary parts in the Fourier series of $\ymd_\beta$
cancel each other.
\eunum
\epf
\brem
The just proven continuity statement can also be gained by means
of the so-called heat-kernel. One can show 
(\cite{Stein}, see also \cite{e49}), that the heat-kernel
$K : \LG \kreuz (0,\infty) \nach \R$ for the 
diffusion operator $\del_t - \lapl$ 
is a $C^\infty$-function and fulfills the equation
\zgl{K(g,t) = \sum_\vnu d_\vnu \: \e^{-c_\vnu t} \charakt_\vnu(g).}\noindent
Hence we can identify $\ymd_\beta$ with the heat-kernel
at time $t = \einhalb\kopp^2\FIL\beta$. This way 
we could have used the asymptotics
$\tr\: \e^{-t \lapl} \rund t^{-\einhalb\dim\LG} p(t)$
with some power series $p(t)$ \cite{Taylor2} for the 
proof of the second statement.

However, since we are interested in more general 
assertions on $\ymd_\beta$, we decided despite those general results
for the direct proof above.
\erem
Even when calculating the expectation values of the Yang-Mills measure
one sees that the integrations w.r.t.\ non-overlapping flag factorize.
This is confirmed by 
\bprop
\label{prop:ymd_graph}
For every graph $\GR$ and every moderately independent
flag world $\gb = \{\beta_1,\ldots,\beta_n\}$ of $\GR$ 
we have
\zgl{\mu_{\YM,\gb} = \ymd_\gb \kp \mu_{0,\gb} \:\:\: \text{ with } \:\:\: 
\ymd_\gb (\vec g) := \ymd_{\beta_1}(g_1) \cdots \ymd_{\beta_n}(g_n).}\noindent
In particular, $\ymd_\gb : \LG^n \nach \C$ is an $\Ad$-invariant and
continuous (even $C^\infty$) function.
\eprop
\bpf
Together with the single $\ymd_{\beta_\nu}$, also 
$\ymd_\gb$ is continuous and $\Ad$-invariant.
By Proposition \ref{prop:ymd_fahne},
$\ymd_{\beta_\nu} = 
   \sum_{\darst_\nu\in\allirrdarst\LG} \dim\darst_\nu \: 
         \e^{-\einhalb \kopp^2 c_{\darst_\nu} \FIL{\beta_\nu}} 
           \: \charakt_{\darst_\nu}$
is absolutely and uniformly convergent for every flag $\beta_\nu$.
Hence, all the subsequent rearrangements are allowed:
\bgl
\ymd_\gb (\vec g) 
 & = & \prod_{\nu = 1}^n \ymd_{\beta_\nu}(g_\nu) \\
 & = & \sum_{\vec\darst\in{\allirrdarst\LG}^n} 
       \prod_{\nu = 1}^n 
               \dim\darst_\nu \: 
               \e^{-\einhalb \kopp^2 c_{\darst_\nu} \FIL{\beta_\nu}} \:
               \charakt_{\darst_\nu}(g_\nu) \\
 & = & \sum_{\vec\darst\in{\allirrdarst\LG}^n} 
       \sum_{\darst\in\vec\darst} 
       \sum_{\veci,\vecj} 
          \kontr_{\vec\darst,\darst}^{\vecj\veci}
          \prod_{\nu = 1}^n 
               \dim\darst_\nu \: 
               \e^{-\einhalb \kopp^2 c_{\darst_\nu} \FIL{\beta_\nu}}
               \darst_\nu^{i_\nu j_\nu}(g_\nu) \\
 &   & \erl{$\prod_\nu \charakt_{\darst_\nu} (g_\nu) 
             = \sum_{\veci,\vecj} 
               \EINS_{\vec\darst}^{\vecj\veci}
               \prod_\nu 
                     \darst_\nu^{i_\nu j_\nu}(g_\nu)$ and 
            $\EINS_{\vec\darst} 
             = \sum_{\darst\in\vec\darst} \kontr_{\vec\darst,\darst}$}\pktplatz \\ 
\egl
\pagebreak
\bgl
\phantom{\ymd_\gb (\vec g)}
 & = & \sum_{\vec\darst\in{\allirrdarst\LG}^n} 
       \sum_{\darst\in\vec\darst} 
        \Bigl(\sqrt{\dim\darst}
           \prod_{\nu = 1}^n \bigl(
           \sqrt{\dim\darst_\nu} \:
               \e^{-\einhalb \kopp^2 c_{\darst_\nu} \FIL{\beta_\nu}}\bigr)\Bigr)
           \:\:\kreuz \\
 &   & \phantom{\sum_{\vec\darst\in{\allirrdarst\LG}^n} 
       \sum_{\darst\in\vec\darst}}
       \Bigl(\inv{\sqrt{\dim\darst}}	
       \sum_{\veci,\vecj} 
          \kontr_{\vec\darst,\darst}^{\vecj\veci}
          \prod_{\nu = 1}^n 
               \sqrt{\dim\darst_\nu} \: 
               \darst_\nu^{i_\nu j_\nu}(g_\nu) \Bigr)\\
 & = & \sum_{\vec\darst\in{\allirrdarst\LG}^n} 
       \sum_{\darst\in\vec\darst} 
        \Bigl(\int_\Ab \nchar_{\vec\darst,\darst} \circ \pi_\gb \:
              \dd\mu_{\Ab,\YM}\Bigr)
           \nchar_{\vec\darst,\darst} 
       \erl{Corollary \ref{corr:erwwlnws}}\pktplatz \\
 & = & \sum_{\vec\darst\in{\allirrdarst\LG}^n} 
       \sum_{\darst\in\vec\darst} 
        \Bigl(\int_{\LG^n} \nchar_{\vec\darst,\darst} \:
              \dd\mu_{\YM,\gb}\Bigr)
           \nchar_{\vec\darst,\darst} \erl{Definition of $\mu_{\YM,\gb}$}.
\egl
Here, in part, we used $\vec\darst$ and $\bigtensor_\nu \darst_\nu$ 
synonymously. So, e.g.,  
$\darst\in\vec\darst$ just denotes an irreducible representation $\darst$
contained in $\bigtensor_\nu \darst_\nu$.

Consequently, 
\bgl
\int_{\LG^n} \nchar_{\vec\darst,\darst} \: \ymd_\gb \: \dd\mu_{0,\gb}
 & = & \lzweihaarn{\ymd_\gb}{\nchar_{\vec\darst,\darst}} 
        \erl{$\mu_{0,\gb} = \mu_\Haar^n$ and $\ymd_\gb$ real}\\
 & = & \sum_{\vec\darst',\darst'} 
        \Bigl(\int_{\LG^n} \nchar_{\vec\darst',\darst'}
                   \: \dd\mu_{\YM,\gb}\Bigr)
        \lzweihaarn{\nchar_{\vec\darst',\darst'}}{\nchar_{\vec\darst,\darst}} \\
 & = & \int_{\LG^n} \nchar_{\vec\darst,\darst} \: 
                          \dd\mu_{\YM,\gb} \\
 &   & \erl{Orthonormalization of the $n$-characters
              (Theorem \ref{thm:peter+weyl_verallg})}
\egl
for all $n$-characters $\nchar_{\vec\darst,\darst}$.
But, because these by the Peter-Weyl theorem span a dense subspace of
$C_\Ad(\LG^n)$, we have 
\zgl{\int_{\LG^n} f \: \ymd_\gb \: \dd\mu_{0,\gb}
       = \int_{\LG^n} f \: \dd\mu_{\YM,\gb}}\noindent
for all $f\in C_\Ad(\LG^n)$ as in the proposition above,
hence $\mu_{\YM,\gb} = \ymd_\gb \kp \mu_{0,\gb}$.
\qed
\epf
\bcorr
\label{corr:faktorint_nichtueberlapp}
Let $\GR$ be a graph and $\gb = \{\beta_1,\ldots,\beta_n\}$ 
be a moderately independent flag world in $\GR$.

Then we have
$\mu_{\YM,\gb}(U_1 \kreuz \cdots \kreuz U_n) 
 = \mu_{\YM,\beta_1}(U_1) \cdots \mu_{\YM,\beta_n}(U_n)$
for all measurable and $\Ad$-invariant $U_i \teilmenge \LG$,
$i = 1, \ldots, n$.
\ecorr
\bpf
\bglklein
 &   & \mu_{\YM,\gb}(U_1 \kreuz \cdots \kreuz U_n) \\
 & = & \int_{\LG^n} \charfkt{U_1 \kreuz \cdots \kreuz U_n} 
                    \: \ymd_\gb \: \dd\mu_{0,\gb} \\
 & = & \int_{\LG^n} \charfkt{U_1}(g_1) \cdots \charfkt{U_n}(g_n)
          \: \ymd_{\beta_1}(g_1) \cdots \ymd_{\beta_n}(g_n) 
          \: \dd\mu_{0,\beta_1}(g_1) \cdots \dd\mu_{0,\beta_n}(g_n) \\
 & = & \bigl(\int_{\LG} \charfkt{U_1}(g_1) \ymd_{\beta_1}(g_1)
                  \: \dd\mu_{0,\beta_1}(g_1)\bigr) \cdots 
       \bigl(\int_{\LG} \charfkt{U_n}(g_n) \ymd_{\beta_n}(g_n) 
                  \: \dd\mu_{0,\beta_n}(g_n)\bigr) \\
 & = & \mu_{\YM,\beta_1}(U_1) \cdots \mu_{\YM,\beta_n}(U_n)
\eglklein
\qed
\epf

Finally we show that the integration of 
(not necessarily continuous) cylindrical functions
always can be reduced to the analysis of absolutely convergent Fourier series.
\bprop 
\label{prop:konv_fourier_ym2}
Let $f\in L^2(\LG)$ be an $\Ad$-invariant function. Then 
we have for all flags $\beta\in\hg$ 
\zgl{
     \int_\LG f \: \dd\mu_{\YM,\beta}
     = \sum_{\vnu\in\allirrdarst\LG}
         \lzweihaar{\charakt_\vnu}{f} d_\vnu \e^{-\einhalb\kopp^2 c_\vnu \FIL\beta},}\noindent
whereas the rhs always converges absolutely.
\eprop
\bpf
By Corollary \ref{corr:fourier_L2(LG)},
$\lzweihaar{\charakt_\vnu}{f} \charakt_\vnu$
converges to $f$ in $L^2(\mu_\Haar) \ident L^2(\LG)$.
Since the function $\ymd_\beta$ with 
$\mu_{\YM,\beta} = \ymd_\beta \kp \mu_{0,\beta}$ from 
Proposition \ref{prop:ymd_fahne}
is continuous, hence bounded,
we have $L^2(\mu_\Haar)\teilmenge L^2(\mu_{\YM,\beta})$, i.e.,
$\lzweihaar{\charakt_\vnu}{f} \charakt_\vnu$ converges in 
$L^2(\mu_{\YM,\beta})$ to $f$ as well. Hence, 
\bglklein
\int_\LG f \: \dd\mu_{\YM,\beta}
 & \ident & \lzweispez{1}{f}{\YM,\beta} \\
 & = & \sum_{\vnu\in\allirrdarst\LG} 
            \lzweihaar{\charakt_\vnu}{f} \lzweispez{1}{\charakt_\vnu}{\YM,\beta} \\
 & = & \sum_{\vnu\in\allirrdarst\LG} 
            \lzweihaar{\charakt_\vnu}{f} 
            d_\vnu \e^{-\einhalb\kopp^2 c_\vnu \FIL\beta}.
\eglklein
By $\betrag{\lzweihaar{\charakt_\vnu}{f}} 
     \leq \lzweihaarnorm{\charakt_\vnu} \lzweihaarnorm{f} 
     = \lzweihaarnorm{f}$,
hence
\zgl{\bigbetrag{\lzweihaar{\charakt_\vnu}{f} 
        d_\vnu \e^{-\einhalb\kopp^2 c_\vnu \FIL\beta}}
     \leq \lzweihaarnorm f \const_\LG \norm\vnu^{\dim\LG_\he - l} 
              \e^{-\einhalb\kopp^2 c_- \FIL\beta \norm\vnu^2},}\noindent
and by Corollary \ref{corr:konvkritNZ_spez} 
the series converges even absolutely:
\zgl{g(r) := r^{\dim\LG_\he - l} \e^{-\einhalb\kopp^2 c_- \FIL\beta r^2}}\noindent
decreases for sufficiently large $r$ monotonically, and 
$g(r) \: r^{k+l-1}$ is integrable on $[0,\infty]$.
\qed
\epf
Completely analogously we get
\bprop
\label{prop:konv_fourier_ym2_n}
Let $\GR$ be a graph and $f\in L^2(\LG^{\rh(\GR)})$
be an $\Ad$-invariant function. Then we have 
for all moderately independent flag worlds $\gb$ in $\GR$
\zgl{ \!\!
\int_{\LG^n} f \: \dd\mu_{\YM,\gb}
     = \sum_{\vec\darst,\darst}
         \lzweihaarn[\rh(\GR)]{\nchar_{\vec\darst,\darst}}{f} 
         \sqrt{\dim\darst} \:
         \prod_{I=1}^{\rh(\GR)} 
             \Bigl( 
             \sqrt{\dim\darst_I} \:
             \e^{-\einhalb \kopp^2 c_{\darst_I} \FI{G_I}}
             \Bigr).}\noindent
\eprop

\brem
The definition of the expectation values of $\mu_\YM$ started
with the Wilson action, hence with a quantity 
directly gained from the standard Yang-Mills action $\einhalb(F,F)$.
This, however, has the disadvantage that 
the existence of the continuum limit 
in Proposition \ref{prop:ex(erww_lim)} had to be proven laboriously.

There is another possibility to process: One can put the 
calculated expectation values directly into the definition of
the regularization of $\ymwirk$. One simply defines for all graphs $\GR$ and
moderately independent flag worlds $\FW$ in $\GR$
\zglnum{
\ymwirk[\GR] := - \ln 
    \Bigl(\sum_{\vec\darst,\darst}
           \sqrt{\dim\darst} \:
           \prod_{I=1}^{\rh(\GR)} 
               \Bigl( 
               \sqrt{\dim\darst_I} \:
               \e^{-\einhalb \kopp^2 c_{\darst_I} \FI{G_I}}
               \Bigr) \: 
           \quer\lnws_{(\FW,\vec\darst,\darst)} 
    \Bigr)}{gl:heatkernel-action}\noindent
and defines then
\zgl{\erw (f_\GR \circ \pi_\GR) 
      := \int_\agb \e^{-\ymwirk[\GR]} 
                         f_\GR \circ \pi_\GR \: \dd\mu_0.}\noindent
This way one avoids the problem of the limit
and gets the ``correct'' expectation values directly.
This variant has been used, e.g., by Aroca and Kubyshin \cite{iosa2}.
(However, there only some flags has been considered what is insufficient 
for the determination of the full measure.)
Typically the action \eqref{gl:heatkernel-action}
is also called heat-kernel action
or Villain action \cite{g19}.
Of course, this method has the disadvantage that the relation to the 
standard Yang-Mills theory is not as close as in the case of the 
Wilson approximation; the problem is only shifted.
\erem

\section{Support of the Yang-Mills Measure}
\label{abschn:sing(YM2)}
In this section we are going to prove that the Yang-Mills measure
is purely singular w.r.t.\ the Ashtekar-Lewandowski measure. Moreover,
we present an explicit (however, of course, non-unique) decomposition of 
$\AbGb$ into disjoint subsets that support the one and the other 
measure, respectively. Finally, we investigate the impact of smooth
connections and of the Gribov problem.

\subsection{General Remarks on Singularity Proofs}
To prove the singularity of a (finite Borel) measure $\mu$ w.r.t.\ 
to the Ashtekar-Lewandowski measure 
$\mu_0$ we need to show that there is no $L^1(\AbGb,\mu_0)$-measureable 
function $f$ fulfilling
$\mu = f \kp \mu_0$ (or $\dd\mu = f \: \dd\mu_0$). 
However, there is a very simple criterion for that:
\bprop
\label{prop:singkrit(lewand)}
Let $\mu$ be some (normalized) regular Borel measure on $\AbGb$.
Then $\mu$ is singular w.r.t.\ to $\mu_0$ if there are 
uncountably many non-zero spin-network expectation values.
\eprop
Obviously, there is no $L^2(\mu_0)$-function $f$ with
$\mu = f \kp \mu_0$. Namely, if this were not the case, then this
$f$ could be expanded into a spin-network series \cite{e9}
with uncountably many non-vanishing ``Fourier'' coefficients which is
impossible in a Hilbert space. The idea for the proof in the $L^1$-case 
is due to Jerzy Lewandowski:
\bpf
Suppose $\mu$ would be absolutely continuous w.r.t.\ $\mu_0$. Then
there would be an $L^1(\mu_0)$-function with $\mu = f \kp \mu_0$.
Since the cylindrical functions form a dense subalgebra of the 
continuous functions in $\AbGb$, they form a dense subspace in the
Banach space $L_1(\mu_0)$. Hence, there is a sequence $(f_n)$ of
(w.l.o.g.\ real) cylindrical functions with $f_n \gegen f$ in $L^1$. Obviously,
then $T f_n \gegen T f$ in $L^1$, i.e.\
\zglnum{ \overline{\lzweispez{T}{f_n}{\mu_0}} \ident  \int_\AbGb T f_n\:\dd\mu_0 
    \gegen \int_\AbGb T f\:\dd\mu_0 
         = \int_\AbGb T \dd\mu \ident \erww T}{eq:conv}\noindent
for every continuous function $T$ on $\AbGb$.
Since every cylindrical
function can be written as a sum of countably many spin-networks,
there are at most countably many spin-network states $T$ with non-vanishing
$\lzweispez{T}{f_n}{\mu_0}$ for some $n$. But, by assumption there
are uncountably many spin-network states $T$ having non-vanishing 
$\mu$-expectation value $\erww T$. We get a contradiction to \eqref{eq:conv}.
\qed
\epf
Now, since Wilson loops can be regarded as special spin networks,
the Yang-Mills measure fits to the condition of the proposition above.
Consequently, $\mu_\YM$ is not absolutely continuous w.r.t.\ to $\mu_0$.
Note, more general, that measures having 
a continuous symmetry typically fulfill the condition above.

It remains now the proof of the pure singularity. This means that 
there is a $\mu_0$-zero subset having the full $\mu$-measure $1$.
General arguments are provided, i.e., by the notion of ergodicity. 
More precisely, let $H$ be some (not necessarily continuous) 
transformation group on the measure space $X$. Then every two $H$-ergodic
and $H$-(quasi-)invariant measures are equivalent or 
purely singular to each other \cite{Ya}. 
In the Yang-Mills case we know that the
measure is invariant under the action of area and analyticity preserving
automorphisms. It is well-known that $\mu_0$ is not only invariant under
these transformations as well, but even ergodic.
Therefore, if we were able to prove the
corresponding ergodicity of $\mu_\YM$, we would get the pure singularity.
However, up to now, we do not know whether the ergodocity is given (although
we guess it is). Moreover, it would not immediately yield the 
partition of $X = \AbGb$ into the supports of $\mu_\YM$ and $\mu_0$ that
will be given below.

Finally, we note that using ergodicity the inequivalence of the 
Fock measures and the Ashtekar-Lewandowski measure 
has been proven \cite{d46,b18}.

\subsection{Idea}
The proof of the inequivalence of $\mu_\YM$ and $\mu_0$
is based on the following two facts:
\bnum{2}
\item
If the flag $\beta$ shrinks, the measure $\mu_{\YM,\beta}$
concentrates around $e_\LG$.
\item
The measure w.r.t.\ non-overlapping flags 
is the product of the measures for the single flags.
\enum
More precisely, for the first item we know that 
the Radon-Nikodym derivative $\ymd_\beta$ of the Yang-Mills measure 
equals 
$\sum_{\vnu\in\allirrdarst\LG} d_\vnu \: 
         \e^{-\einhalb \kopp^2 c_\vnu \FIL\beta} \: \charakt_\vnu$
for a single flag. Obviously, the Fourier coefficients of this expansion
fall the slowlier, the smaller $\FIL\beta$ is. The slow falling of a
Fourier series typically corresponds to a strong concentration
of the original function in a point.
(Remember the extremal cases
$\delta$-``function'' with 
$\delta(\varphi) = \sum_{n\in\Z} 1 \cdot \e^{\I n\varphi}$ and
1-function with
$1(\varphi) = \sum_{n\in\Z} \delta_{n0} \cdot \e^{\I n\varphi}$.)
This way it should be possible to find a neighbourhood $U$ of $e_\LG$ 
with Haar measure $\varepsilon$ whose (pushed-forward) Yang-Mills measure 
$\mu_{\YM,\beta} (U) \rund \ymd_\beta(e_\LG) \: \mu_\Haar(U)$
goes to $1$ for shrinking flags.
Using the second item above and an increasing number of shrinking flags
we let on the one hand the Haar measure of $U^n\teilmenge\LG^n$ go
to $0$ with $\varepsilon^n$ and 
let on the other hand the corresponding Yang-Mills measure 
go to a non-vanishing value. For instance, we can always choose some flag
$\beta_n$ with $\mu_{\YM,\beta_n}(U) = \e^{-2^{-n}}$. Namely, here
$\prod \e^{-2^{-n}} = \e^{-1} > 0$.
Hence, the Yang-Mills measure is definitely not absolutely continuous 
w.r.t.\ $\mu_0$. 
This argument can be used even for the proof of the 
full singularity. After a refinement of the indicated estimates,
we will show that the Yang-Mills measure for $U^n$
can adopt for appropriate flags $\beta_n$ every value different from $1$.
Hence we reach the full Yang-Mills measure $1$.

We will assume thoughout this section that $\LG$ is nontrivial.
Otherwise, $\ag = \AbGb$, i.e.\ every generalized connection
would be regular, and thus $\mu_\YM(\ag) = \mu_\YM(\AbGb) = 1$.
Additionally, $\mu_\YM = \mu_0$, i.e.\ $\mu_\YM$ would be absolutely 
continuous w.r.t.\ $\mu_0$.

\subsection{Pure Singularity of $\mu_\YMueberschrift$ w.r.t.\ $\mu_0$}
\label{uabschn:ym2:singYM}
Let us choose some fixed $\varepsilon\in(0,1)$.
\blem
\label{lem:sing_YM_0:1}
There is a constant $c \geq 0$ depending only on $\varepsilon$ and $\LG$ 
and an open $\Ad$-invariant subset $U\teilmenge\LG$, such that 
we have for all flags $\beta\in\hg$
\bunum
\item
$\mu_{0,\beta}(U) \leq \varepsilon$ and
\item
$\mu_{\YM,\beta}(U) \geq 1 - c \FIL\beta$.
\eunum
\elem
\bpf
Let us choose for $U$ some open $\Ad$-invariant neighbourhood of the 
identity $e_\LG$ with $\mu_\Haar(U)\leq \varepsilon$. Obviously,
$\mu_{0,\beta} (U) \ident \mu_\Haar(U) \leq \varepsilon$; so we are left
with the proof of $\mu_{\YM,\beta}(U) \geq 1 - c \FIL\beta$.
\bunum
\item
Let $f : \LG \nach [0,1]$ be some 
$\Ad$-invariant $C^\infty$-function with $\supp f \teilmenge U$ and 
$f(e_\LG) = 1$.
\item
By Corollary \ref{corr:glm+abs_konv(fourier)} the Fourier series 
$f = \sum_{\vnu\in\allirrdarst\LG} \lzweihaar{\charakt_\vnu}{f} \charakt_\vnu$
converges absolutely and uniformly. In particular, 
\zglnum{f(e_\LG) = \sum_{\vnu\in\allirrdarst\LG} 
                 \lzweihaar{\charakt_\vnu}{f} d_\vnu = 1}{gl:four_einselem}
converges absolutely. 
\item
Hence we get
\bgl
\mu_{\YM,\beta}(U) 
 & \geq & \int_\LG f \: \dd\mu_{\YM,\beta} \\
 &  =   & \sum_{\vnu\in\allirrdarst\LG} \lzweihaar{\charakt_\vnu}{f}
             d_\vnu \e^{-\einhalb\kopp^2 c_\vnu \FIL\beta} \\
 &  =   &   \sum_{\vnu\in\allirrdarst\LG} \lzweihaar{\charakt_\vnu}{f} d_\vnu \\
 &      & - \sum_{\vnu\in\allirrdarst\LG, \vnu\neq\vec 0} 
               \lzweihaar{\charakt_\vnu}{f} d_\vnu \:
               \einhalb\kopp^2 c_\vnu \FIL\beta \:
	       \frac{1-\e^{-\einhalb\kopp^2 c_\vnu \FIL\beta}}%
                    {\einhalb\kopp^2 c_\vnu \FIL\beta} \\
\egl
\pagebreak
\bgl
\phantom{\mu_{\YM,\beta}(U)}
 &  =   & 1
          - \FIL\beta\:
            \sum_{\vnu\in\allirrdarst\LG, \vnu\neq\vec 0}
	    \klammerunten
	     {
                 \lzweihaar{\charakt_\vnu}{f} \: d_\vnu \:
                 \einhalb\kopp^2 c_\vnu \:
	         \frac{1-\e^{-\einhalb\kopp^2 c_\vnu \FIL\beta}}%
                      {\einhalb\kopp^2 c_\vnu \FIL\beta}}
             {=:\xi_{\vnu,\FIL\beta}}.
\egl
In the first step we used $f\leq \charfkt U$ and in the second  
the integration formula from Proposition \ref{prop:konv_fourier_ym2}.
For the third step we used the absolute convergence of both 
series ($\sum_{\vnu\in\allirrdarst\LG} 
\betrag{\lzweihaar{\charakt_\vnu}{f}} d_\vnu$
is a convergent majorant) and in the fourth 
Equation \eqref{gl:four_einselem}.
\item
For the study of the convergence of the $\xi$-series we need 
the following estimates ($\vnu\neq\vec 0$):
First, $\frac{1-\e^{-x}}{x} \leq 1$ for all $x\in \R_+$, and second
$\betrag{\lzweihaar{\charakt_\vnu}{f}} \leq \const_{s,f} \: c_\vnu^{-s}$ 
for all $s\in\N_+$ by $f\in C^\infty(\LG)$ and
by Proposition \ref{prop:asymp(fourierkoeff)}.
Hence, for all $s\in\N$
\zgl{
\bigbetrag{\xi_{\vnu,\FIL\beta}} 
 \leq \einhalb \kopp^2 \const_{s,f} c_\vnu^{1-s} d_\vnu.
 }

In particular, for $2s = \einhalb(\dim\LG + k + l) + 3$ the relation 
$\einhalb(\dim\LG_\he - l) + 2 (1-s) = -(k+l+1)$ is fulfilled.
By the convergence criterion in Lemma \ref{lem:sum_dim+casimir} we have
\zgl{\sum_{\vnu\in\allirrdarst\LG,\vnu\neq\vec 0} \xi_{\vnu,\FIL\beta}
 \leq \einhalb \kopp^2 \const_{s,f}
      \sum_{\vnu\in\allirrdarst\LG,\vnu\neq\vec 0} c_\vnu^{1-s} d_\vnu
  =: c < \infty,}
independent of $\FIL\beta$.
Finally, we get
\zgl{\mu_{\YM,\beta} (U) \geq 1 - c \FIL\beta.}
\qed
\eunum
\epf

\blem
\label{lem:sing_YM_0:2}
Let $F\in(0,1)$ and let $(\beta_i)_{i\in\N_+} \teilmenge \hg$ 
be a sequence of flags, such that 
\bnum{3}
\item
$\FIL{\beta_i} \leq \frac{F}{2^i c}$ for all $i\in\N_+$, where
$c$ is the constant of Lemma \ref{lem:sing_YM_0:1}, and
\item
for every $\Lambda\in\N_+$ the set
$\gb_\Lambda := \{\beta_1,\ldots,\beta_\Lambda\}$ 
is a moderately independent flag world 
in the graph $\GR_\Lambda$ spanned by $\gb_\Lambda$.%
\footnote{Descriptively, the flags $\beta_i$ are just non-overlapping.}
\enum
Then there are open $\Ad$-invariant $V_\Lambda\teilmenge\LG^\Lambda$,
such that we have for all $\Lambda\in\N_+$
\bunum
\item
$\mu_{0,\gb_\Lambda}(V_\Lambda) \leq \varepsilon^\Lambda$,
\item
$\mu_{\YM,\gb_\Lambda}(V_\Lambda) \geq 1 - F$ and
\item
$\pi^{-1}_{\gb_{\Lambda+1}}(V_{\Lambda+1}) \teilmenge 
 \pi^{-1}_{\gb_\Lambda}(V_\Lambda)$.
\eunum
\elem
\bpf
Choose the $\Ad$-invariant $U\teilmenge\LG$ of the preceding lemma
and define $V_\Lambda:=U^\Lambda\teilmenge\LG^\Lambda$.
Obviously, $V_\Lambda$ is always open and $\Ad$-invariant.
Moreover, we have for all $\Lambda\in\N_+$: 
\bunum
\item
$\mu_{0,\gb_\Lambda} (V_\Lambda) = \mu_\Haar^\Lambda(U^\Lambda)
     = (\mu_\Haar(U))^\Lambda \leq \varepsilon^\Lambda.$
\item
By Corollary \ref{corr:faktorint_nichtueberlapp} we have 
\bglklein
\mu_{\YM,\gb_\Lambda} (V_\Lambda) 
 & = &  \mu_{\YM,\gb_\Lambda} (U^\Lambda) 
 \breitrel=   \prod_{i=1}^\Lambda \mu_{\YM,\beta_i} (U) \\
 & \geq & \prod_{i=1}^\Lambda (1 - c \FIL{\beta_i}) 
 \breitrel\geq \prod_{i=1}^\Lambda (1 - \frac{F}{2^i}) 
 \breitrel\geq 1 - F.
\eglklein
Here we used $c \FIL{\beta_i} \leq \frac{F}{2^i} < \einhalb$ and 
(in the last step) the relation
$\prod_{i=1}^\Lambda (1-\frac{F}{2^i}) \geq 1 - F$ valid for
all $F\in(0,1)$.%
\footnote{Using the absolute convergence of the Taylor series
of $\ln(1-F)$ we have for $0 < F < 1$:
\bglklein
\ln\prod_{i=1}^{\Lambda} \bigl(1 - \frac{F}{2^i}\bigr)
 & = & \sum_{i=1}^{\Lambda} \ln\bigl(1 - \frac{F}{2^i}\bigr) 
 \breitrel= - \sum_{i=1}^{\Lambda} \sum_{j=1}^\infty
         \inv j \Bigl(\frac{F}{2^i}\Bigr)^j 
 \breitrel= - \sum_{j=1}^\infty \sum_{i=1}^{\Lambda} 
         \inv j \Bigl(\frac{F}{2^i}\Bigr)^j \\
 & = & - \sum_{j=1}^\infty \frac{F^j}{j}
         \Bigl(\klammerunten{\textstyle \sum_{i=1}^{\Lambda} \inv{2^{ij}}}
	                    {< 1}\Bigr)
 \breitrel\geq - \sum_{j=1}^\infty \frac{F^j}{j} 
 \breitrel=  \ln(1-F).          
\eglklein}
\item
By
\bglklein
\qa\in\pi_{\gb_\Lambda}^{-1} (V_\Lambda)
 & \aequ & \bigl(h_{\beta_1}(\qa),\ldots,h_{\beta_\Lambda}(\qa)\bigr)
           = \pi_{\gb_\Lambda}(\qa) \in V_\Lambda = U^\Lambda \\
 & \aequ & h_{\beta_i}(\qa) \in U \fueralle i = 1,\ldots,\Lambda
\eglklein
we have
$\pi^{-1}_{\gb_{\Lambda+1}}(V_{\Lambda+1}) \teilmenge 
 \pi^{-1}_{\gb_\Lambda}(V_\Lambda)$.
\qed
\eunum
\epf

\blem
\label{lem:sing_YM_0:3}
For every $F\in(0,1)$ there is a measurable $W_F \teilmenge \AbGb$
with
\zgl{\mu_0(W_F) = 0 \text{ and } \mu_\YM(W_F)\geq 1 - F.}
\elem
\bpf
Obviously there is a sequence 
$(\beta_i)_{i\in\N_+} \teilmenge \hg$ of flags with the properties 
listed in Lemma \ref{lem:sing_YM_0:2}.
Let again 
$\gb_\Lambda := \{\beta_1,\ldots,\beta_\Lambda\}$.
According to the lemma above choose for every $\Lambda\in\N_+$ 
some $\Ad$-invariant subset $V_\Lambda\teilmenge\LG^\Lambda$
and define 
$W_F := \bigcap_{\Lambda\in\N_+} \pi(\pi_{\gb_\Lambda}^{-1}(V_\Lambda))$.
By the $\Ad$-invariance of $V_\Lambda$ we have 
$\pi^{-1}(W_F) = 
  \bigcap_{\Lambda\in\N_+} \pi_{\gb_\Lambda}^{-1}(V_\Lambda)$.
Hence, by 
$\pi^{-1}_{\gb_{\Lambda+1}}(V_{\Lambda+1}) \teilmenge 
 \pi^{-1}_{\gb_\Lambda}(V_\Lambda)$ and by the theorem
on monotone convergence of measures
\bnum{2}
\item
$\mu_0(W_F) 
  = \lim_{\Lambda\gegen\infty} \mu_0(\pi_{\gb_\Lambda}^{-1}(V_\Lambda))
  = \lim_{\Lambda\gegen\infty} \mu_{0,\gb_\Lambda}(V_\Lambda)
  = \lim_{\Lambda\gegen\infty} \varepsilon^\Lambda
  = 0$ and
\item
$\mu_\YM(W_F) 
  = \lim_{\Lambda\gegen\infty} \mu_{\Ab,\YM}(\pi_{\gb_\Lambda}^{-1}(V_\Lambda))
  = \lim_{\Lambda\gegen\infty} \mu_{\YM,\gb_\Lambda}(V_\Lambda)
 \geq 1 - F$.
\qed
\enum
\epf
Consequently, $\mu_\YM$ is not absolutely continuous w.r.t.\ $\mu_0$.
The pure singularity comes from

\bthm
\label{thm:sing_YM_0}
There is a measuable $W\teilmenge\AbGb$ with
\zgl{\mu_0(W) = 0 \text{ and } \mu_\YM(W) = 1.}
\ethm
\bpf
Define $W := \bigcup_{n\in\N, n>1} W_{\inv n}$, where 
$W_{\inv n}$ is defined as in Lemma \ref{lem:sing_YM_0:3}. Then we have
\bnum{2}
\item
$0 \leq \mu_0(W) \leq \sum_{n\in\N, n>1} \mu_0(W_{\inv n}) = 0$ and
\item
$1 \geq \mu_\YM(W) \geq \sup_{n\in\N, n>1} \mu_\YM(W_{\inv n}) 
              =  \sup_{n\in\N, n>1} \{1 - \inv n\} = 1.$
\qed
\enum
\epf

\subsection{$\mu_\YMueberschrift$-Almost Global Triviality of the Generic Stratum}
In \cite{paper5} the impact of the Gribov problem on the kinematical level,
i.e.\ w.r.t.\ $\mu_0$,
has been investigated. This is strongly related to the 
existence of so-called almost global trivializations of the generic stratum%
\footnote{The generic stratum contains exactly those connections that
have minimal stabilizer w.r.t.\ the action of gauge transforms 
\cite{paper2+4,paper5}. It has
been proven that a stabilizer is minimal iff it contains precisely the constant
center-valued gauge transforms. Moreover, the generic stratum has induced
Haar measure $1$.}.
This means that there is a covering of the generic stratum
consisting of gauge-invariant sets having full measure $1$. This way it
has been shown that there are sections in the fibering $\Ab \nach \AbGb$ 
being continuous almost everywhere -- the Gribov problem is
concentrated on a zero subset. In this subsection we will see that this is
true dynamically as well, i.e.\ for the Yang-Mills measure.
\bthm
\label{thm:Abgen_YMfastglobtriv}
The generic stratum of $\Ab$ is $\mu_{\Ab,\YM}$-almost globally trivial.
\ethm
\bpf
We simply show that the $\mu_0$-almost global trivialization from 
\cite{paper5} is also a $\mu_{\Ab,\YM}$-almost global trivialization.
We recall that there every element of the covering of the generic
stratum was the preimage $\pi_\ga^{-1}(V)$ of some
$\Ad$-invariant set $V\teilmenge\LG^{\elanz\ga}$ 
with Haar measure $1$ where $\ga$ can be chosen to be a moderately independent 
flag world. 

Now we have 
$\mu_{\Ab,\YM}(\pi_\ga^{-1}(\LG^n\setminus V)) = 
 \mu_{\YM,\ga} (\LG^n \setminus V) = 0$ with $n:=\elanz\ga$,
hence
$\mu_{\Ab,\YM} (\pi_\ga^{-1}(V)) = 1$,
for all 
$\Ad$-invariant $V\teilmenge\LG^{n}$ with
$\mu_\Haar^{n} (\LG^{n} \setminus V) = 0$, since
$\mu_{\YM,\ga}$ is absolutely continuous w.r.t.\ $\mu_{0,\ga} = \mu_\Haar^{n}$.
\qed
\epf

We get immediately 
\bthm
\label{thm:mu_YM(Abgen)=1}
We have $\mu_\YM(\AbGbgen) = 1$.
\ethm

\subsection{Smooth Connections}
Finally we are going to show that smooth connections are not only contained
in a $\mu_0$-zero subset \cite{a42}, but also in a $\mu_\YM$-zero subset.
\bthm
\label{thm:muYM(agb)=0}
$\ag$ is contained in subset of $\mu_\YM$-measure $0$.
\ethm
The idea of the proof extends that of the proof (see \cite{a42}) in the 
$\mu_0$-case.
\bpf
\bunum
\item
Consider $\LG$ again as a subset of some 
$U(N) \teilmenge \Gl_\C(N) \teilmenge \C^{N\kreuz N}$, hence  
$\Lieg\teilmenge \gl_\C(N) = \C^{N\kreuz N}$.
Choose some $\Ad \LG$-invariant
norm $\algnorm\cdot$ on $\C^{N\kreuz N}$ and define
$B_\varepsilon(e_\LG) := \{g\in\LG \mid \algnorm{g - e_\LG} < \varepsilon\}$ 
for all $\varepsilon\in\R_+$.
Obviously, $B_\varepsilon(e_\LG)$ is always an $\Ad$-invariant set.

Next we choose some bounded domain $U \teilmenge \R^{2}$ given the 
Euclidian metric. We assume that the image of every $\alpha\in\hg$ 
used in this proof is contained in $U$. 
\item
Now we define for all $\alpha$
and all real $r\in\R_+$ the (by the $\Ad$-invariance of $B_\varepsilon(e_\LG)$) 
$\Gb$-invariant set
\zgl{U_{\alpha,r} := \pi_\alpha^{-1}(B_{r\FIL\alpha}(e_\LG)) \teilmenge \Ab.} 
We have
\bgl
\mu_{\Ab,\YM}(U_{\alpha,r}) 
 &  =   & \mu_{\YM,\alpha}(B_{r\FIL\alpha}(e_\LG)) \\
 & \leq & \supnorm{\ymd_\alpha} \: \mu_{0,\alpha}(B_{r\FIL\alpha}(e_\LG)) 
          \erl{$\mu_{\YM,\alpha} = \ymd_\alpha \kp \mu_{0,\alpha}$}\\
 & \leq & \Bigl(\sum_{\nu=0}^{\dim\LG} \const_\nu \FIL\alpha^{-\einhalb\nu}\Bigr)
          \: c (r\FIL\alpha)^{\dim\LG} \\
 &      & \erl{Proposition \ref{prop:ymd_fahne} 
               and Lemma \ref{lem:absch(haarmasz)}},
\egl
where the last term is a polynom in $\sqrt{\FIL\alpha}$ whose
lowest order equals $-\dim\LG + 2\dim\LG = \dim\LG \geq 1$.
($\LG$ has been assumed nontrivial and connected.)
Hence,
$\mu_{\YM}(U_{\alpha,r})$ goes to $0$ for $\FIL\alpha\konvrunter 0$.
\item
Now let $(\alpha_i)_{i\in\N}$ be some sequence of circles in $U$ with
$\FIL{\alpha_i}\konvrunter 0$, where each two circles have $m$ as unique 
common point. We define
\zgl{U_r := \bigcap_{i\in\N} U_{\alpha_i,r}.}

Obviously,
$\mu_\YM(U_r) \leq \inf_i \{\mu_\YM(U_{\alpha_i,r})\} = 0$.

\item
On the other hand, for every $A\in\A$ there is a $c_A\in\R_+$ with
$A\in U_{\alpha,c_{A}} \ident \pi_\alpha^{-1}(B_{c_{A}\FIL\alpha}(e_\LG))$
for all circles $\alpha$ (see Corollary \ref{corr:regzusklholo}).
Hence $A\in U_{c_{A}}$.
Consequently, $U := \bigcup_{r\in\N_+} U_r$ is obviously 
a $\mu_\YM$-zero subset containing $\A$. Since $U$ is $\Gb$-invariant as well,
$U/\Gb$ is again a $\mu_\YM$-subset containing now $\ag$.
\qed
\eunum
\epf

\section{Generalization}
Originally, we expected $\mu_\YM$ to be absolutely continuous w.r.t.\
$\mu_0$. This presumption has been induced by the observation that 
although $\ymwirk$ cannot be extended from $\ag$ to $\agb$, i.e.\ 
a direct definition via $\dd\mu_\YM := \e^{-\ymwirk} \dd\mu_0$ 
is impossible, after exchanging limit and integral the expectation values
are indeed completely well-defined. Moreover,
it has been doubtful whether 
$\ymwirk$ on $\ag$ can serve as a starting point for the 
definition of such an action on $\agb$ because $\ag$ is
contained in a $\mu_0$-zero subset.
However, as we have seen in the last section,
the Yang-Mills measure and the Ashtekar-Lewandowski measure 
{\em are}\/ inequivalent. This has the following fundamental
consequence:
\begin{center}
The interaction measure $\mu_\YM$ cannot be constructed from $\mu_0$ 
using the action method.
\end{center}
This means, there is no measurable function 
$\quer\ymwirk$ on $\agb$, such that
$\mu_\YM = \e^{-\quer\ymwirk} \kp \mu_0$.

This immediately raises the question, whether the Ashtekar-Lewandowski measure
were simply the wrong 
starting point for the construction of 
the Yang-Mills measure. This argumentation is indeed 
entitled 
because in the naive limit of an infinite coupling
($\kopp \gegen \infty$) both measures are identical 
or --~in other words~-- the Ashtekar-Lewandowski measure 
$\mu_0$ is simply the Yang-Mills measure for infinite coupling.
Physically it is obvious that both cases of finite 
and infinite coupling have to be essentially different.
To underpin this measure-theoretically the corresponding measures 
are to be inequivalent.
But, why should one take the measure of the rigid theory
as a kinematical measure? Typically, one starts with the free theory anyway,
hence with vanishing coupling ($\kopp = 0$).
In this case, however, one sees that $\mu_\YM$ is simply the 
Dirac measure in $e_\LG$ which obviously is singular w.r.t.\ to the 
Yang-Mills measure of finite coupling as well.
Not only that is why we consider $\mu_0$ as a kinematically destined 
measure. On the one hand, $\mu_0$ personifies (in complete contrast 
to a point measure)
by its $\Gb$-invariance and its even larger invariance on the graph level
the principle of equal a-priori probability: Only the dynamics
should tell us which configurations are favoured by the system.
On the other hand, via $\mu_0$ one can easily construct 
all continuum measures that are at the lattice level 
absolutely continuous w.r.t.\ the Haar measure.
This is true as we already noticed even if 
the continuum measure is purely singular.
Could not the continuum limit be the 
real deeper reason for the singularity of the interaction measure?

In order to study this question (on a very preliminary stage, of course)
we review the proofs for the singularity in the Yang-Mills case
and try to understand their basic ideas. 
First we have proven that the lattice measures for single flags are
absolutely continuous (Proposition \ref{prop:ymd_fahne}). 
For this it was crucial that the Fourier series
$\ymd_\beta := \sum_\vnu \erww{\charakt_\vnu}_\beta \: \charakt_\vnu$
converges because then $\ymd_\beta$ 
can be considered as a well-defined density function of 
$\mu_{\YM,\beta}$ w.r.t.\ $\mu_{0,\beta} \ident \mu_\Haar$.
The convergence itself followed from the fast, 
here even exponential falling of $\erww{\charakt_\vnu}_\beta$ for 
increasing $\norm\vnu$. The absolute continuity of arbitrary lattice measures
now came from the fact that certain (here precisely the non-overlapping)
flags are independent random variables (Proposition \ref{prop:ymd_graph}). 
For the singularity of the continuum measure 
the for decreasing $\beta$ increasing concentration of the density function 
$\ymd_\beta$ around the identity of $\LG$ is responsible
(Lemma \ref{lem:sing_YM_0:1}).
Altogether we see that the presence of 
absolutely continuous lattice measures and purely singular continuum measures 
depends less on the concrete model under consideration, but more 
on three general properties of the expectation values.
In the following we will deduce these three properties from
three (physically relatively plausible) criteria \cite{paper7}.
For this we assume that we are given some physical theory that can
be described within the Ashtekar approach using some (possibly unknown)
measure $\mu$ and that provides us with some appropriate expectation values. 
Here neither the compact structure group $\LG$ is fixed, nor 
the dimension of $M$ is restricted to $2$. 
Now we are going to explain the three mentioned properties.

\subsection{Principle 1: Universality of the Coupling Constant}
We are aiming at the following statement:
If the theory considered has a (in a certain sense) universal 
coupling constant that by itself describes the coupling strength between
the elementary (matter) particles of that theory, then
$\erww{\tr\: \darst(h_\beta)}$ is determined 
completely by $\erww{\tr\: h_\beta}$ and the representation $\darst$.
Here $\erww f$ always denotes the physical expectation value of a function $f$.%
\footnote{In the following, we always assume $\erww{\tr\: h_\beta} \geq 0$.}

Let us consider the simplest case of a 
Yang-Mills theory with structure group $U(1)$.
The elementary matter particles are the single-charged 
particles; the coupling constant be $\kopp = e$.
Classically, the interaction, i.e.\ the potential between
a particle and its antiparticle, is obviously proportional to $\kopp^2$.
Now we call the coupling constant to be {\em universal}\/ if
it yields immediately the (classical) interaction between
{\em arbitrarily}\/ charged particles:
In particular, for composed particles with charges $n$ and $-n$, resp.,
it is proportional $(n \kopp)^2$.
In general, one assumes that also the Wilson-loop expectation values 
$\erww{h_\beta}$ describe the potential between two 
oppositely charged static particles \cite{Wilson,Seiler}.
Namely, if $\beta$ is a rectangular loop running in space
between $\vec x$ and $\vec y$ and in time between $0$ and $\Delta t$,
then the potential between the elementary particles resting in $\vec x$ and
$\vec y$, resp., is given by
\[ 
 V_1(\vec x - \vec y) 
    = - \lim_{\Delta t \gegen \infty} \inv{\Delta t} \ln \erww{h_\beta}.
\]
A Wilson loop so just carries the interaction between an elementary
particle-antiparticle pair; consequently, $n$ loops should 
yield the interaction between a pair of an 
$n$-times charged particle and its antiparticle.
On the other hand, (by the assumed universality of the coupling constant)
the corresponding potential $V_n$ is to be $n^2 V_1$.
Hence, we have
\[
   n^2 V_1(\vec x - \vec y) 
 = V_n(\vec x - \vec y) 
 = - \lim_{\Delta t \gegen \infty} \inv{\Delta t} \ln \erww{h_\beta^n}.
\]
Translating these two equations to the level of 
Wilson-loop expectation values, we get
(at least in the limit $\Delta t \gegen \infty$)
\begin{equation}
\erww{h_\beta^n} = \erww{h_\beta}^{n^2}.%
\label{gl:wlew_n=(wlew_1)quad:U(1)}
\end{equation}

Indeed, the Wilson-loop expectation values of the $U(1)$ theory 
for $d=2$ dimensions in the Ashtekar framework fulfill
equation \eqref{gl:wlew_n=(wlew_1)quad:U(1)} -- and namely not only for 
loops being large w.r.t.\ the time, but for all loops.
Hence, it is by no means unrealistic to identify the validity
of \eqref{gl:wlew_n=(wlew_1)quad:U(1)} for {\em all}\/ loops 
with the existence of a universal coupling constant.

Let us now turn to gauge theories having general compact structure group $\LG$.
Using the following translation table
\begin{center}
\begin{tabular}{l|ccc}
                           & $U(1)$ & $\auf$ & $\LG$ \\ \hline 
irreducible representation & $n$    & $\auf$ & $\darst$ \\
dimension                  & $1$    & $\auf$ & $d_\darst$ \\
normalized character       & $g^n$  & $\auf$ & $\inv{d_\darst} \charakt_\darst(g)$ \\
Casimir eigenvalue         & $n^2$  & $\auf$ & $c_\darst$
\end{tabular} $\:$ $\:$ $\:$,
\end{center}
Equation \eqref{gl:wlew_n=(wlew_1)quad:U(1)} becomes
\begin{equation}
\frac{\erww{\charakt_\darst(h_\beta)}}{d_\darst} =
     \Bigl(\frac{\erww{\charakt_{\darst_1}(h_\beta)}}{d_{\darst_1}}\Bigr)^{\frac{c_\darst}{c_1}},
\label{gl:wlew_n=(wlew_1)quad:LG}
\end{equation}
where $\darst_1$ denotes some nontrivial representation of $\LG$, 
e.g., the standard one of $\LG\teilmenge U(N)$ on $\C^N$.
Therefore, we will call a theory {\em having a universal coupling constant}\/ 
iff Equation \eqref{gl:wlew_n=(wlew_1)quad:LG} is fulfilled for all nontrivial
irreducible representations $\darst$ and all ``non-selfoverlapping'' 
loops $\beta$.

\enlargethispage{0.6cm}
From the physical point of view such an assumption has a very interesting
consequence: If a theory describes confinement (in the sense of an area law)
between the elementary particles, {\em all other}\/ charged
particle-antiparticle pairs are confined as well. In the case of QCD this just
explains why only quark-product particles consisting exclusively of baryons and mesons
are freely observable; they are simply those particles whose 
total color charge $\sqrt{{c_\darst}}$ equals zero, i.e.\ whose
quark product state transforms according the trivial $SU(3)$ representation.
We remark that this discussion is not new because
already about twenty years ago Yang-Mills theories 
with non-elementary charges have been considered (cf., e.g., \cite{Seiler})
and it has been shown that there occurs an area law as well.
However, there one started with the action 
$\inv2(\darst(F),\darst(F))$ specially taylored to those charges, such that
a comparison between differently charged particles is not
possible within {\em one}\/ model -- in contrast to our
description.

The measure-theoretical implication of a universal coupling constant
is now summarized in the following
\bprop
Let us given a theory with a universal coupling constant. Then we have:
\zgl{
\begin{tabular}{c|c@{ \: \: $\impliz$ \: \: }l}
   & expectation value & measure \\ \hline
equality
 & $0=\erww{\charakt_1}_\beta\phantom{{}<d_1}$
 & $\mu_\beta = \mu_\Haar$ \\
absolute continuity
 & $0<\erww{\charakt_1}_\beta < d_1$
 & $\mu_\beta = \ymd_\beta \kp \mu_\Haar$
 \\
singularity
 & $\phantom{0 <{}}\erww{\charakt_1}_\beta = d_1$
 & $\mu_\beta = \delta_{e_\LG}$.
 \\
\end{tabular}
}\noindent
$\ymd_\beta$ is again some smooth function.
\eprop
Here, $\erww\cdot_\beta$ denotes the expectation value
w.r.t.\ the image measure $\mu_\beta \ident \pi_\beta{}_\ast \mu$.
For brevity, we write $\charakt_1$ instead of $\charakt_{\darst_1}$ etc.\
\bpf
The cases $\erww{\charakt_1}_\beta$ equals $0$ or $d_1$ are clear.

Let now $0 < \erww{\charakt_1}_\beta < d_1$. 
Define 
$\ymd_\beta := \sum_\vnu \erww{\charakt_\vnu}_\beta \: \charakt_\vnu$
and $b := - \inv{c_1} \ln(\inv{d_1}\erww{\charakt_1}_\beta) > 0$.
The absolute convergence of $\ymd_\beta$ now follows as in 
Proposition \ref{prop:ymd_fahne} from
\zgl{
\betrag{\ymd_\beta(g)} 
  \leq  \sum_\vnu \erww{\charakt_\vnu}_\beta \supnorm{\charakt_\vnu} 
   =    \sum_\vnu d_\vnu \Bigl(\frac{\erww{\charakt_1}_\beta}{d_1}\Bigr)%
                   ^{\frac{c_\vnu}{c_1}} \: d_\vnu 
   =    \sum_\vnu \e^{- b \: c_\vnu}
                    \: d_\vnu^2
}\noindent
for all $g\in\LG$
and the standard estimates for $c_\vnu$ and $d_\vnu$.
As above $\ymd_\beta$ is even smooth.
The relation $\mu_\beta = \ymd_\beta \kp \mu_\Haar$ comes again 
as in Proposition \ref{prop:ymd_fahne}.
\qed
\epf

Finally, we note that just the universality of the coupling constant
might be a desirable property of unified theories.

\subsection{Principle 2: Independence Principle}
It is well-known that non-overlapping loops yield
independent random variables in the two-dimensional Yang-Mills theory. 
This means, 
for all finite sets $\beta_1, \ldots, \beta_n$ 
of such loops 
we have 
\begin{equation}
        \erww{\charakt_{\darst_1} (h_{\beta_1}) \: \cdots \:
              \charakt_{\darst_n} (h_{\beta_n})} = 
        \erww{\charakt_{\darst_1} (h_{\beta_1})} \: \cdots \:
        \erww{\charakt_{\darst_n} (h_{\beta_n})}
\label{gl:unabhprinzip1}
\end{equation}
for all representations 
$\darst_1, \ldots, \darst_n$ of the structure group $\LG$ 
or --~more precisely in terms of loop-networks~--
\zglnum{\erww{\nchar_{\vec\darst,\darst}}_\gb = 
       \sqrt{d_\darst} \: \prod_{\nu} 
           \frac{\erww{\charakt_{\darst_\nu}}_{\beta_\nu}}%
                {\sqrt{d_{\darst_\nu}}}}{gl:unabhprinzip2}\noindent
for all $\vec\darst$ and $\darst$.
However, to demand Equation \eqref{gl:unabhprinzip2} 
being satisfied for general theories
is too restrictive physically because then every quantum state 
will be ultralocal and the Hamiltonian vanishes \cite{schlingemann}.
Actually we do not need such a general statement for all non-overlapping
loops. What we rather need is a sufficiently large number of ``small''
loops fulfilling the relations above. 
Of course, non-overlapping loops remain natural candidates
for this although their precise definition is worth discussing -- in particular 
from dimension $3$ on.
As a minimal version one could view a set of loops as
non-overlapping if there is a surface in the space-time
such that these loops form a set of non-overlapping loops.
However, this condition seems to be too restrictive.
Perhaps one could resort to the knot theory instead; maybe there are
physically interesting measures where Equation \eqref{gl:unabhprinzip2}
is fulfilled for all sets of loops that have Gauss winding number $0$.

Stopping this discussion here, 
we now just define a set of loops to be measure-theoretically
independent iff it fulfills Equation \eqref{gl:unabhprinzip2}
for all $\vec\darst$ and $\darst$. We have analogously to
Proposition \ref{prop:ymd_graph}
\bprop
The lattice measure $\mu_\gb$ for a measure-theoretically independent
weak fundamental system $\gb$
is absolutely continuous if
all single lattice measures $\mu_{\beta_i}$ are absolutely continuous.

The density function of $\mu_\gb$ w.r.t.\ $\mu_\Haar^{\elanz\gb}$
then equals $\ymd_\gb = \ymd_{\beta_1} \cdots \ymd_{\beta_n}$.
\eprop
Finally we declare a theory to obey the {\em independence principle}\/
if there is an infinite number of loops 
of decreasing geometrical size (see below) that are 
independent both graph-theoretically and measure-theoretically.

\subsection{Principle 3: Geometrical Regularity}
After we have discussed two principles on the level of a fixed lattice, we
are now going to discuss the continuum limit.
If a theory is to have a continuum limit, then the holonomy along a loop
should go to the identity when shrinking the loop to a point.
In other words, since a measure
in general encodes the distribution of certain objects, this suggests that 
the smaller the loop -- the more the corresponding lattice measure
should concentrate around the identity \cite{d26}.
One could even demand that the lattice measure 
goes to the $\delta$-distribution.
Hence, it should be clear 
that the continuum limit naturally leads to singular measures.

In order to retrace this effect also quantitatively,
we transfer it to the level of expectation values.
First it is obvious that $\erww{\charakt_\darst}_\beta$ should go to 
the dimension $d_\darst$ of the representation $\darst$, 
if the (non-selfoverlapping) loop $\beta$ becomes small.
In the case of the two-dimensional Yang-Mills theory, one can even
prove that $d_\darst - \erww{\charakt_\darst}_\beta < \const \FIL{\beta}$ holds,
i.e.\ the expectation values are H\"older continuous w.r.t.\ the
area $\FIL{\beta}$ enclosed by the loop $\beta$.
Therefore we will call a theory {\em geometrically regular}\/
iff there is a non-negative real function $\sigma(\beta)$ 
such that first
\begin{equation}
   \frac{d_\darst - \erww{\charakt_\darst}_\beta}{\sigma(\beta)}
\label{gl:geomregul}
\end{equation}
is bounded as a function of $\beta$ and second $\sigma$
goes to $0$ for shrinking $\beta$. For technical reasons
we assume here that $\darst$ is the representation having smallest
non-zero Casimir eigenvalue.
Examples of conceivable functions $\sigma(\beta)$ are the area
$\FIL\beta$ enclosed by $\beta$ or the length $L(\beta)$
of $\beta$. Now we have
\bprop
\label{prop:prinz123=sing}
In a theory with universal coupling constant, independence principle
and geometrical regularity the continuum measure $\mu$
is always purely singular w.r.t.\ to the Ashtekar-Lewandowski measure $\mu_0$.
\eprop
We note that the geometrical regularity implies immediately
the convergence of the density function $\ymd_\beta$ to 
the $\delta$-distribution in $e_\LG$ for $\sigma(\beta) \gegen 0$.
\bpf
We can assume $\erww{\charakt_\darst}_\beta \neq d_\darst$
for all these independent $\beta$. Otherwise even the 
lattice measure would be singular and the continuum measure all the more.
The possibility $\erww{\charakt_\darst}_\beta = 0$ is excluded 
by Equation \eqref{gl:geomregul}.

The proof now follows mostly the proofs in 
Subsection \ref{uabschn:ym2:singYM},
such that we present here the modifications only.
First all special expectation values are to be substituted by
$\erww{\charakt_\vnu}_\beta$ and 
then $\FIL\beta$ by the more general geometrical
function $\sigma(\beta)$. Moreover, we have to 
observe that from $c_\vnu \geq c_\darst$ for all $\vnu\neq\vec 0$
always 
\zgl{
   \frac{1 - \inv{d_\vnu}\erww{\charakt_\vnu}_\beta}{c_\vnu \sigma(\beta)}
 = c_\darst \frac{1 - (\inv{d_\darst}\erww{\charakt_\darst}_\beta)^{\frac{c_\vnu}{c_\darst}}}
        {\frac{c_\vnu}{c_\darst} \sigma(\beta)}
 \leq c_\darst \frac{1 - \inv{d_\darst}\erww{\charakt_\darst}_\beta}{\sigma(\beta)}
}
follows. Hence the first term is uniformly bounded w.r.t.\ $\vnu$
as a function of $\beta$. This suffices to transfer 
Lemma \ref{lem:sing_YM_0:1}. Lemma \ref{lem:sing_YM_0:2}
follows from the independence principle, 
and Lemma \ref{lem:sing_YM_0:3} is obvious.
The present proof now follows from that of Theorem \ref{thm:sing_YM_0}.
\qed
\epf
We remark that not only the singularity statement itself is true, but also the 
construction of the partition of $\AbGb$ into disjoint supports of
$\mu_0$ and $\mu$ can be reused.

\subsection{Examples}
The first example for a purely singular measure has been
already studied in the sections before -- the Yang-Mills measure for $\R^2$. 
There are also striking hints that the same results can be gained
for the other Yang-Mills theories on two-dimensional spaces as well.
Namely, Sengupta \cite{d30} could prove on the classical level that 
in certain graphs (e.g.\ simple and small graphs that only contain 
homotopically trivial loops) the lattice measures are
given by heat-kernel measures as in the $\R^2$-case.
It can be expected that these results can be transferred to the 
Ashtekar approach as for $\R^2$ because 
holonomies outside a graph have been unimportant for the continuum
limit in $\R^2$.
In contrast to this, calculations of 
Aroca and Kubyshin \cite{iosa2}
indicate for compact space-time that the  
area of the complement of a graph influences the expectation values
by its finiteness.
Hence, the universality
of the coupling constant is given only approximatively.
However, the interpretation of our principles has to be handled with care
for compact space-times anyway: A limit $\Delta t \gegen \infty$ 
is hard to define.
Nevertheless, in general one can expect purely singular continuum measures, 
hence a failure of the action method for $d = 2$.

To get a larger class of theories with purely singular continuum measure 
observe that 
the geometrical regularity is given for every theory with an area
law $\erww{\tr\:\darst(h_\beta)} = d_\darst \: \e^{-\const \FIL\beta}$ or 
a length law $\erww{\tr\:\darst(h_\beta)} = d_\darst \: \e^{-\const L(\beta)}$.
The former one is regarded as an indicator for confinement, and the latter one 
for deconfinement. Since among our three criteria 
just the geometrical regularity is the most important one for the singularity
of the continuum measure, one could expect for both classes of theories
that the action method fails. However, we have to mention
that both the deconfinement and the confinement criterion need 
the corresponding laws for loops that are {\em large}\/ in the time
direction, but we actually need loops of {\em small}\/ size 
to prove the singularity of the measure (at least in two dimensions).
Both requirements can be matched together only in the area-law case:
Here one can still generate 
loops with small area by choosing very narrow loops that are large w.r.t.\
the time which is impossible in the length-law case.   
Therefore, up to now, 
we can only claim that the appearance of an area law is a convincing
indicator for a purely singular continuum measure.

However, if we are looking only for a failure of the action method
(i.e.\ only for singular, not for purely singular measures),
Proposition \ref{prop:singkrit(lewand)} implies that probably almost no theory
can be gained using the action method on the continuum level.

\section{Concluding Remarks}
In the present article we have shown that a theory having a 
universal coupling constant and obeying an independence principle
has absolutely continuous lattice measures, but a singular continuum
measure if the theory is even geometrically regular. That is why
neither non-generic connections nor the Gribov problem play 
any r\^ole in such a theory -- provided one only looks at phenomena 
that can be discussed using the physical measure.
However, it comes to a significant concentration of the continuum measure
in a neighbourhood of the singular strata, since for small $\beta$
the concentration of the density function $\ymd_\beta$ increases in a 
neighbourhood of the ``most'' singular element $e_\LG \in \LG$.%
\footnote{At the first glance, this seems to be a contradiction to 
$\ymd_\beta \nach \delta_{e_\LG}$. But, this is not correct because 
such a limiting process runs over different lattices and does therefore
not yield a comparably 
convergent process after lifting to the level of $\AbGb$.}
This (qualitative) observation strengthens the conjecture of 
Emmrich and R\"omer \cite{f21} that 
singularities typically lead to 
concentrations of the wave functions.
Maybe that this way the singular strata indeed get some influence
although it cannot be described measure-theoretically.

However, from our point of view
much more important is the realization that
the singularity of the full interaction measure $\mu$
can be regarded as a typical property of the continuum. Hence, in particular, 
regular continuum limit and action method exclude each other: 
Assuming regularity, the definition of the interaction measure via
$\mu := \e^{-S} \kp \mu_0$ is {\em impossible}.  
If one uses the action method, one can at most ``approximate'' it
by lattice measures constructed this way. For all that 
it is mostly tried to get $\mu$ via the action method 
on the continuum level. Maybe that just this sticking
to the action method is a deeper reason for the problems 
with the continuum limit or quantizations occuring permanently up to now.
The desired absolute continuity seems to be a deceptfully simple tool,
since it hides important physical phenomena. But, the singularity of a measure 
{\em per se}\/ is completely harmless. There is no singularity in the 
dual picture, i.e.\ for the expectation values. Moreover, strictly speaking, 
the measure is no physically relevant quantity; only expectation
values are detectable. So far it is to be evaluate absolutely positive
that the interaction measure $\mu$ has not been used in our principles, 
but rather some of its expectation values. 
It has been completely sufficient to know
that $\mu$ does {\em exist} at all for extracting properties of $\mu$ 
from our physical principles in a mathematically rigorous way.
Thus, a measure is only the mathematical arena where anything happens.
To know it might be superfluous from the physical point of view; however,
one must be able to rely on it.

\section*{Acknowledgements}
The author thanks Gerd Rudolph, Eberhard Zeidler and, in particular,
Jerzy Lewandowski for very fruitful discussions.
The author is supported by the Reimar-L\"ust-Sti\-pen\-dium of the 
Max-Planck-Gesellschaft.

\anhangengl

\section{Estimates for the Fourier Analysis}
\label{appabschn:fourierreihen}
In this appendix we give some criteria needed for the 
convergence proofs of series over $\N^l$.
Before doing this we introduce some notation.
The set
$\wuerfel-l{\vn} := \{\vx\in\R^l\mid n_i-1 < x_i \leq n_i \: \forall i\}$ 
describes a half-open cube with edge length $1$ in $\R^l$
that is determined by the two corners $\vn-\vec 1$ and $\vn$.
Analogously we define 
$\wuerfel{}l{\vn} := \{\vx\in\R^l\mid n_i -\einhalb \leq x_i < n_i+\einhalb 
 \: \forall i\}$. 
\bprop
\label{prop:konvkritN1}
Let $f:\R_+\nach\R$ be some function, $l\in\N_+$ and 
$\nu_\vn := \elanz\{i\mid n_i \neq 0\}$.

If there exists some $\rho\in\R_{\geq 0}$ and some function
$g:\R_+\nach\R_{\geq 0}$ with
\bunum
\item
$\int_\rho^\infty g(r) \: r^{\laufi-1} \: \dd r < \infty$ for all
$\laufi\in\N$, $1\leq \laufi\leq l$, and
\item
$\betrag{f(\norm\vn)} \leq g(\norm\vx)$ for all 
$\vn\in\N^l$ with $\norm\vn\geq\rho+1$ 
and for all $\vx\in\wuerfel-l\vn\cap\R_{\geq 0}^l$,
\eunum
then $\sum_{\vn\in\N^l\setminus\{\vec 0\}} f(\norm\vn)$ 
converges absolutely and we have 
\zgl{
 \sum_{\substack{\vn\in\N^l \\ \vn\neq\vec 0 \\ \norm\vn\geq\rho+\sqrt{\nu_\vn}}}
   \betrag{f(\norm\vn)}
 \leq \sum_{\laufi=1}^l 
         \binom l \laufi 
         \frac{\pi^{\frac \laufi 2}}{2^{\laufi-1} \Gamma(\frac \laufi 2)}
         \int_\rho^\infty g(r) \: r^{\laufi-1} \: \dd r.}
Here, $\Gamma$ is the Gamma-function.
\eprop
\bpf
We only consider functions with $f\geq 0$.
(Here, convergence equals absolute convergence.)
\bunum
\item
We divide the sum over all $\vn\in\N^l\setminus\{\vec 0\}$ 
into $2^l-1$ partial sums.
For this, $I\teilmenge\{1,\ldots,l\}$ be some non-empty subset.
We define the partial sum $S_I$ belonging to $I$ by
$S_I := \sum_{\vn\in\N^l, n_i\neq 0 \aequ i\in I} f(\norm\vn)$.
Since $\N^l\setminus\{\vec 0\}$ is the disjoint union of all
$\{\vn\in\N^l\mid n_i\neq 0 \aequ i\in I\}$ with running $I$, it suffices to
show the (absolute) convergence of $S_I$ for every $I$.
\item
Obviously,
\bgl
 S_I   & = & \sum_{\vn\in\N^l, n_i\neq 0 \aequ i\in I} f(\norm\vn) 
\breitrel=   \sum_{\vn'\in\N^{\elanz I}, n'_i\neq 0\: \forall i} f(\norm{\vn'}).
\egl     
Hence we are left to prove the convergence of
$S_\laufi := \sum_{\vn\in\N_+^\laufi} f(\norm{\vn})$
for all $1\leq \laufi \leq l$.
We set
\zgl{S_{I,\rho} := \sum_{\substack{\vn\in\N^l \\ n_i\neq 0 \aequ i\in I \\
                        \norm\vn\geq\rho+\sqrt{\nu_\vn}}} f(\norm\vn)
\text{ \: \: \: and \: \: \:}
     S_{\laufi,\rho} := \sum_{\substack{\vn\in\N_+^\laufi \\ \norm\vn\geq\rho+\sqrt \laufi}}
                         f(\norm\vn).}

We have $S_I = S_{\elanz I}$ and $S_{I,\rho} = S_{\elanz I,\rho}$.
\item
It is clear that the union
\zgl{\bigcup_{\vn\in\N_+^\laufi} \wuerfel-\laufi\vn = \R_+^\laufi}
is disjoint, and we have
$\bigcup_{\vn\in\N_+^\laufi,\norm\vn<\rho+\sqrt \laufi} \wuerfel-\laufi\vn 
 \obermenge B_\rho^\laufi \cap \R_+^\laufi$:
Let $\vx\in B_\rho^\laufi\cap\R_+^\laufi$,
and let $\vn$ be that element in $\N_+^\laufi$ with $\vx\in\wuerfel-\laufi\vn$;
then $\norm\vn \leq \norm{\vn-\vx}+\norm{\vx} < \sqrt \laufi + \rho$.
Here, $B_\rho^\laufi$ is the ball around the origin $\vec 0$ with radius $\rho$.
Now we have
\bgl 
S_{\laufi,\rho} 
    &  =   & \sum_{\vn\in\N_+^\laufi, \norm\vn\geq\rho+\sqrt \laufi} f(\norm{\vn}) \\
    & \leq & \sum_{\vn\in\N_+^\laufi, \norm\vn\geq\rho+\sqrt \laufi} 
                    \int_{\wuerfel-\laufi{\vn}}
		          g(\norm{\vx}) \: \dd^\laufi\vx \\
    &      & \erl{$f(\norm\vn) \leq g(\norm\vx)$ for all 
                  $\vn\in\N_+^\laufi$, $\norm\vn\geq\rho+\sqrt \laufi$,
		  and all $\vx\in\wuerfel-\laufi\vn$}\\
    & \leq & \int_{\R_+^\laufi\setminus B_\rho^\laufi}
	           g(\norm{\vx}) \: \dd^\laufi\vx 
             \erl{by 
                  $\bigcup_{\norm\vn\geq\rho+\sqrt \laufi}\wuerfel-\laufi\vn
		   \teilmenge \R_+^\laufi\setminus B_\rho^\laufi$}\\
    &  =   & \inv{2^\laufi}
             \int_{\R^\laufi\setminus B_\rho^\laufi}
                    g(\norm{\vx}) \: \dd^\laufi\vx \\
    &  =   & \inv{2^\laufi} \: \vol(\del B_1^\laufi) \:
             \int_\rho^\infty g(r) \: r^{\laufi-1} \: \dd r
	     \\
    &  <   & \infty
\egl
by assumption.
Since the set of all $\vn\in\N_+^\laufi$ with $\norm\vn<\rho+\sqrt \laufi$
is finite, also
$S_\laufi \ident \sum_{\vn\in\N_+^\laufi, \norm\vn<\rho+\sqrt \laufi} f(\norm{\vn}) 
            + S_{\laufi,\rho}$
is finite.
\item
Therefore 
$\sum_{\vn\in\N^l\setminus\{\vec 0\}} f(\norm\vn)$ converges (absolutely).
\item
Moreover, we have
\bgl
\sum_{\vn\in\N^l, \norm\vn\geq\rho+\sqrt{\nu_\vn}} f(\norm\vn)
 &  =   & \sum_{I\teilmenge\{1,\ldots,l\}, I\neq\leeremenge} S_{I,\rho}
\breitrel=\sum_{\laufi=1}^l \binom l \laufi S_{\laufi,\rho} \\
 & \leq & \sum_{\laufi=1}^l \binom l \laufi \: \inv{2^\laufi} \: \vol(\del B_1^\laufi) \:
               \int_\rho^\infty g(r) \: r^{\laufi-1} \: \dd r.
\egl
\item
The assertion follows from
$\vol(\del B_1^\laufi) = \frac{2 \pi^{\frac \laufi 2}}{\Gamma(\frac \laufi 2)}$.
\eunum
The case of an arbitrary function $f$ is now clear.
\qed
\epf

\bcorr
\label{corr:konvkritNZ_rho=0}
Let $k,l\in\N$, $k+l\neq 0$, and let
$f:\R_+\nach\R$ be some function. If there exists a function
$g:\R_+\nach\R_{\geq 0}$ with
\bunum
\item
$\int_0^\infty g(r) \: r^{\nu-1} \: \dd r < \infty$ for all $\nu\in\N$,
$1\leq \nu\leq k+l$, and
\item
$\betrag{f(\norm\vnu)} \leq g(\norm\vx)$ for all $\vnu\in\N^l\kreuz\N^k$,
$\vnu\neq\vec 0$, and
all $x\in\wuerfel-{k+l}\vnu \cap \R_{\geq 0}^{l+k}$,
\eunum
then $\sum_{\vnu\in\N^l\kreuz\Z^k, \vnu\neq\vec 0} f(\norm\vnu)$
converges absolutely and we have in this case 
\zgl{
 \sum_{\substack{\vnu\in\N^l\kreuz\Z^k \\ \vnu\neq\vec 0}}
   \betrag{f(\norm\vnu)} 
 \leq 2^k \sum_{\laufi=1}^{k+l} 
         \binom{k+l}\laufi 
         \frac{\pi^{\frac \laufi 2}}{2^{\laufi-1} \Gamma(\frac \laufi 2)}
         \int_0^\infty g(r) \: r^{\laufi-1} \: \dd r.}
\ecorr
\bpf
For all $\vnu\in\N^l\kreuz\N^k$ we have obviously 
$\norm\vnu\geq\sqrt{\nu_\vnu}$.
Hence by Proposition \ref{prop:konvkritN1} the series
$\sum_{\vnu\in\N^l\kreuz\N^k,\vnu\neq\vec 0} \betrag{f(\norm\vnu)}$
converges absolutely with 
\zgl{
 \sum_{\substack{\vnu\in\N^l\kreuz\N^k \\ \vnu\neq\vec 0}}
   \betrag{f(\norm\vnu)} 
 \leq \sum_{\laufi=1}^{k+l} 
         \binom{k+l}\laufi 
         \frac{\pi^{\frac \laufi 2}}{2^{\laufi-1} \Gamma(\frac \laufi 2)}
         \int_0^\infty g(r) \: r^{\laufi-1} \: \dd r.}
Consequently,
$\sum_{\vnu\in\N^l\kreuz\Z^k,\vnu\neq\vec 0} \betrag{f(\norm\vnu)} \leq 
 2^k \sum_{\vnu\in\N^l\kreuz\N^k,\vnu\neq\vec 0} \betrag{f(\norm\vnu)}$ 
converges absolutely as well.
\qed
\epf

If one is not interested in a concrete estimate, but only in a 
convergence statement, one can get help from
\bcorr
\label{corr:konvkritNZ_spez}
Let $k$, $l$ and $f$ be as above.

If there is a $\rho\in\R_{\geq 0}$ and a function 
$g:\R_{\geq\rho}\nach\R$,
such that
\bunum
\item
$\int_\rho^\infty g(r) \: r^{{k+l}-1} \: \dd r < \infty$,
\item
$g$ is monotonically decreasing on $[\rho,\infty)$ and
\item
$\betrag{f(r)} \leq g(r)$ on $[\rho,\infty)$,
\eunum
then $\sum_{\vnu\in\N^l\kreuz\Z^k, \vnu\neq\vec 0} f(\norm\vnu)$ 
is absolutely convergent.
\ecorr
\bpf
Set $\rho':=\rho+\sqrt {k+l}$.

By $\rho'\geq 1$ we have first $g(r) r^{\laufi-1} \leq g(r) r^{{k+l}-1}$ 
for all $r\geq\rho'$ and $\laufi\leq {k+l}$. Hence, also 
$\int_{\rho'}^\infty g(r) \: r^{\laufi-1} \: \dd r \leq
 \int_{\rho'}^\infty g(r) \: r^{{k+l}-1} \: \dd r < \infty$ 
 for all $\laufi\leq {k+l}$.
 
Second we have 
$\norm\vx \geq \norm\vnu - \norm{\vnu-\vx} 
 \geq \rho' + 1 - \sqrt {k+l} = \rho + 1$
for all $\norm\vnu\geq\rho'+1$ and 
$\vx\in\wuerfel-{k+l}\vnu\cap\R_{\geq 0}^{k+l}$.
For monotonicity reasons we have 
$\betrag{f(\norm\vnu)} \leq g(\norm\vnu) \leq g(\norm\vx)$ 
for all $\vnu\in\N^l\kreuz\N^k$ with
$\norm\vnu\geq\rho'+1$ and all
$\vx\in\wuerfel-{k+l}\vnu\cap\R_{\geq 0}^{k+l}$.

Proposition \ref{prop:konvkritN1} yields the assumption for the sum over
$\N^l\kreuz\N^k$. For the sum over $\N^l\kreuz\Z^k$ 
one argues as in Corollary \ref{corr:konvkritNZ_rho=0}.
\qed
\epf

\bcorr
\label{corr:konvkritNZ_poly}
Let $k,l\in\N$ with $k+l\neq 0$ and $\rho\in\R_+$ be arbitrary, and
let $f:\R_+\nach\R$ be some function with
$\betrag{f(x)}\leq\const\inv{x^{k+l+1}}$ for all $x\geq\rho$.

Then $\sum_{\vnu\in\N^l\kreuz\Z^k, \vnu\neq\vec 0} f(\norm\vnu)$ 
converges absolutely.
\ecorr
\bpf
The function $g(x) := \const \inv{x^{k+l+1}}$ is obviously 
monotonically decreasing on the whole $\R_+$. 
Moreover, we have 
$\int_\rho^\infty g(r) \: r^{k+l-1} \: \dd r = 
 \const \int_\rho^\infty r^{-2} \: \dd r < \infty$. 
Corollary \ref{corr:konvkritNZ_spez} gives the assertion.
\qed
\epf

\section{Small Holonomies}
\label{abbabschn:klholo}
In this appendix we study the behaviour of holonomies for small loops
not restricting ourselves to two-dimensional manifolds.
$\LG$ is always considered as a subset of some 
$U(N) \teilmenge \Gl_\C(N)$; hence,
 $\Lieg\teilmenge \gl_\C(N) = \C^{N\kreuz N}$.
Additionally, $\algnorm\cdot$ is some algebra norm on $\gl_\C(N)$.
To compute holonomies locally we choose a local chart 
$U\teilmenge M$ with the chart mapping 
$\karte : U \nach \karte(U)$ and the
coordinate functions $x^\mu$. 
We arrange for every positive $c$ 
\bdf
Let $\alpha:[0,T] \nach U$ be some path in $U$ and
$x := \karte \circ \alpha$ its image on $\karte(U)$.

We call $\alpha$ $(\mu\nu,c)$-round (or shortly \df{$c$-round}) iff
\bunum
\item
$\im\alpha$ is contained completely in the 
surface spanned by the coordinates $x^\mu$ and $x^\nu$,
\item
$\alpha$ is a closed Jordan curve in that surface,
\item
$\sum_i \betrag{\dot x^i(t)}^2 = 1$ for all $t\in[0,T]$ and
\item
$T^2 \leq c \FIL{\alpha}_U$.
\eunum
Here, $G_\alpha$ is the domain in the $\mu\nu$-surface enclosed by $\alpha$. 

$\FIG{\alpha}{U} := 
      \int_{\karte(G_\alpha)} \dd x^\mu \keil \dd x^\nu$
is its ``oriented'' and
$\FIL{\alpha}_U := 
      \betrag{\int_{\karte(G_\alpha)} \dd x^\mu \keil \dd x^\nu}$
its ``absolute'' area. ($\FIL{\alpha}_{U}$ is just the Euclidian
area of $G_\alpha$.)
\edf
We have chosen the term ``round'' because of the last condition
$T^2 \leq c \FIL{\alpha}_{U}$.
Due to the isoperimetrical inequality we have always
$T^2 \geq 4\pi \FIL{\alpha}_{U}$ with equality precisely for the circle.
Additionally, $\alpha$ has to become similar to a circle
if we let $c$ decrease.
Nevertheless, for $c > 4\pi$ there is a huge number of paths $\alpha$
fulfilling the conditions above.

Now we have 
\bprop
\label{prop:kl_holo}
Let the image of $U$ in $\R^{\dim M}$ be convex
and let $\betrag{x^\mu}$ for all $\mu$ be bounded on $U$ 
by some $C\in\R_+$. Moreover, let $c \geq 4 \pi$ be arbitrary, but fixed.

Then for all $A\in\A$ there is a constant $\const_A\in\R$
(depending only on $A$, $U$, $c$ and the algebra norm $\algnorm\cdot$),
such that 
\zgl{\algnorm{h_A(\alpha) - (\EINS - F_{\mu\nu}(m_0)\FIG \alpha U)} 
            \leq \const_A (\FIL{\alpha}_{U})^{\frac32}}
for all $\mu$, $\nu$ and all $(\mu\nu,c)$-round $\alpha$ in $U$ 
with base point $m_0\in U$.

Here, $F_{\mu\nu} := \del_{[\mu} A_{\nu]} + [A_\mu, A_\nu]$ is 
the curvature for $A$.
\eprop
The proof is not very difficult, but quite technical, and is therefore 
dropped here. It can be found in \cite{diss}.
\bcorr
\label{corr:regzusklholo}
For all $A\in\A$ there is a constant $c_A \in\R$ depending on $A$,
such that 
$\algnorm{h_A(\alpha) - \EINS} \leq c_A \FIL\alpha_U$
for all $c$-round paths $\alpha\in\hg$ in $U$.
\ecorr
\bpf
By Proposition \ref{prop:kl_holo} there is a constant $\const_A \in\R$
for every $A\in\A$, such that
\zglklein{\algnorm{h_A(\alpha) - (\EINS - F_{\mu\nu}(m)\FIG{\alpha}{U})} 
  \leq \const_A (\FIL\alpha_U)^{\frac32},}
provided $\alpha\in\hg$ is a $c$-round path contained in the 
coordinate surface $(x^\mu, x^\nu) \teilmenge U$.
Since $\FIL\alpha_U$ is bounded, there is in each case some
$c_A\in\R$
with $\algnorm{h_A(\alpha) - \EINS} \leq c_A \FIL\alpha_U$
for all round $\alpha\in\hg$.
\qed
\epf

\section{Haar Measure Estimate}
We estimate the Haar measure of all $g\in\LG$ whose distance to
$e_\LG$ is smaller than $\varepsilon$.
Again we consider $\LG$ as a subset 
of some $U(N) \teilmenge \Gl_\C(N)$, hence 
$\Lieg\teilmenge \gl_\C(N) = \C^{N\kreuz N}$, 
and choose some algebra norm $\algnorm\cdot$ on $\gl_\C(N)$.
Correspondingly we set
\bunum
\item
$B_\varepsilon(e_\LG) := \{g\in\LG \mid \algnorm{g - e_\LG}<\varepsilon\}$,
\item
$B_\varepsilon(\EINS) := \{g\in\Gl_\C(N) \mid \algnorm{g - \EINS}<\varepsilon\}$
and 
\item
$B_\varepsilon(\NULL) := \{X\in\gl_\C(N) \mid \algnorm{X}<\varepsilon\}$
\eunum
for $\varepsilon\in\R_+$.
Note that $e_\LG = \EINS$, but $e_\LG$ 
is used for $\LG$ and $\EINS$ for $\Gl_\C(N)$. 
\blem
\label{lem:absch(haarmasz)}
There is a constant $c$ 
with $\mu_\Haar(B_\varepsilon(e_\LG)) \leq c \: \varepsilon^{\dim\LG}$
for all $\varepsilon>0$.
\elem
\bpf
\bunum
\item
We consider the log-function \cite{HilgNeeb}
\zgl{\ln g = \sum_{n = 1}^\infty (-1)^{k+1} \frac{(g-\EINS)^k}{k}.}
For $g\in B_1(\EINS)$ this series converges absolutely and fulfills
$\exp(\ln g) = g$. Additionally, we have $\ln(\exp X) = X$ for all
$X \in B_{\ln 2}(\NULL)$.
Hence, for $g\in B_\einhalb(\EINS)$ we get
\zgl
{\algnorm{\ln g} 
  \leq \algnorm{g - \EINS} \: 
          \sum_{n = 1}^\infty \frac{\algnorm{g-\EINS}^{k-1}}{k}
  \leq (2 \ln 2) \algnorm{g - \EINS},}
i.e.\
$B_\varepsilon(\EINS) = \exp(\ln(B_\varepsilon(\EINS)))
 \teilmenge \exp(B_{(2\ln 2)\varepsilon}(\NULL))$
for all $\varepsilon < \einhalb$.
\item
$\exp : \Lieg \nach \LG$ is
a local diffeomorphism. Hence, there is an $\varepsilon_0 > 0$
(w.l.o.g.\ $\varepsilon_0 < 1$),
such that 
$\exp^{-1} : B_{\varepsilon_0}(e_\LG) \nach \exp^{-1}(B_{\varepsilon_0}(e_\LG))$ 
is a diffeomorphism, hence a chart mapping. 

Since the Haar measure on $\LG$ is a Lebesgue measure ,
there is a nowhere vanishing $C^\infty$-function
$f$ with
$(\exp^{-1})_\ast \mu_\Haar = f \kp \lambda^{\dim\Lieg}$
on $\exp^{-1}(B_{\varepsilon_0}(e_\LG))$. 
Here, $\lambda^{\dim\Lieg}$ is the Lebesgue measure on $\Lieg$. 

By the compactness of
$\exp^{-1}(\quer B_{\einhalb\varepsilon_0}(e_\LG))$,
$\betrag f$ has a maximum $f_0 < \infty$ there;
hence
\bgl
\hspace*{-0.4em}\mu_\Haar(B_\varepsilon(e_\LG)) 
  & \leq &      \mu_\Haar(\exp(B_{(2 \ln 2)\varepsilon}(e_\LG)))
\breitrel\ident (\exp^{-1})_\ast \mu_\Haar(B_{(2 \ln 2)\varepsilon}(e_\LG)) \\
  & \leq &      f_0 \vol(B^{\dim\Lieg}) (2 \ln 2)^{\dim\Lieg} \varepsilon^{\dim\Lieg}
\breitrel{=:}   \widetilde c \: \varepsilon^{\dim\Lieg}
\egl
for all $\varepsilon \leq \einhalb\varepsilon_0$. Here, 
$\vol(B^n)$ is the volume of the unit ball of the 
$n$-dimensional Euclidian space.
\item
Setting $c := \max\{\widetilde c, (\frac{2}{\varepsilon_0})^{\dim\Lieg}\}$
we get
\bnum{2}
\item
$\mu_\Haar(B_\varepsilon(e_\LG)) \leq \widetilde c \: \varepsilon^{\dim\Lieg}
                              \leq c \: \varepsilon^{\dim\Lieg}$ 
for $\varepsilon\leq\einhalb\varepsilon_0$ and
\item
$\mu_\Haar(B_\varepsilon(e_\LG)) 
    \leq 1 
    \leq (\frac{2\varepsilon}{\varepsilon_0})^{\dim\Lieg}
    \leq c \: \varepsilon^{\dim\Lieg}$
for $\varepsilon\geq\einhalb\varepsilon_0$
by the normalization of $\mu_\Haar$.
\qed
\enum
\eunum
\epf


\end{document}